\documentclass{jfm}
\def\drwln#1#2{\raise 2.5pt\vbox{\hrule width #1pt height #2pt}}
\def\solid{\ \drwln{24}{1.0}\ }
\def\spc#1{\hskip #1pt}
\def\dashed{\ \hbox {\drwln{4}{1.0}\spc{2}
                   \drwln{4}{1.0}\spc{2}\drwln{4}{1.0}}\nobreak\ }
\def\dashdot{\ \hbox {\drwln{8}{1.0}\spc{2}
                   \drwln{2}{1.0}\spc{2}\drwln{8}{1.0}}\nobreak\ }
\def\dotted{\hbox {\drwln{1}{1.0}\spc{2}
                   \drwln{1}{1.0}\spc{2}\drwln{1}{1.0}}\nobreak\ }

\usepackage{graphicx}
\usepackage{epstopdf, epsfig}
\usepackage{xcolor}
\usepackage{lscape}

\def\drwln#1#2{\raise 2.5pt\vbox{\hrule width #1pt height #2pt}}
\def\solid{\ \drwln{24}{1.0}\ }
\def\spc#1{\hskip #1pt}
\def\dashed{\ \hbox {\drwln{4}{1.0}\spc{2}
                   \drwln{4}{1.0}\spc{2}\drwln{4}{1.0}}\nobreak\ }
\def\dashdot{\ \hbox {\drwln{8}{1.0}\spc{2}
                   \drwln{2}{1.0}\spc{2}\drwln{8}{1.0}}\nobreak\ }
\def\dotted{\hbox {\drwln{1}{1.0}\spc{2}
                   \drwln{1}{1.0}\spc{2}\drwln{1}{1.0}}\nobreak\ }

\newcommand{\dpr}[2] {\frac{\partial{#1}}{\partial {#2}}} 

\newcommand{\uu}[0] {$\overline{u'u'}$ }
\newcommand{\vv}[0] {$\overline{v'v'}$ }
\newcommand{\ww}[0] {$\overline{w'w'}$ }
\newcommand{\uv}[0] {$\overline{u'v'}$ }

\usepackage{todonotes}
\usepackage{pdflscape} 

\shorttitle{Effects of rotation on flow separation}
\shortauthor{B. S. Savino and W. Wu}

\title{Impact of Spanwise Rotation on Flow Separation and Recovery Behind a Bulge in Channel Flows}

\author{Benjamin S. Savino
 \and Wen Wu \corresp{\email{wu@olemiss.edu}}}

\affiliation{Department of Mechanical Engineering, University of Mississippi,
Oxford, MS 38677, USA}

\begin{document}

\maketitle

\begin{abstract}
Direct numerical simulations of spanwise-rotating turbulent channel flow with a parabolic bump on the bottom wall are employed to investigate the effects of rotation on flow separation. Four rotation rates of $Ro_b := 2\Omega H/U_b = \pm 0.42, \; \pm 1.0$ are compared with the non-rotating scenario. The mild adverse pressure gradient induced by the lee side of the bump allows for a variable 
pressure-induced separation. 
The separation region is reduced (increased)  when the bump is on the anti-cyclonic (cyclonic) side of the channel,
compared with the non-rotating separation. The total drag is reduced in all rotating cases. 
Through several mechanisms, rotation alters the onset of separation, reattachment, and wake recovery. The mean momentum deficit is found to be the key. A physical interpretation of the ratio between the system rotation and mean shear vorticity, $S:=\Omega/\Omega_s$, 
provides the mechanisms regarding stability thresholds of $S=-0.5$ and $-1$. 
The rotation effects are explained accordingly with reference to the dynamics of several flow structures.   
For anti-cyclonic separation, particularly, the interaction between the Taylor-G\"ortler vortices and hairpin vortices of wall-bounded turbulence is proven to be responsible for the breakdown of the separating shear layer. A generalized argument is made regarding the essential role of near-wall deceleration 
and resultant ejection of enhanced hairpin vortices
in destabilizing an anti-cyclonic flow.
This mechanism is anticipated to have broad impacts on other applications in analogy to rotating shear flows, such as thermal convection and boundary layers over concave walls.

\end{abstract}


\section{Introduction}

\label{sec:Intro}
\begin{figure}
    \centering
    \includegraphics[width=0.6\textwidth]{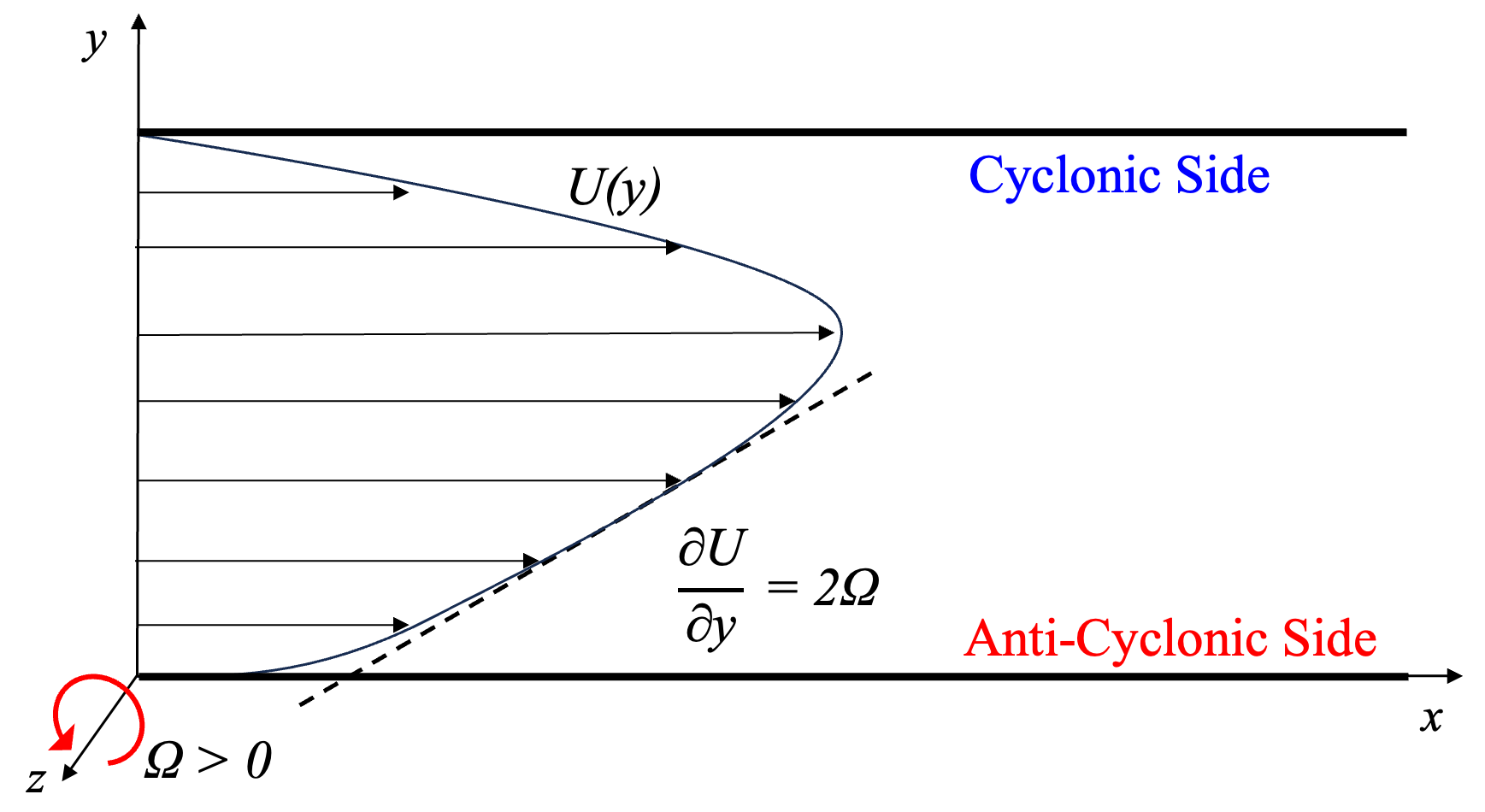}
    \caption{Sketch of spanwise-rotating channel flow. \solid mean streamwise velocity profile; \dashed $\partial U/\partial y = 2\Omega$. Anti-cyclonic and cyclonic walls for counter-clockwise (positive) rotation are marked accordingly.}    
    \label{fig:RoSketch}
\end{figure}

Rotation characterizes many turbulent flows, both in nature ({\it e.g.}, geophysical flows) and in engineering applications (\textit{e.g.}, turbines, pumps, cyclone separators, radar cooling system flows, and so on). The literature on the subject of rotating flow is quite extensive. For a channel that is rotating about its spanwise axis ($z$), the Coriolis force appears as terms $2\Omega v$ and $-2\Omega u$ in the streamwise ($x$) and wall-normal ($y$) momentum equations, respectively. $\Omega$ is the rotation rate. Its influence tends to stabilize the flow when the rotation has the same sign as the mean shear vorticity and destabilize it when the two have opposite signs. The two sides of the channel corresponding to these regions are described as anti-cyclonic (cyclonic), unstable (stable), or pressure (suction) in different studies. For reference, schematic of a rotating channel is shown in figure \ref{fig:RoSketch}.

Numerous efforts have been put in characterizing the turbulent generation/suppression mechanisms in rotating flows \citep{Johnston72, Kristoffersen93, Andersson95, Johnston98, Nakabayashi05} and modeling them~\citep{Launder87, Piomelli95, Lamballais98, Jakirlic02, Grundestam08b, Yangeetal12, Jiang18, Huang19, Zhang19}, both experimentally \citep{Johnston72, Rothe76, Alfredsson89, Nakabayashi96, Maciel03, Visscher11b} and numerically \citep{Tafti91, Kristoffersen93, Lamballais96, Nakabayashi05,Liu07, LiuLu07, Grudenstam08, Yang12, Xia16, Brethouwer17, Wu19}. 
Several key features of rotating plane channel are: 1) a linear region of constant velocity gradient, $U = 2\Omega y + C$, in the anti-cyclonic side of the channel. The scaling of $C$ has been related to the ratio between the rotation rate and the friction velocity on the anti-cyclonic wall \citep{Johnston72, Nakabayashi96, Nickels00, Nakabayashi05, Hamba06, Yang20}.
2) The formation of large-scale, streamwise-oriented roll cells that are reminiscent of Taylor-G\"ortler (TG) vortices in the constant velocity gradient region \citep{Gortler59, Tani62, Hart71, Johnston72, Alfredsson89, Nakabayashi05, Liu07, Grudenstam08, Dai16, Brethouwer17, Zhang22}. These rollers are analogous to those in thermally convective and stratified flows \citep{Hart71, Lezius76, Zhang19, Zhang22}, as well as boundary layers over concave walls \citep{Tani62, Bradshaw69, Moser87}. 
3) Turbulence is enhanced at low to moderate rotation rates on the anti-cyclonic side, corresponding to augmented hairpin vortices, however attenuated as rotation rate further increases \citep{Grudenstam08, Wallin13}. 
4) On the cyclonic side, the flow tends towards relaminarization at high rotation rate. Oblique waves and $\Lambda$-shaped vortices with turbulent spots appear on this side~\citep{Kim83, Kristoffersen93, Andersson95, Xia16, Brethouwer14, Brethouwer16, Brethouwer17}. 

\subsection{Flow separation in rotating flows}
A flow phenomenon that often occurs and has dramatic influence on the dynamics of rotating flows is flow separation. 
It can be induced by either an abrupt geometrical expansion and/or an adverse pressure gradient (APG) in the rotating flow. Flow separation itself is a complex phenomena due to the deviation from equilibrium boundary layer mechanisms. The stabilizing/destabilizing influence of the Coriolis force and the dynamic structures reviewed above may promote or delay flow separation and reattachment, leading to further complexity.  However, the interaction between rotation and flow separation is much less investigated than the plane rotating channel. 
In many rotating turbulent flows, APGs and flow separation are inevitable due to the curvature of the solid surface. Turbines and propellers used in centrifugal pumps, hydro-turbines, and impellers are just a few examples. Flow separation results in the degradation of their performance and efficiency \citep{Horlock73, Cheah07}. 
Geophysical flows that are characterized by the Coriolis force, such as atmospheric and oceanic boundary layers, also often experience separation (\textit{i.e.}, behind islands, mountains, buildings, \textit{etc.}) which alters weather patterns and oceanic currents \citep{Plate71, Heywood90, Boegman09, Omidvar20, Morgans22}. Hence, an in-depth understanding of the multiphysics interaction between rotation and flow separation is of vital importance to a variety of applications. This is the focus of the present study.

While limited in number, investigations into rotating separating flows have provided valuable insight into various flow problems regarding the interaction of the Coriolis force and separating flows. \cite{Rothe79} experimentally analyzed a spanwise-rotating turbulent backward-facing step (BFS). For rotation numbers $Ro_b := 2\Omega H/U_b = 0 - 0.15$ (where $H$ and $U_b$ are the channel height and bulk velocity upstream of the BFS), they showed that enhanced mixing caused by augmented 3D turbulent structures resulted in earlier reattachment when the step was on the anti-cyclonic side. Meanwhile when separation was cyclonic, the stabilized 2D spanwise vortices led to delayed reattachment. \cite{Barri10} performed a direct numerical simulation (DNS) of the same configuration with the separation occurring on the anti-cyclonic side, testing moderate rotation rates up to $Ro_b = 0.4$. They corroborated the findings of \cite{Rothe79} for low rotation rates, yet found that the anti-cyclonic reattachment length was not further decreased at highest rotation rates. 
Analyzing the Reynolds stresses and their budgets, they highlighted that production and redistribution of turbulent kinetic energy (TKE) in the anti-cyclonic separating shear layer (SSL) differed from the conventional SSL. 
This statistical mechanism was supported by \cite{Visscher11} in their experiments of rotating BFS up to $Ro_b = 0.8$. 
Unlike the one-side separation in BFSs, \cite{Lamballais14} used DNS to investigate a rotating sudden expansion in which both anti-cyclonic and cyclonic separation occurred simultaneously. Rotation numbers based on quantities upstream of the expansion up to $Ro_b = 1.0$ were tested.
The reattachment length on the anti-cyclonic (cyclonic) side decreased (increased) monotonically until the high rotation rates, at which the separation length on both sides plateaued, agreeing qualitatively with the hypothesis of \cite{Barri10}. 

An important metric that has been used in characterizing rotating flows is the `absolute vorticity ratio', defined as the ratio of system rotation to mean shear vorticity ($\Omega_s$), $S:={\Omega}/{\Omega_s}$.
Stability analysis of rotating constant-shear, mixing layer, and wake flows \citep{Hart71, Yanase93, Bidokhti92, Cambon94, Metias95, Salhi97, Brethouwer05} have shown that the effects of rotation can be segmented into three regimes: the destabilized regime ($-1<S<0$), the neutral-stability regime ($S=-1$), and the stabilized regime ($S<-1$ and $S>0$). Besides, $S=-0.5$ was found to be corresponded to the maximum destabilization, quantified by growth rates of TKE and three-dimensional disturbances. These regimes have been used to explain the observations in rotating channel flows~\citep{Johnston72, Tafti91, Kristoffersen93, Andersson95, Brethouwer17, Wu18}. Specifically, the linear region in the anti-cyclonic side represents where perturbations will be neutrally stable. 
Spatial variations of this region in separating rotating flows have been reported in \cite{Barri10} and \cite{Lamballais14}. The separation was found to drive $S$ less negative (\textit{i.e.}, closer to 0).

\subsection{Motivation and objectives}
The present study aims to provide further insights on the interaction between separation and rotation. The fixed-point separation in the existing geometry-induced separation studies limits the freedom of the separation point. In engineering applications, the onset of separation may be caused by an APG over a flat or mildly curved surface. In such pressure-induced flow separation, the separation is capable of changing with the flow or control condition \citep{Simpson89, You08, Ceccacci22, Wu22}. It can also show inherent natural unsteadiness \citep{Na98, Kaltenbach99, Taifour16, Wu20}. We believe that such freedom is necessary to understand the interaction between the separated shear layer (specifically its onset) and rotation. In this study, the separation will be introduced by a mildly curved bump on one of the walls of a turbulent channel to increase the variability of the separation point.

The analysis of the present flow will focus on several perspectives that have received less attention in the previous studies, aiming at revealing the mechanisms underlying the separation-rotation interaction and providing insights on the control, optimization and modeling of separating rotating flows.
First, we will provide evaluations on performance modulation, \textit{i.e.}, variation in drag, which are of great importance to engineering applications.
Second, the stability regimes have shown great success in revealing the mechanisms of turbulence modulation in rotating flows with 1D velocity gradients. However, previous studies on separating rotating flows have primarily reported the 2D distribution without establishing a connection to the spatial distribution of Reynolds stresses. 
Our investigation seeks to bridge this gap.
Third, although several types of flow structures have been identified and analyzed in conventional rotating flows, their roles in a separating setting are not well characterized nor understood. There are four key structures which have the potential to interact in a rotating separating flow: 1) TG vortices due to rotation; 2) spanwise-oriented roller vortices generated by Kelvin-Helmholtz (KH) instability in the separating shear layer \citep{Comte92, Rogers92}; 3) hairpin vortices known to exist in wall-bounded turbulence \citep{Zhou99, Adrian07}; and 4) oblique waves and associated $\Lambda$-shaped vortices present on the cyclonic side \citep{Brethouwer14, Brethouwer16}. The interactions of these structures are physically responsible for the modulation of the stability regimes and Reynolds stresses. \citeauthor{Lamballais14} examined the instantaneous structures and vortex lines to characterize these vortices. Streamwise-oriented, elongated structures were considered to be responsible for enhanced mixing across the shear layer and the observed early reattachment. On the cyclonic side, organization of two-dimensional, spanwise-oriented structures were considered as signs of stabilized SSL resistant to three-dimensional breakdown~\citep{Lamballais14}.
We aim to characterize the structural evolution of these vortices in more depth to provide physical insights on stability and variation of the turbulent statistics. 

The following manuscript is organized as follows: first, we 
describe the flow configuration and numerical methods in 
\S\ref{sec:method}. Modulation of the mean flow, separation point, and separation region is analyzed in \S\ref{sec:mean}, and associated changes in form drag and skin-friction are assessed in \S\ref{sec:perf}. Mean momentum budget is used in \S\ref{sec:mombud} to justify the effects of rotation on the onset of the separation. The spatial variation of the stability regimes in the wake of the bump and their impacts are discussed in \S\ref{sec:stabreg}. 
The reattachment of the separating shear layer and the recovery of the flow are analyzed using Reynolds stresses and their budgets in \S\ref{sec:RS}. 
Finally, the characterization of turbulent structures, their interaction, and the effect of the interaction on the observed flow behavior is performed in \S\ref{sec:struct}.

\section{Methodology} \label{sec:method}

\subsection{Simulation Configuration}
\begin{figure}
\centering
  \includegraphics[width=0.45\textwidth]{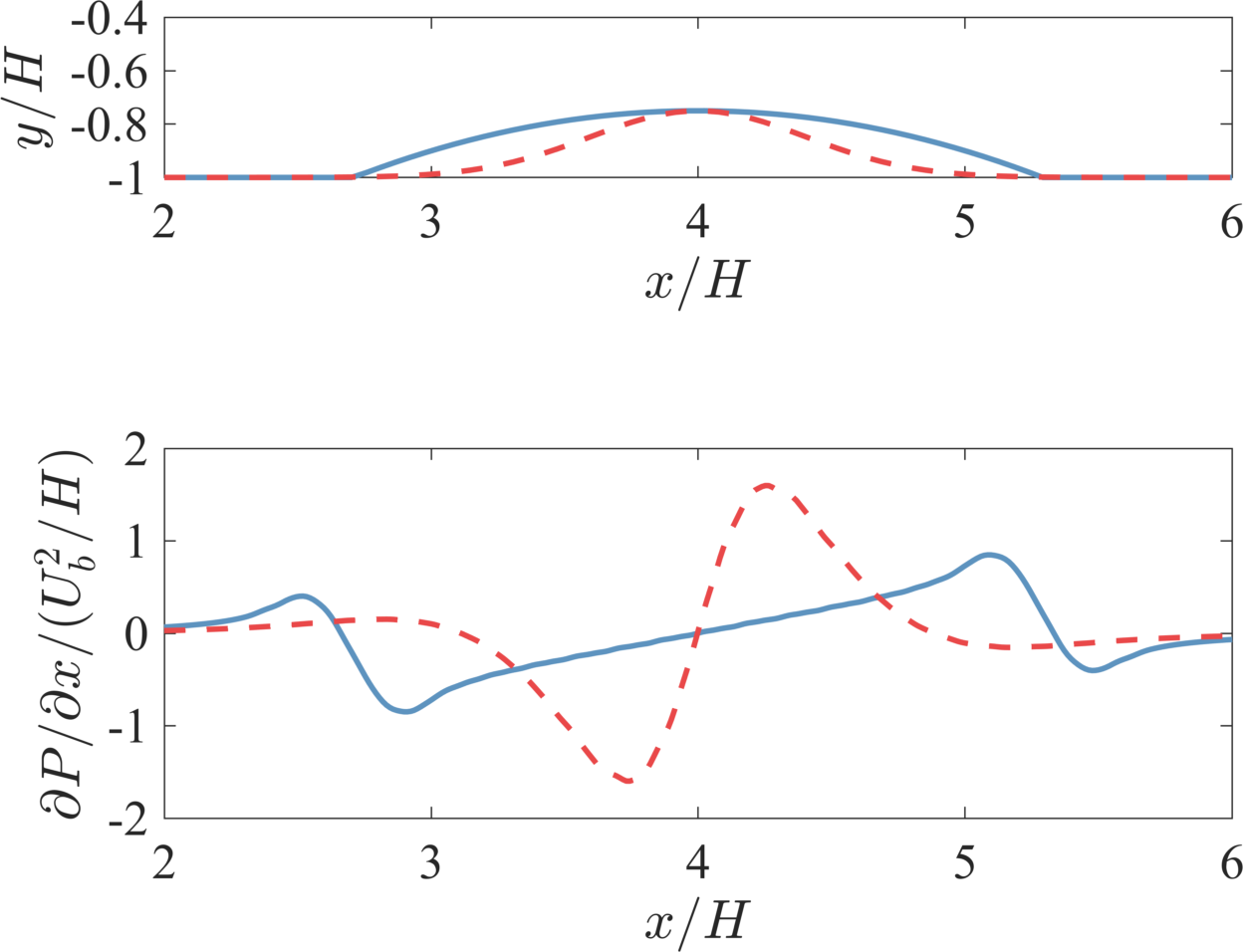}
  \caption{Bump profile (top) and inviscid streamwise pressure gradient along the surface (bottom). The profiles for the Gaussian-shaped bump used in \cite{Balin21} and \cite{Uzun22} are scaled to the same height as the current bump for comparison. \solid Current parabolic bump; \dashed Gaussian bump.}
  \label{fig:dpdx}
\end{figure}

Turbulent channel flows rotating in the spanwise direction at Reynolds number $Re_b := U_b H/\nu = 2500$ (where $H$ is the channel half height and $U_b$ the bulk velocity) are simulated by direction numerical simulation (DNS). The friction Reynolds number, $Re_\tau := u_\tau H/\nu$, is 160 when the channel is not rotating. 
A two-dimensional bump defined by the following parabolic formula (normalized by $H$):
\begin{equation}
y = \max[-a(x-4)^2 + h,0] - 1 
\end{equation}
is placed on the bottom wall of the channel at $y=-H$ (see figure \ref{fig:dpdx}\textit{a}). 
Depending on the sign of the rotation rate, this side is either anti-cyclonic ({\it i.e.}, 
$Ro_b := 2\Omega H/U_b >0$ where $\Omega$ is the rotation rate) or cyclonic ($Ro_b<0$). 
The height of the bump is set to be $h=0.25H$. The parameter $a=0.15$ yields a bump with a streamwise length of $2.58H$ along the wall (\textit{i.e.}, $x/H=[2.71,5.29]$). These dimensions are chosen for two reasons. First, the blockage in the wall-normal direction is relatively low. Compared with the Gaussian-shaped Boeing bump \citep{Balin21, Uzun22}, for example, the relatively large length-to-height ratio of the current bump has several advantages for this study: 1) the favorable pressure gradient (FPG) caused by the contraction at the windward side of the bump is mild such that flow is not relaminarized \citep{Yuan11}; 2) the APG at the aft part of the bump is mild (figure \ref{fig:dpdx}\textit{b}) such that the separation point is non-fixed. 
The second reason for using $h=0.25H$ is that the flow up to the height of the bump will be in the destabilized (stabilized) region under anti-cyclonic (cyclonic) rotation. This will be shown in \S\ref{sec:stabreg}.

Five cases corresponding to rotation number $Ro_b = 0$, $\pm0.42$ and $\pm1.0$ were performed. 
The cases are named as `P/NXX' where P or N denotes positive or negative $Ro_b$, while XX $=04$ denotes $|Ro_b|=0.42$ and XX $=10$ denotes $|Ro_b|=1.0$. The non-rotating case is denoted as case 00. Cases P04 and P10 will be referred to as the `positive-rotating' or `anti-cyclonic' cases interchangeably. Conversely, cases N04 and N10 will be referred to as `negative-rotating' or `cyclonic' cases. The simulation parameters for each case are summarized in table \ref{tab:cases}.

A computational domain of 39$H \times 2H \times 6H$ in the streamwise ($x$), wall-normal ($y$), and spanwise ($z$) directions is employed. The long computational domain in the streamwise direction is used to ensure that the channel is fully recovered near the outlet, as a streamwise-periodic boundary condition is employed. This allows for a complete examination of the recovery of the wake flow and avoidance of auxiliary boundary conditions for generating physical rotating turbulence at the inflow.

In previous studies, it is reported that a flow without spanwise rotation is well recovered after 30 times the height of an obstacle \citep{Castro79, Le97, Song00, Mollicone17}. In the rotating BFS study performed by \cite{Barri10}, they showed the mean skin-friction coefficient gradually approached a constant level but was not completely recovered by 32 step heights. The length of the domain in this study is greater than 130 bump heights. It is carefully justified that not only the first-order mean quantities but also the higher-order statistics are fully developed when the flow approaches the outflow boundary. A periodic boundary condition is also employed in the spanwise direction. For the remaining boundary conditions, the no-slip condition is enforced along the bottom and top walls, including the bump surface. 

\begin{table}
  \begin{center}
    \def~{\hphantom{0}}
    \begin{tabular}{lcccccccccccc}
      Cases & $Re_b$ & $Ro$ & Bump side & $Re_{\tau,c}$ & $Re_{\tau,u}$ & $Re_{\tau,s}$ & $Re_{\tau}$ & $Ro_{\tau}$ & $x_\textrm{sep}$ & $y_\textrm{sep}$   & $L_\textrm{sep}$ & $F_D$ \\[3pt]
      00       & 2500 & 0     & -        & 213 & 160 & 159 & 160 & 0    & 4.61 & 0.20 &  1.71  & 0.37\\
      P04     & 2500 & 0.42  & anti-cyclonic & 205 & 191 & 117 & 159 & 6.6 & 4.78 & 0.16 &  0.85 & 0.34\\   
      P10     & 2500 & 1.0   & anti-cyclonic & 181 & 150 & 106 & 130 & 19.2 & 4.97 & 0.11 &  0.44 & 0.23\\   
      N04     & 2500 & -0.42 & cyclonic   & 211 & 193 & 118 & 160 & 6.6 & 4.53 & 0.21 &  3.70 & 0.36\\  
      N10     & 2500 & -1.0  & cyclonic   & 192 & 152 & 105 & 132 & 19.0 & 4.53 & 0.21 &  4.73 & 0.25\\  
      \end{tabular}
   \caption{Simulation parameters. $Re_{\tau,c}$, friction Reynolds number at the bump crest. $Re_{\tau,u}$ ($Re_{\tau,s}$), the friction Reynolds number for the anti-cyclonic (cyclonic) side of the channel. $Re_{\tau}$, the friction Reynolds number in the fully recovered channel section. $Ro_{\tau} := 2\Omega H/u_\tau$, the friction rotation number in the fully recovered channel section. $x_\textrm{sep}$, streamwise location of mean separation point. $y_\textrm{sep}$, wall-normal location of mean separation point (relative to bottom wall). $L_\textrm{sep}$, streamwise length of mean separation bubble. $F_D$, total mean drag per unit span over the entire channel.}
    \label{tab:cases}
  \end{center}
\end{table}

\subsection{Numerical Methods}
Incompressible Navier-Stokes equations for a Newtonian fluid, non-dimensionalized by $U_b$ and $H$:
\begin{equation}
\frac{\partial u_i}{\partial x_i} = 0, 
\label{eq:ns1}
\end{equation}

\begin{equation}
\frac{\partial u_i}{\partial t} + \frac{\partial}{\partial x_j}(u_ju_i) = -\frac{\partial p}{\partial x_i} + \frac{1}{Re_b}\nabla^2u_i - Ro_b \epsilon_{i3k} u_k + f_i - \delta_{i1}
\frac{\Delta P}{L},
\label{eq:ns2}
\end{equation}
are solved by DNS. Here, $-\Delta P/L$ is the mean pressure gradient required to drive the flow. $p$ is the modified pressure. $Ro_b$ is the bulk rotation number. The term $f_i$ is used to enforce the no-slip boundary conditions on the bump, which is achieved by an immersed boundary method based on a volume-of-fluid (VOF) approach \citep{Peskin72, Scotti06}. The fraction of cell volume that is occupied by the fluid, denoted as $\phi$, is calculated analytically in a pre-processing calculation using the simulation grid. During the simulation, the velocity in the cells that are occupied partially or fully by the solid is weighted by $\phi$ through term $f_i$. This method has been extensively used~\citep{Scotti06, Yuan14, Yuan14b, Wu16, Wu18, Wu19, Savino23} in the investigations of flow around embedded objects. 

A Cartesian grid is designed such that the bump surface and the wake of the bump are well resolved. The grid is uniform in the $x$ direction around the bump and in the far wake (\textit{i.e.}, $\Delta x/H = 0.011$ for $x/H=[2.09,6.68]$ and $\Delta x/H = 0.053$ for $x=[12, 39]$), while being stretched between $x/H = [0, 2.09]$ and $x/H =[6.68, 12]$ to transition from the two uniform spacings. The grid is also uniform in the $y$ direction below the crest with a hyperbolic-tangent stretching towards the centerline. The grid is uniform in the $z$ direction. 
The maximum stretch ratio is less than 3\% in all directions.

\begin{figure}
\centering
  \includegraphics[width=0.7\textwidth]{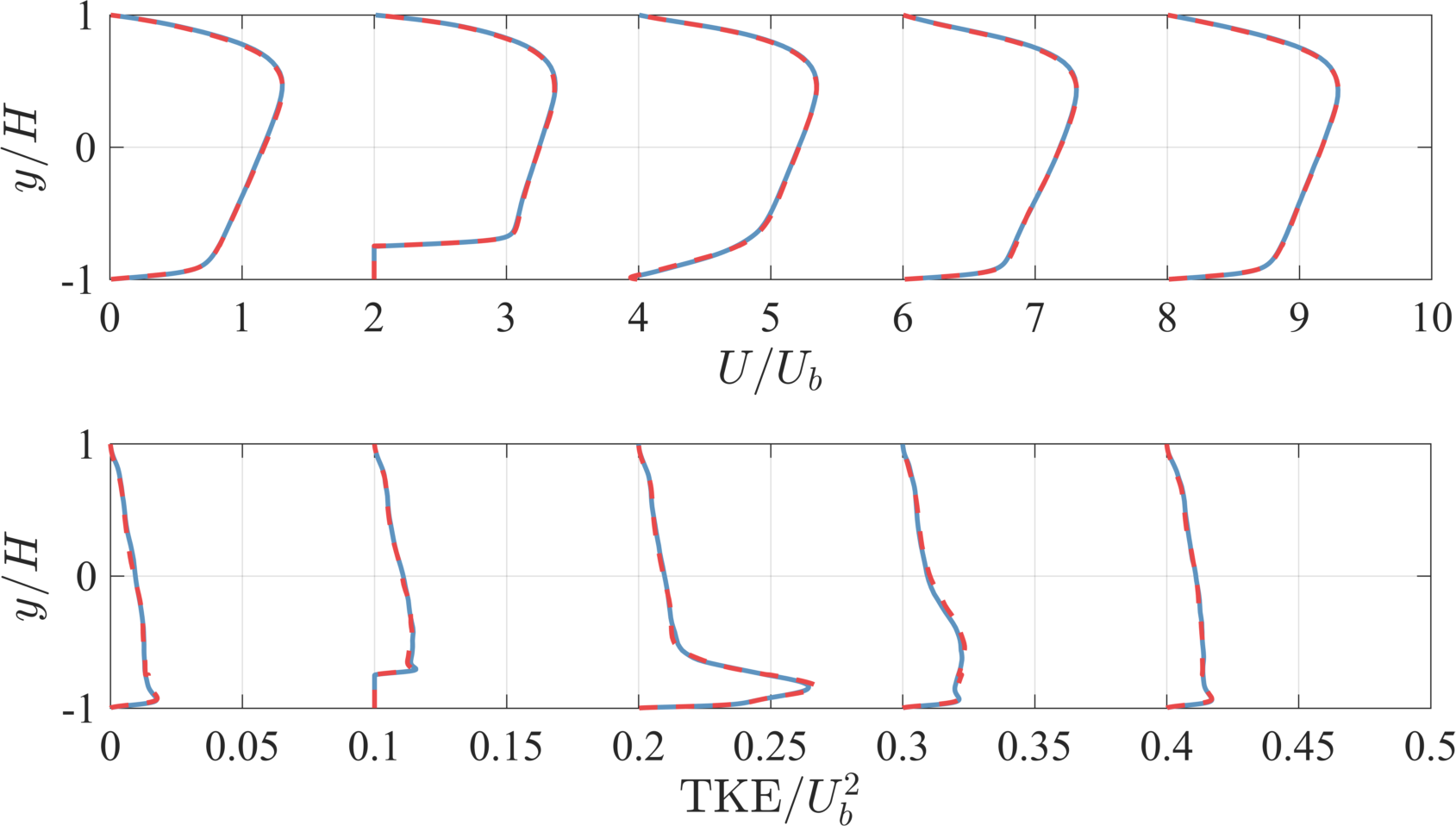}
  \caption{Comparison of mean streamwise velocity and turbulent kinetic energy (TKE) at $x/H=2, 4, 5.5, 8$ and 20, Case P04. \solid Grid I (coarser one); \dashed Grid II (finer one). Each profile is shifted to the right by 2 units for $U$ and 0.1 units for TKE for clarity. }
  \label{fig:gridconv}
\end{figure}

Two grids are tested for grid convergence. The first grid, denoted as Grid I, consists of $1196 \times 192 \times 184$ grid points in the streamwise, wall-normal, and spanwise directions, respectively. For all the five cases using Grid I, the maximum $\Delta x^+$ and $\Delta z^+$ near the two walls are 9.3 and 6.5. Away from the wall, Grid I gives $\Delta h/\eta \le 6$, which is much smaller than the length scale at which the maximum dissipation occurs: $24\eta$ \citep{Pope2000}. Thus, Grid I is expected to be capable of resolving a substantial portion of the dissipation spectrum. Nevertheless, a finer grid (Grid II) is employed for the P04 case to verify the grid convergence. This case is chosen because the moderate rotation rate leads to the greatest increase in turbulence intensity on the anti-cyclonic wall. Grid II consists of $1584\times239\times288$ grid points in the $x$, $y$ and $z$  directions, which corresponds to a refinement factor of $32\% \times 25\% \times 57\%$. The mean streamwise velocity and turbulent kinetic energy are compared at selected streamwise locations in Fig. \ref{fig:gridconv}. The agreement indicates that the solutions on both grids are grid-independent. In the following only the results calculated using Grid I are shown.

The equations of motion are solved using a well-validated finite difference code \citep{Keating04, Wu16, Wu18, Wu19, Savino23b, Wu24} which solves equations (\ref{eq:ns1}) and (\ref{eq:ns2}) on a staggered grid. The code is second-order accurate both in time and space: a second-order accurate central differencing scheme is used for all spatial derivatives. A second-order accurate semi-implicit time advancement method is employed in which the Crank–Nicolson scheme is used for the wall-normal diffusion terms, while the Adams–Bashforth scheme is applied to all remaining terms. Tthe Poisson equation is solved directly via a pseudo-spectral method \citep{Moin2010} using the blktri matrix solver from the FISH-PACK software library in which a generalized cyclic reduction algorithm is employed~\citep{Sweet74, Swarztrauber79}. The code is parallelized using the message-passing interface (MPI) protocol.  After each case reaches its statistically steady state, the three-dimensional flow field data is saved at time interval $\delta t = 5H/U_b$ over a total of $400H/U_b$ (1200$h/U_b$) for statistical averaging. Two-dimensional slides of the flow field are also sampled at several planes every 115 time steps ($\sim0.17H/U_b$), and history profiles of the velocity are monitored at every time step at selected locations. Statistical averages are performed in time and over the homogeneous spanwise direction. The mean quantities are denoted by capital letters or by operator $\overline{(.)}$. The superscript $(.)^+$ denotes quantities scaled with the wall units. Results in the near wake of the bump will be focused on and the region far downstream will not be shown unless necessary. 

\subsection{Validation}
\begin{figure}
\centering
  \includegraphics[width=0.49\textwidth]{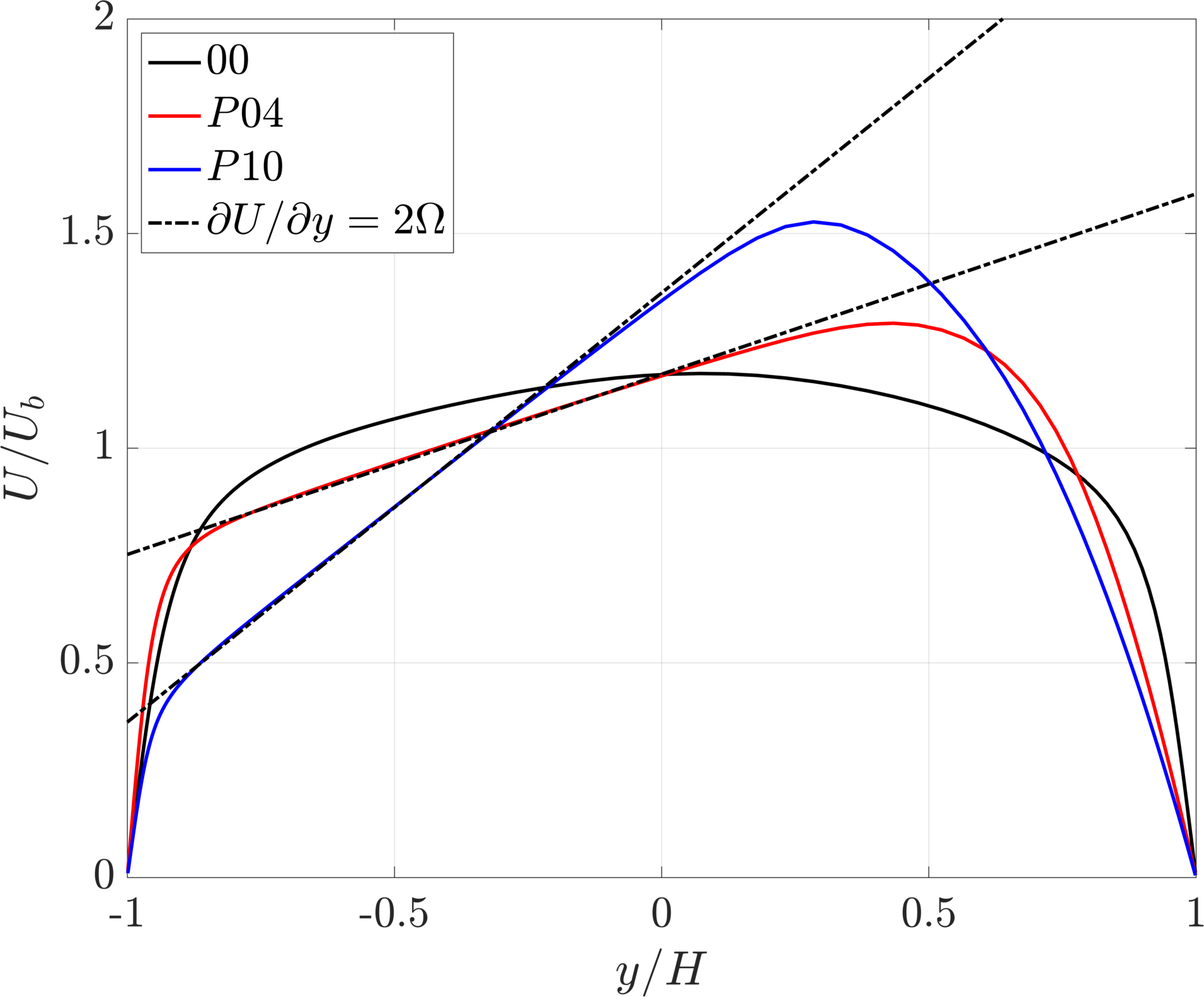}
  \includegraphics[width=0.49\textwidth]{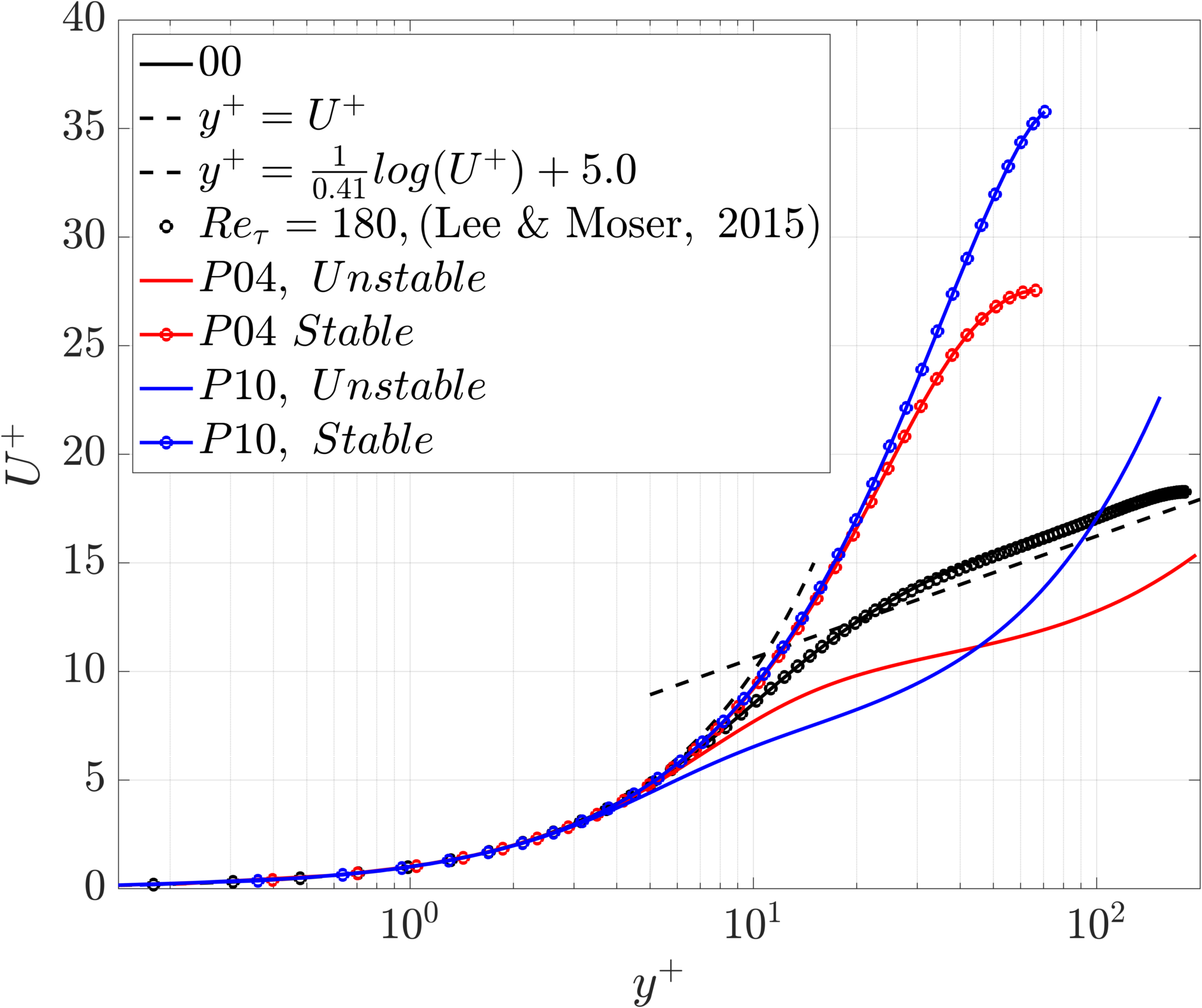}
  \caption{(Left) Mean streamwise velocity scaled with outer units. Dashed lines representing $\partial U/\partial y$ are shown for each case. (Right) Mean streamwise velocity scaled with inner units. Note for the rotating cases, the anti-cyclonic and cyclonic walls are scaled with their respective friction velocity. Case 00 is compared with DNS data from \cite{Lee15}. In both panels, profiles are averaged in the streamwise direction in the range $x/H\in [28,31]$.}
  \label{fig:validation}
\end{figure}

The convergence of the statistics is ensured by checking that the mean velocity and Reynolds stresses obtained using only half of the sample are within 1\% of that calculated using the entire sample (analysis not shown for brevity). 
The mean velocity profiles in the fully recovered region ($x\in[28,31]$) are shown in figure \ref{fig:validation}. The left panel shows the mean streamwise velocity scaled in outer units. As shown by the dot-dashed lines, the expected region of $\partial U/\partial y = 2\Omega$ is captured for both moderate and high rotation rates. The right panel shows the mean streamwise velocity scaled with inner units. Case 00 is compared with the DNS of \cite{Lee15} at $Re_{\tau} = 180$. Despite slight differences in Reynolds number, the data collapses well.  Note that $Re_{\tau}$ differs on the anti-cyclonic and cyclonic walls for the rotating cases (see table \ref{tab:cases}). Thus, the top and bottom portions of the channel are normalized with $Re_{\tau,s}$ and $Re_{\tau,u}$, respectively. The near-wall flow collapses on the linear law of the wall for the viscous sub-layer, indicating the sufficient resolution of near-wall turbulence regardless of modulation in turbulence intensity. Additionally, in the outer layer, the anti-cyclonic side displays a down-shift compared to the canonical logarithmic relation, while the cyclonic side displays an parabolic laminar profile. This is consistent with the behavior observed in the studies of \cite{Watmuff85, Tafti91, Kristoffersen93, Wu19}.

\section{Mean flow modulation: separation bubble}
\label{sec:mean}
\begin{figure}
    \centering
    \includegraphics[width=0.8\textwidth]{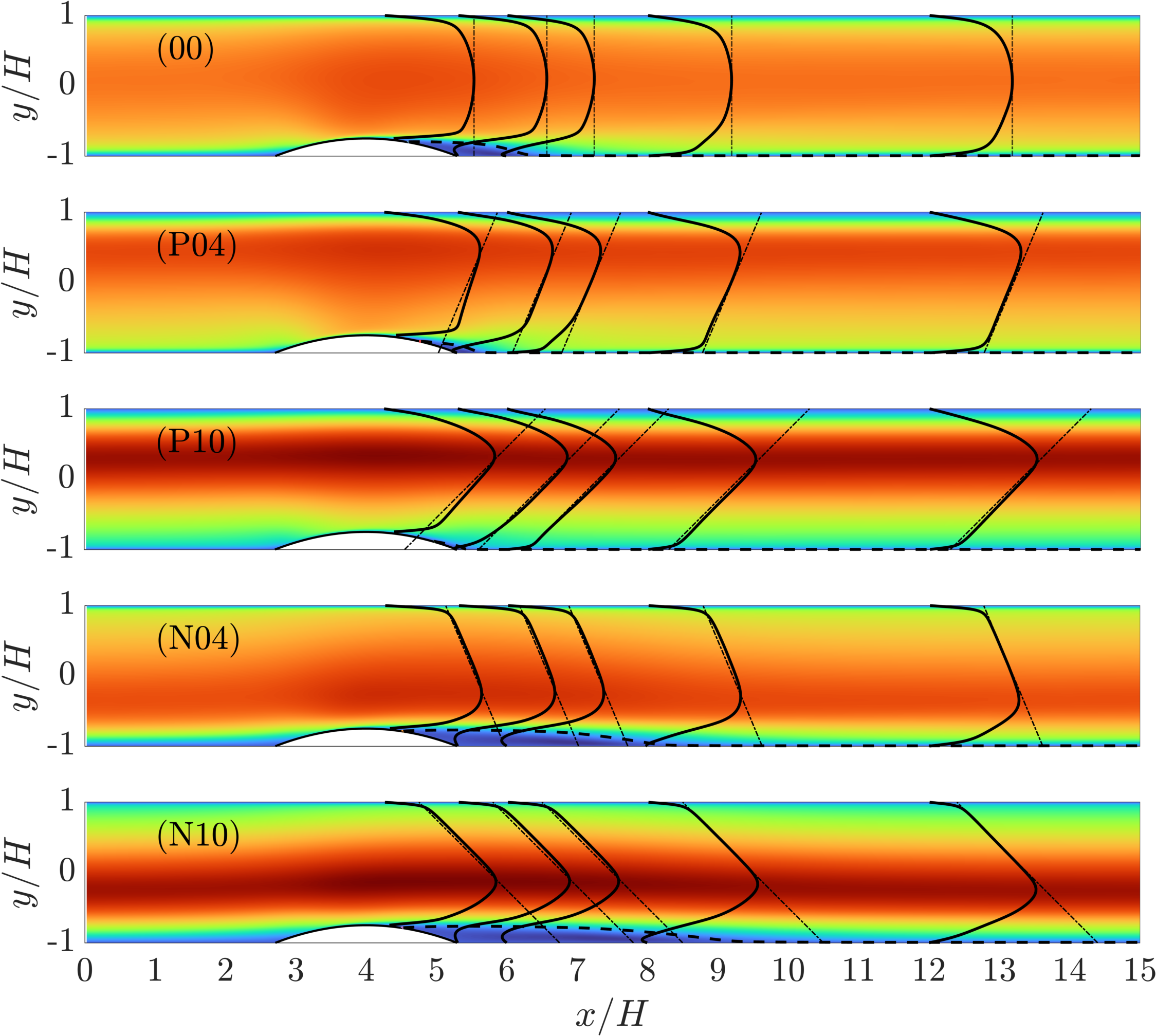}
    \includegraphics[width=0.2\textwidth]{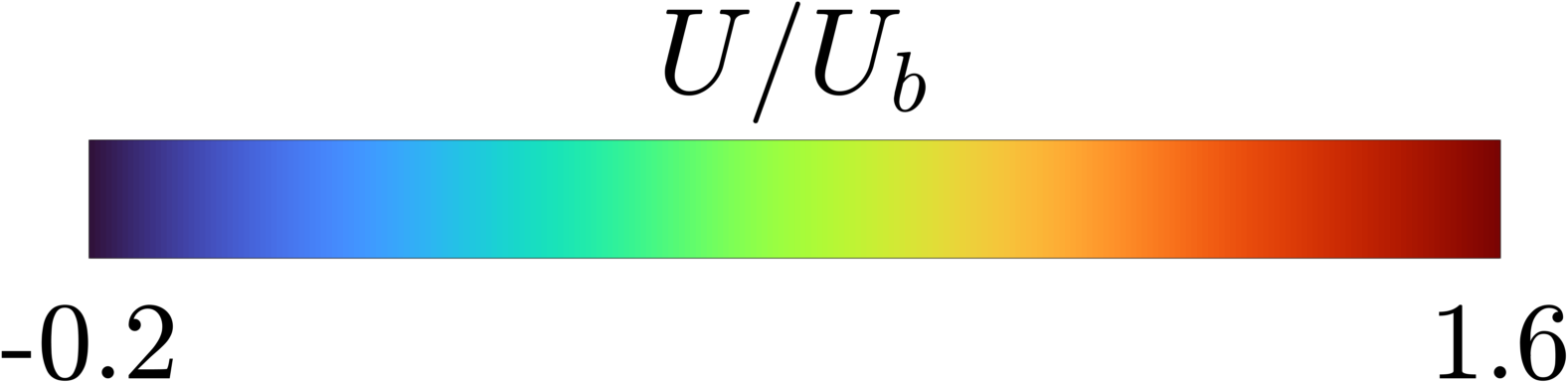}
    \caption{Mean streamwise velocity contours. \solid Mean streamwise velocity profiles at $x/H = $4, 5.5, 6, 8, and 12; \dashed separating streamlines; \dashdot velocity gradient ($\partial U/\partial y$) corresponding to $2\Omega$. From top to bottom: $Ro_b=0,0.42,1.0,-0.42,-1.0$. }
    \label{fig:Ucont}
\end{figure}

Mean streamwise velocity contours are shown in figure \ref{fig:Ucont}. Mean separating streamlines, velocity profiles at select streamwise locations ($x/H = 4.0, 5.5, 6.0, 8.0$ and  $12.0$), and linear plots corresponding to $\partial U / \partial y = 2\Omega$ are superimposed. The separation (reattachment) point is defined as the streamwise location on the bump or wall where $C_f = 0$ and $\partial C_f / \partial x < 0$ ($\partial C_f / \partial x > 0$).
When subject to rotation, the velocity profiles display asymmetry about the centerline of the channel with the development of a clear linear region. As expected, the peak velocity shifts nearer to the cyclonic side of the channel, and the magnitude of the peak velocity increases with the rotation rate. The linear region ($\partial U/\partial y \approx 2\Omega$) develops on the respective anit-cyclonic side of the channel. For the positive rotating cases, the greatest deviation of the linear region occurs at the bump crest.
 For the negative rotating cases, the bump is exposed to quasi-laminar flow represented by the parabolic velocity profile. The constant velocity gradient region now on the opposite side of the bump appears minimally affected by the bump and separation region.

\begin{figure}
    \centering
    \includegraphics[width=0.99\textwidth]{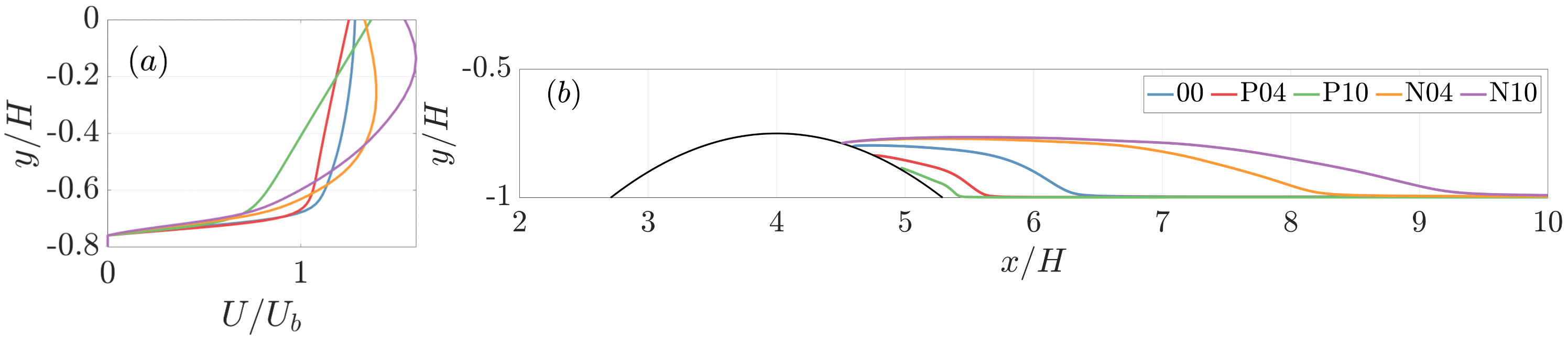}
    \caption{(\textit{a}) Near-wall mean streamwise velocity profiles upstream of separation ($x=4.25H$). Note the wall is located at $y=-0.75H$ at this streamwise location. (\textit{b}) Mean separating streamlines plotted from separation point. The $y$-axis is stretched by a factor of 2 for clarity. Note the difference between $y$-axis scales in panels \textit{a} and \textit{b}. The legend correlating line color and case is given in panel \textit{b}.}
    \label{fig:SepSL}
\end{figure}

The skewed velocity profiles indicate that the mean flow which is subjected to the APG and separation differs between cases. A reduced velocity, \textit{i.e.}, mean momentum deficit (MDD), in the vicinity of the bump in the rotating cases is observed when compared to case 00. This is evident in observing the near-wall region of the velocity profiles at $x=4.25H$ (prior to separation) as shown figure \ref{fig:SepSL}\textit{a}. Because separation occurs when the near-wall fluid is decelerated to zero velocity, the increased MDD indicates the rotating cases may separate earlier than case 00. 
However, our data shows that this intuitive assumption is not sufficient for predicting the deceleration, separation onset, and separation size. As clearly observable in figure \ref{fig:Ucont}, when on the anti-cyclonic side, the separation region experiences a notable reduction in size, accompanied by a decrease in the magnitude of reverse flow as compared to the non-rotating separation bubble. Conversely, when the separation bubble is on the cyclonic side, both the size of the bubble and the magnitude of the reverse flow exhibit an increase compared to the non-rotating case. These trends exhibit monotonic behavior as the rotation rate increases. 

The changes at the lower rotation rate are consistent with \cite{Barri10}, \cite{Visscher11}, and \cite{Lamballais14}. However at higher rotation rates, \cite{Lamballais14} reported a non-monotonic behavior in that, compared with moderate rotation rates, the size of the separation bubble increases on the anti-cyclonic side and decreases on the cyclonic side. A significant discrepancy between \cite{Lamballais14} and the current work is that the separation in their sudden expansion channel was triggered by the fixed corner of the expansion. 
Therefore, the inertia of the incoming fluid dictated a minimum separation region size despite modulation in flow conditions due to rotation. In the current study, the mild surface curvature (without an abrupt geometric change) and associated mild APG allow for a variable separation point, resulting in increased freedom of the separation region to be affected by the rotation. Therefore, the suppression of the separation region in cases P04 and P10 of this study represents the effect of anti-cyclonic rotation on pressure-induced separation.

To assess how the variable separation point is affected by rotation, the mean separating streamlines of the five cases are superposed in figure \ref{fig:SepSL}\textit{b}. For reference, separation point, reattachment point, and separation bubble length are listed in table \ref{tab:cases}. Compared with the non-rotating case, cases P04 and P10 show a delayed onset of separation. Conversely, the flows in cases N04 and N10 separate earlier. When subject to positive rotation, the behavior is monotonic: as the rotation rate increases, the separation point is further delayed.  When subject to negative rotation, however, the separation point does not move further towards the bump crest.
The observed changes in separation point challenge the intuitive assumption that separation onset is directly correlated to MMD, a metric that is often used for pre-separation characterization \citep{Simpson89, Greenblatt00}. The flows in question have varying MMDs which do not indicate the separation location, and thus a single parameter alone is not sufficient to determine the behavior of the separation point. As shown in figure \ref{fig:SepSL}, case P10 displays the largest MMD yet separates the latest. The MMD of case N10 is slightly greater than that of case N04, however their separation points are identical. 

The reattachment point also varies between cases, as shown in figure \ref{fig:SepSL}\textit{b}. The flow reattaches earlier when subject to positive rotation and later when subject to negative rotation. These behaviors are monotonic with an increasing rotation rate. Despite that it appears earlier separation is correlated with later reattachment (and vice versa), this is not the sole relationship; the observed variation of the reattachment point with the rotation rate is present despite separation occurring at the same location in the negative rotating cases. The reattachment behavior differs from the existing literature at high rotation rates, such as \cite{Lamballais14} who reported later (earlier) reattachment when subject to positive (negative) rotation at $Ro_b = 1.0$ compared to $Ro_b = 0.33$.

The preceding observations show that despite an MMD in all rotating cases due to the skewed velocity distributions, this change alone is not sufficient to predict separation and reattachment. Other mechanisms play important roles as well.
When comparing flows with the same MMD, for example, turbulent mixing is often used to justify changes in separation points. Additionally, using MMD to predict separation implicitly assumes that the streamwise pressure gradient ($\partial P/\partial x$) does not change with the separation region, yet this may only be applicable to external flows. In internal flows, the pressure gradient is inevitably altered with the separation through inviscid coupling.
We will further investigate the onset of separation in \S \ref{sec:mombud} and reattachment in \S \ref{sec:RS} and \S \ref{sec:struct}. Before identifying the responsible mechanisms, we first discuss the performance changes of practical relevance to engineering applications.

\section{Performance metrics} \label{sec:perf}
\subsection{Total drag} 
Using the volume-of-fluid (VOF) immersed-boundary method, the force exerted on the fluid to enforce the no-slip condition by the bump is calculated during every time iteration. This force contains both the frictional and form drag components produced by the bump. Integrating this force along with the wall-shear stress on the planar walls provides the total drag of the channel. The conventional notion regarding the relation between flow separation and form drag is that larger separation implies larger form drag, and vice versa. The results below show this intuition can be misleading.

\begin{figure}
    \centering
    \includegraphics[width=0.7\textwidth]{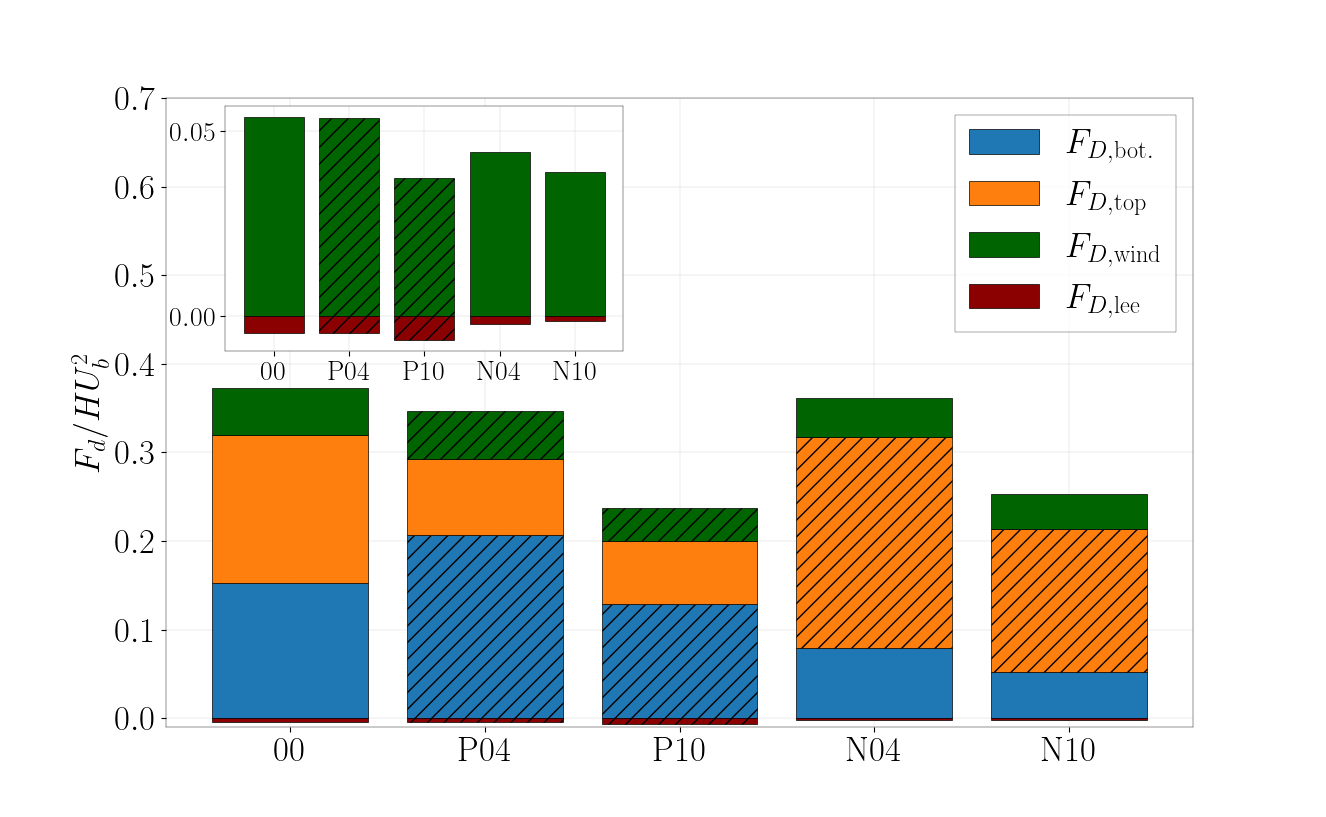}
    \caption{Total drag per unit span. The bars are split into drag produced on the bottom wall ($F_{D,\mathrm{bot.}}$), top wall ($F_{D,\mathrm{top}}$), wind ($F_{D,\mathrm{wind}}$), and lee ($F_{D,\mathrm{lee}}$) sides of the bump, as shown in the legend. The inset compares the drag produced by the wind and lee sides of the bump. In all plots, hatched regions represent drag produced on the anti-cyclonic side of the channel.}
    \label{fig:totalDrag}
\end{figure}

The total drag per unit span is compared in figure \ref{fig:totalDrag}, decomposed into four sources: the friction drag produced by the bottom wall (excluding the bump), 
$F_{D,\mathrm{bot.}} = \int \tau_w(x,-H) dx$; 
the friction drag produced by the top wall, 
$F_{D,\mathrm{top}} = \int \tau_w(x,H) dx$; 
the drag produced on the wind side of the bump, 
$F_{D,\mathrm{wind}} = \int_{-H}^{H}\int_{\mathrm{LE}}^{x_c} F_1 dx dy$; 
and the drag produced on the lee side of the bump, 
$F_{D,\mathrm{lee}} = \int_{-H}^{H}\int_{x_c}^{\mathrm{TE}} F_1 dx dy$. LE, TE, and $x_c$ correspond to the streamwise locations of the bump leading edge, trailing edge, and crest, respectively. $F_1$ is the mean IBM force from Eq. (\ref{eq:ns2}) in the streamwise direction. For cases P04 and P10, the bottom wall skin friction (blue bar) and the drag produced by the bump (green and red bars) sum to the total drag on the anti-cyclonic side (hatched regions). In cases N04 and N10, the friction on the top wall (orange, hatched bar) is solely responsible for anti-cyclonic side drag production. Note that the total drag (sum of these four terms) is given in table \ref{tab:cases}.

All rotating cases exhibit a decrease in total drag compared to case 00. In the current configuration, the modulation of skin friction along the walls appears to contribute most significantly to the change in total drag, given that the length of the channel is considerably longer than the bump (the bump occupies $7\%$ of a single wall). The primary contributor to the total drag decrease is the reduction in skin friction along the cyclonic wall (orange bars for cases P04 and P10, blue bars for cases N04 and N10), a consequence of diminished turbulence intensity due to relaminarization by the cyclonic rotation. Conversely, the skin friction on the anti-cyclonic wall changes non-monotonically with $Ro_b$ (hatched bars). At the moderate rotation rate (P04 and N04), skin friction on the anti-cyclonic wall is increased, counteracting the drag reduction on the cyclonic side and thus leading to a minor net reduction in total drag. At the high rotation rate (P10 and N10), the skin friction on the anti-cyclonic wall is lower than the non-rotating case, resulting in a drastic total drag decrease together with the laminar opposite side. The suppression of turbulence and the associated drag over the anti-cyclonic wall at high rotation rates is consistent with the observations in the attached rotating flow studies of \cite{Johnston72} and \cite{Brethouwer17}. 

Compared to the total drag differences, the drag produced solely by the bump differs little between all cases despite the significant change in the size of the separation region. The bump-produced drag is dominated by the positive wind-side force (drag, green bars) compared to the negative lee-side force (thrust, red bars). This behavior is better illustrated by the inset of figure \ref{fig:totalDrag}, where the bump-produced forces are directly compared. The predominant change by rotation is the decrease in drag on the wind side of the bump rather than the thrust on the lee side. 
It is found that the decrease in wind side drag follows the same trend as the incoming mean momentum deficit. Phenomenological, this is expected, as a higher (lower) incoming velocity will result in a larger (smaller) increase of the stagnation pressure at the wind side of the bump.  
Case 00 displays the largest incoming velocity and, consequently, the largest wind side drag. Case P10, with the lowest incoming velocity, produces the least wind side drag. 
The negative force on the lee side 
is often considered as a `back-pressure' whose recovery depends on the size of the separation region. That is, a smaller recirculation region is expected to result in a greater back-pressure (\textit{i.e.}, more negative). Our results support this trend in general, with the exception being that case P04 has a similar $F_{D,\mathrm{lee}}$ to case 00 yet a smaller separation region. The variation of back-pressure between the four rotating cases, nevertheless, is small compared with that of the wind-side drag, and the latter remains the dominant contributing force. Therefore, the variation of separation size contributes little to the change in bump-produced drag.
These observations indicate that the size of the separation region is not a proper indicator of the drag, even locally around the bump, at least for the current rotating configuration in question.	

\subsection{Skin friction on the channel walls}
\begin{figure}
    \centering
    \includegraphics[width=1.0\textwidth]{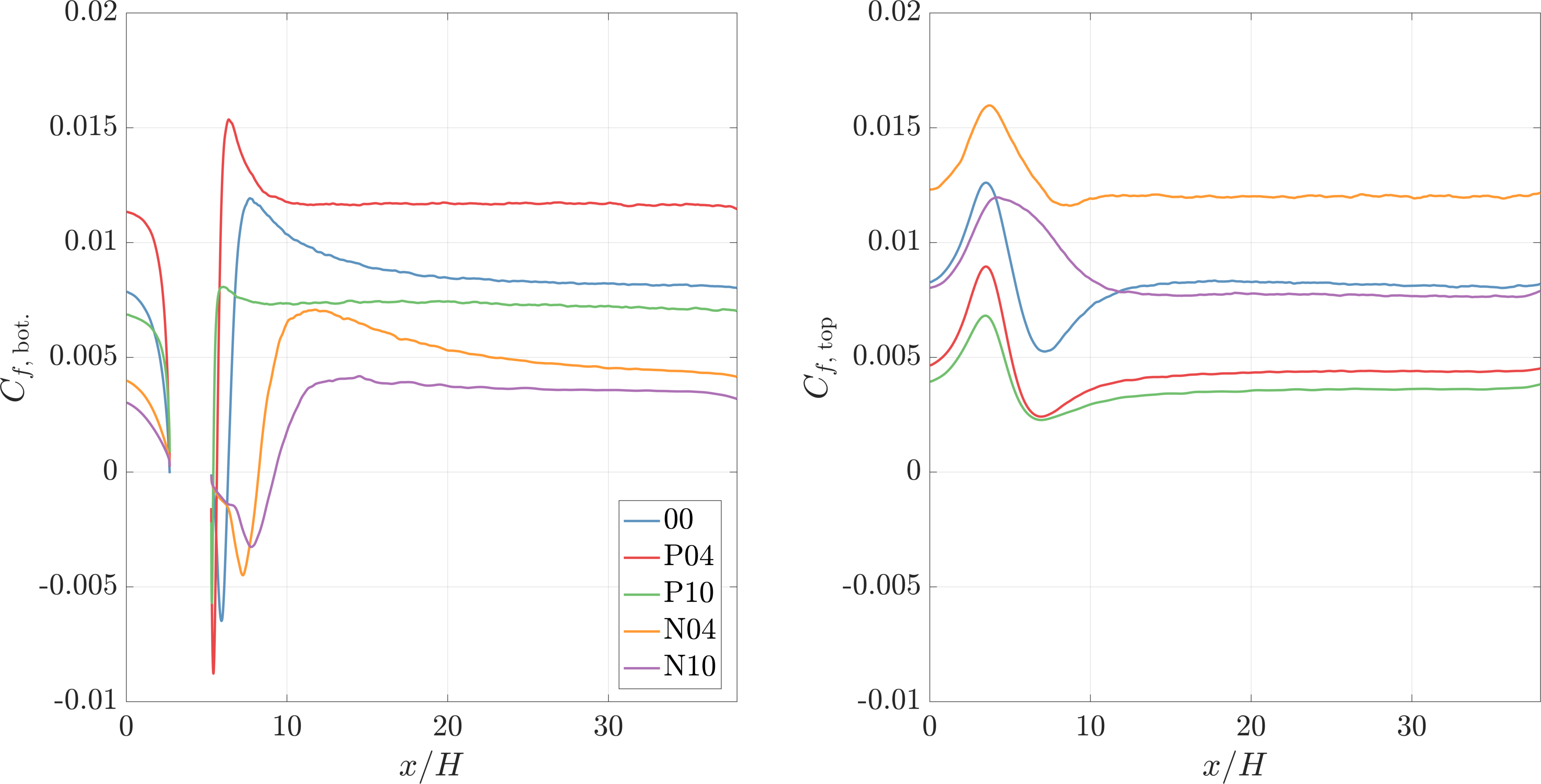}
    \caption{Skin friction coefficient along the bottom wall (left) and top wall (right). The skin friction over the surface of the bump is excluded from the plot.}
    \label{fig:cf}
\end{figure}

Because the skin friction on the channel walls is found to be the dominant term contributing to total drag and its variation with rotation, we now further examine its streamwise evolution associated with the separation region, wake, and flow recovery. The mean skin friction coefficient ($C_f = 2\tau_w/U_b^2$) along the bottom and top walls are shown in figure \ref{fig:cf}. Skin friction on the bump is excluded from the bottom wall drag as in the previous section. Integrating $C_f$ along $x$ on both walls shows that the accumulated skin friction exceeds the drag produced by the bump 6--7$H$ downstream of the bump (or $\sim25$ bump heights). Therefore, unless the wall extends only over such as short distance downstream of the protrusion in physical applications, the skin friction along it during the prolonged recovery of the wake would remain the main source of total drag. This is typical, for example, when the protrusion is located near the leading edge of a turbine.

When the flow is non-rotating (case 00), the skin friction along the bottom wall displays the expected behavior of a separation bubble and reattaching flow. Specifically, a negative peak prior to reattachment signifies strong reverse flow in the recirculation region while a positive peak post reattachment indicates strong forward flow. The latter is characteristic of impinging-type reattachment in which a strong mean downwash and/or rapidly decaying roller vortices strike the surface, as discussed in \cite{Le97} and \cite{Na98}. The steep separating streamline of case 00 near the reattachment point (refer to figure \ref{fig:SepSL}\textit{b}) supports this claim. The recovery of $C_f$ on the bottom wall takes $\sim 20 H$ (80 bump heights). On the top wall, $C_f$ shows a variation due to the change of flow area by the bump. It reaches its peak at $x=4H$ where the crest of the bump is and its minimum at $x \approx 7H$, slightly downstream of the reattachment point, indicating that the `dead fluid zone' within the separation bubble \citep{Na98} reduces the effective flow area of the channel. Following the minimum, the flow gradually recovers, attaining the recovered $C_f$ value $\sim 15H$ (60 bump heights) downstream of the reattachment point.

When subjected to rotation, the wake of the bump on the bottom wall exhibits a similar negative-positive-peak pattern near the reattachment region as the non-rotating case. Positive rotation leads to a shorter transition between the two peaks due to the reduced separation region and rapid recovery to a constant $C_f$ within a few $H$ of the bump trailing edge. 
Note that the reduction of the recovery length is much more significant than that of the separation region. This indicates that rotation not only modulates the mean separation region but also the reattached flow. 
The non-monotonic change of the skin friction on the anti-cyclonic side in figure \ref{fig:totalDrag} is quantified in figure \ref{fig:cf} as a significant plateau at a higher (lower) $C_f$ is attained for case P04 (P10) than that of case 00. This is consistent with \cite{Brethouwer17} (among others) who notes a reduction in skin friction on the anti-cyclonic wall beyond $Ro\approx0.45$.
Negative rotation, on the other hand, attenuates the peaks near reattachment. This indicates that the reattachment at the end of the long separation bubbles of cases N04 and N10 is characterized by diffusion and mild impingement compared to the former cases. As shown in figure \ref{fig:SepSL}\textit{b}, the separating streamlines indeed exhibit more mild curvature than the positive-rotating cases. Comparing cases N04 and N10, the positive peak after reattachment is negligible at the higher rotation rate, thus $C_f$ reaches its asymptomatic value shortly downstream of reattachment. In case N04, conversely, the recovery of $C_f$ is significantly slower such that the wake persists downstream until $x\approx32H$. 
Along the top, planar wall, the streamwise variation of $C_f$ due to the bump blockage remains similar among most cases, with a shift in the asymptotic value which $C_f$ recovers to in each case. Again, $C_f$ is reduced (increased) when the top wall is the anti-cyclonic (cyclonic) wall during rotation. In the study of a rotating backward-facing step by \cite{Barri10}, an additional laminar separation bubble is observed on the cyclonic, planar wall opposite of the step. This does not occur in the positive-rotating cases with the current configuration. Rather, $C_f$ remains positive at its minimum when the flow mildly expands over the bump and the separating shear layer.

\section{Onset of separation: mean momentum budget} \label{sec:mombud}
As previously discussed, separation cannot be characterized by mean momentum deficit of the incoming flow alone, and rather could be influenced by several factors, including turbulent mixing, the adverse pressure gradient, Coriolis effects, etc. We quantify these factors here using the mean momentum budget to gain insights into the cause of the difference of mean separation point among the five cases. From a phenomenological perspective, one can anticipate the mechanisms amplifying acceleration in the streamwise will delay separation. Conversely, additional deceleration to that imposed by the adverse pressure gradient (APG) is likely to promote early separation. The mean momentum equation, considering a homogeneous span, reads:
\begin{equation}
    0 =-\frac{\partial P}{\partial x_i} + A_i + D_i + R_i + G_i  + F_i\; ; \; i=1,2
    \label{eq:meanmom}
\end{equation}
The terms in the right-hand-side are the mean pressure gradient ($-\partial P/\partial x_i$), mean convection ($A_i$) which arises from the material derivative of the mean velocity, mean viscous diffusion ($D_i$), mean Reynolds stress divergence ($R_i$), mean Coriolis force ($G_i$), and mean IBM force ($F_i$). Individual term definitions are provided in the Appendix. Notice that the Coriolis terms are proportional to the mean velocity:
\begin{equation}
    G_x = 2\Omega V,\
    G_y = -2\Omega U.
    \label{eq:G}
\end{equation}

\begin{figure}
    \centering
    \includegraphics[width=1.0\textwidth]{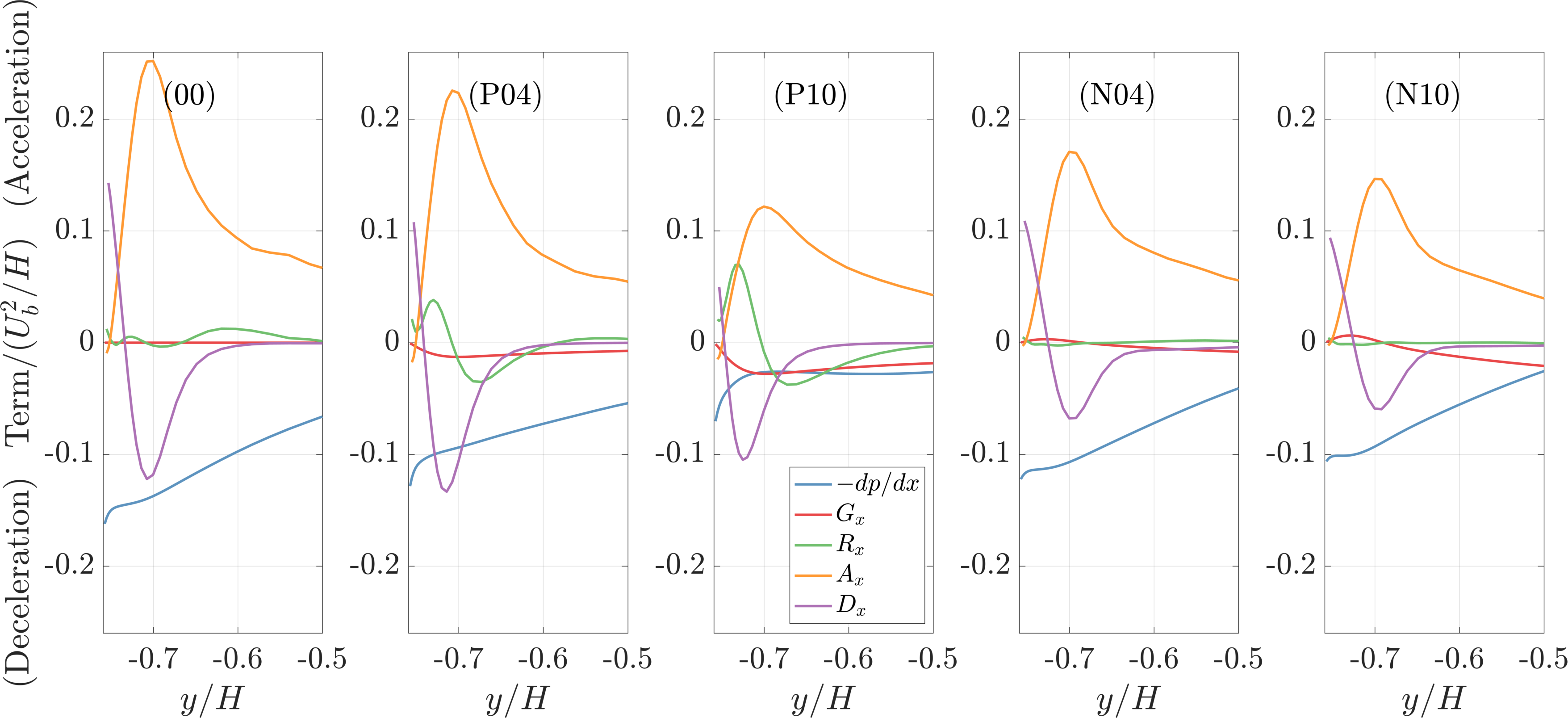}
    \caption{Mean streamwise momentum balance terms at $x=4.25H$. Subplots represent individual cases as denoted by case name labels. $-\partial P/\partial y$, mean wall-normal pressure gradient; $G_y$, mean wall-normal Coriolis force; $R_y$ Reynolds stress divergence; $A_y$ negative mean wall-normal advection; $D_y$ viscous diffusion.  All quantities are normalized by $U_b^2/H$. Note the profiles begin from the height of the bump at $x=4.25H$ ($y=-0.76H$).}
    \label{fig:UBalance}
\end{figure}

The streamwise momentum budgets at $x=4.25H$, just upstream of the separation point for all cases, are shown in figure \ref{fig:UBalance}. For the $x$-momentum budget, positive terms contribute to streamwise acceleration, promoting attached flow. Conversely, negative terms will tend to encourage separation. It must be noted that because the advection term is being calculated on the right-hand side (RHS) of the momentum balance, it represents the \textit{negative} of the flows spatial acceleration. Accordingly, if $A_x>0$, flow is decelerating (and trending toward separation), and if $A_x<0$, flow is accelerating (counteracting the action of the APG). 
We will focus on the wall-normal position of the peak $A_x$ in figures \ref{fig:UBalance}. This occurs $\sim0.06H$ above the bump surface. 
The mean velocity shown in figure \ref{fig:SepSL}\textit{a}, can be used as a reference for the mean momentum deficit.

When the channel is not rotating, the streamwise deceleration is primarily caused by the APG (figure \ref{fig:UBalance} case 00). 
For positive rotation, $G_x <0 $ as the mean flow tends to follow the lee side contour of the bump ($V<0$). Thus, the Coriolis force tends to enhance deceleration and promote separation. However, it is not the major term for the $x-$momentum in this flow. 
The leading change in the decelerating forces of cases P04 and P10, compared with case 00, is a reduction in APG. 
The change of the APG by rotation can be explained by a bulk estimation of the pressure gradient based on Bernoulli's Equation and conservation of mass:
\begin{equation}
\dpr{P}{x} = -U\dpr{U}{x} \propto \frac{U}{A^2} \dpr{A}{x}
\label{eq:dpdx}
\end{equation}
where $A$ denotes the channel area varied along the bump. 
From Eq. (\ref{eq:dpdx}) it is clear that the APG will be lower if the mean $U$ is smaller over the lee side. 
Indeed the variation of APG is correlated to that of the MMD (see figure \ref{fig:SepSL}\textit{a}). Specifically, the stark decrease in velocity in the positive-rotating cases is reflected in the decrease of APG.
The Coriolis term is significantly weaker than the APG at the moderate rotation rate, and comparable to it at the high rotation rate. Furthermore, the Reynolds stress divergence is increased in these two cases and enables the near-wall fluid to counteract the deceleration, especially at the higher rotation rate. 
Note that this term corresponds to both the generation and diffusion of turbulent kinetic energy and thus should not be interpreted solely as a change in the magnitude of turbulent fluctuations. Case P10 actually has a lower turbulent kinetic energy in the near-wall region than case 00 (see Section \S\ref{sec:Restr}) yet it shows an increase in Reynolds stress divergence in the mean momentum budget. 
Therefore, the flow in the positive rotating cases is subjected to a significantly reduced APG, a mild additional deceleration through the Coriolis force, and an augmented Reynolds stress divergence transporting momentum towards the bump. The delayed onset of the separation in these two cases despite their significant MMD can be concluded to be result of significantly lower decelerating forces and an enhanced/more effective turbulent mixing.     

Under negative rotation (cases N04 and N10), the streamwise APG, as the major decelerating force, is reduced compared to the non-rotating case. This again is attributed to the increased MMD. 
The Reynolds stress divergence diminishes as the flow tends to relaminarize. Thus, no turbulence is present to transport high momentum towards the bump as in the other three cases. The Coriolis term ($G_x$), which is expected to be positive (since $\Omega<0$) and delay separation, is significantly weaker in magnitude than that of cases P04 and P10. Further away from the bump, it becomes negative as the mean downward flow along the bump diminishes. The viscous diffusion is responsible for diffusion of high momentum towards the wall, yet is unable to compensate the deceleration in the vicinity of the wall. Overall, the $x$-momentum balance shows that the early separation in the negative rotating cases is because of the lack of turbulence, even despite the weakened decelerating force. The Coriolis term does not have a significant direct impact when a mean downwash is absent. However, it must be kept in mind that the Coriolis force is the key to the relaminarization and the momentum deficit. Therefore, it indirectly promotes the flow separation under cyclonic rotation. However, any flow control approach which imparts a mean $V<0$ will render the Coriolis term as a contributing factor to delay flow separation.

\section{Stability regimes}
\label{sec:stabreg}

\begin{figure}
    \centering
    \includegraphics[width=0.85\textwidth]{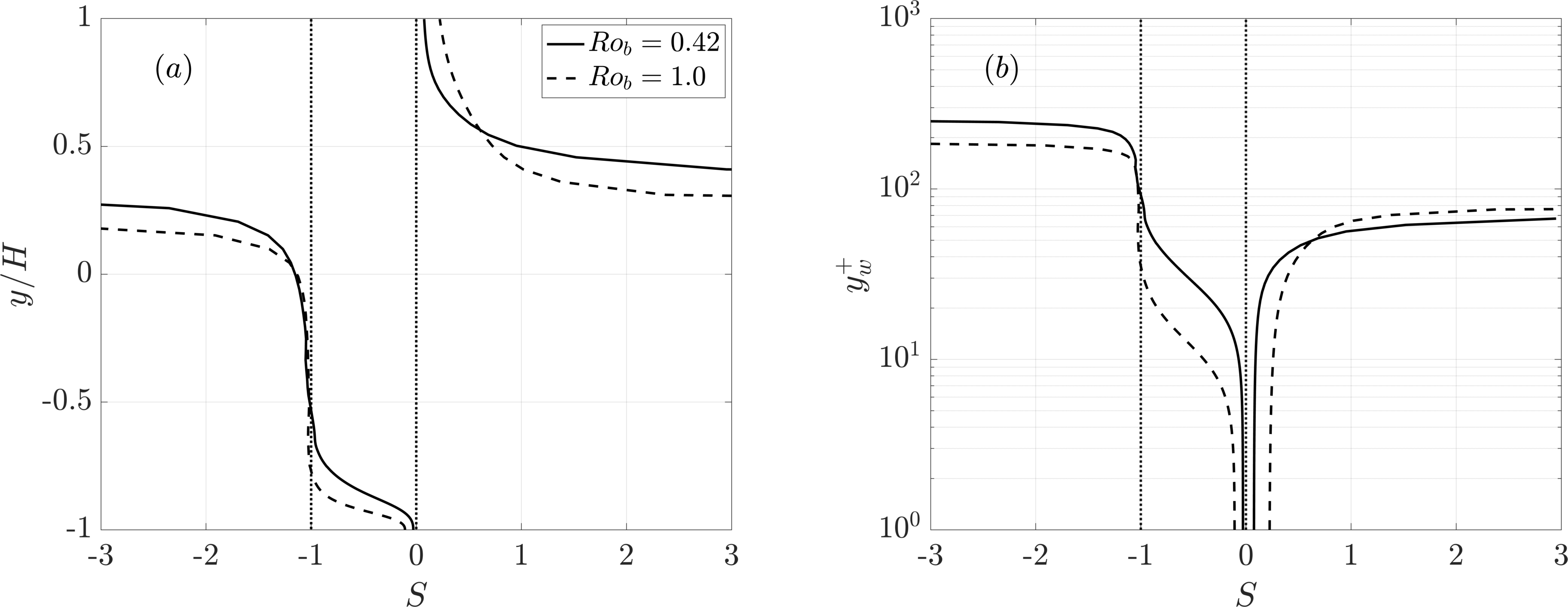}
    \caption{Absolute vorticity ratio ($S$) at the fully recovered region of the channel. (a) as a function of $y/H$. (\textit{b}) $S$ as a function of the distance to the wall in wall units, $y^+ = y u_\tau/\nu$.  Vertical dotted lines mark the stability thresholds of $S=-1$ and $S=0$. }
    \label{fig:SLines}
\end{figure}

{\color{black}Stability analysis of various rotating shear flows \citep{Hart71, Yanase93, Cambon94, Metias95, Salhi97, Brethouwer05} have shown that flow stability is dictated by the ratio of system rotation to shear vorticity, $S:=\Omega / \Omega_s$, as discussed in \S\ref{sec:Intro}.} 
The vorticity ratio in the fully recovered region as a function of $y/H$ and $y^+$ is given figure \ref{fig:SLines}\textit{a} and \textit{b}, respectively. Dotted lines at $S=-1$ and $S=0$ bound the destabilized region. In the vicinity of the anti-cyclonic wall, the flow is destabilized. After the constant gradient layer with $S=-1$, the flow begins to deviate from the neutral regime when it approaches the channel centerline. It is followed by a stabilized region characterized by $S<-1$ below the peak velocity and $S>0$ in the cyclonic side. 
In outer units, the destabilized region extends $0.4H$ from the anti-cyclonic wall at the low $Ro_b$ and $0.2H$ for the high $Ro_b$. 
Therefore, less near-wall flow is destabilized at the higher rotation rate. This leads to reduced turbulence and skin friction as justified in the literature \citep{Johnston72, Xia16, Brethouwer17}.

We found that the physical impact of the reduction of the destabilized region at the high rotation rate is more notable in wall units (figure \ref{fig:SLines}\textit{b}). For the moderate rotation rate, the entire inner layer of the channel including the log-law region is destabilized, with the maximum destabilization ($S\approx0.5$) occurring at the lower log layer at $y^+ = 40$. At the high rotation rate, on the contrary, the destabilization only affects up to the lower limit of the log layer. The peak destabilization shifts down to the buffer layer around $y^+ = 15$. 
As a result, the turbulence in the anti-cyclonic side is augmented for the entire dynamically important inner layer in cases P04 and N04, and is only enhanced near the legs of the embedded hairpin vortices in cases P10 and N10. 

\begin{figure}
    \centering
    \includegraphics[width=0.67\textwidth]{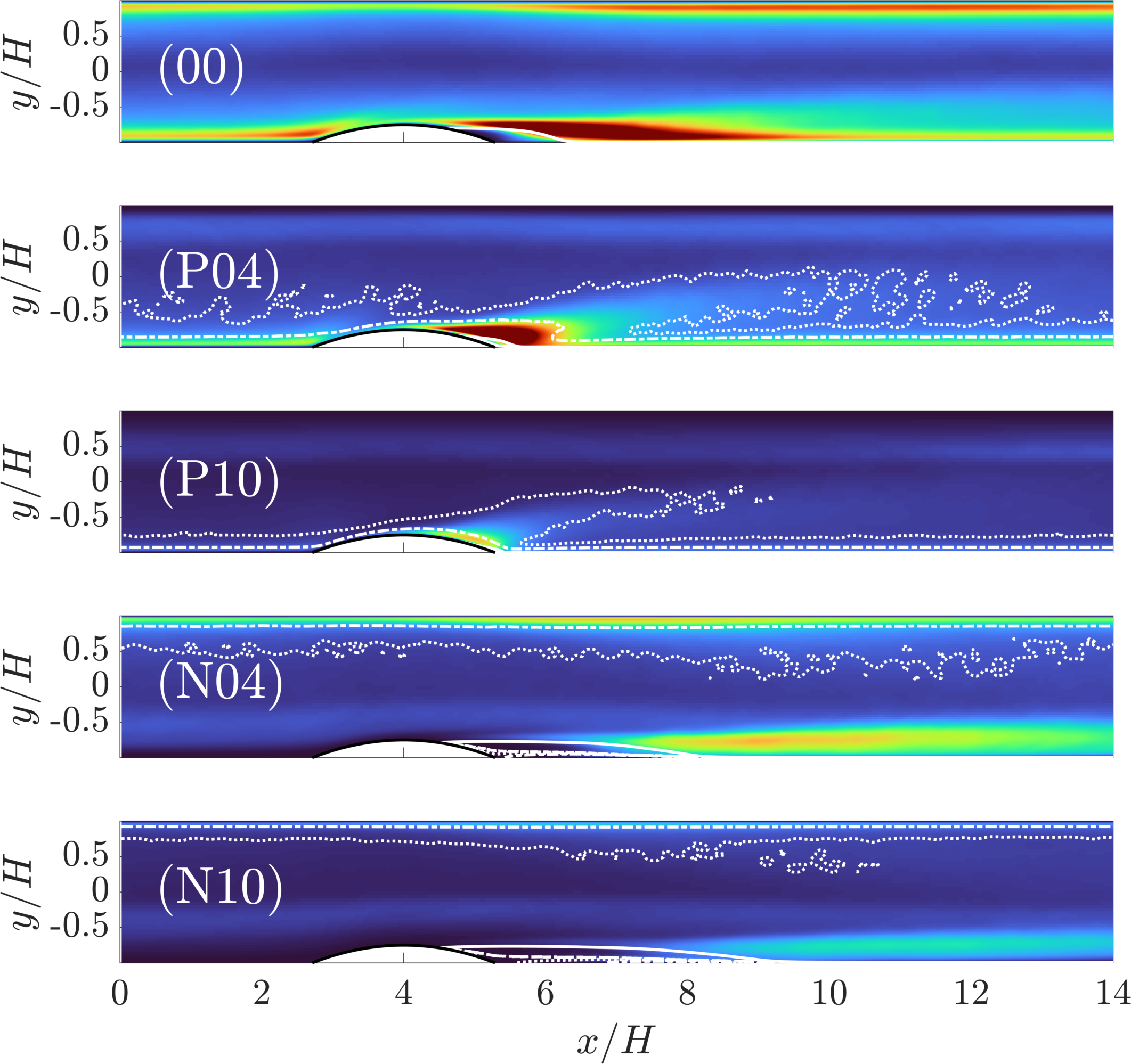}
    \\ 
    \centering
    \hspace{0.44cm}
    \includegraphics[width=0.2\textwidth]{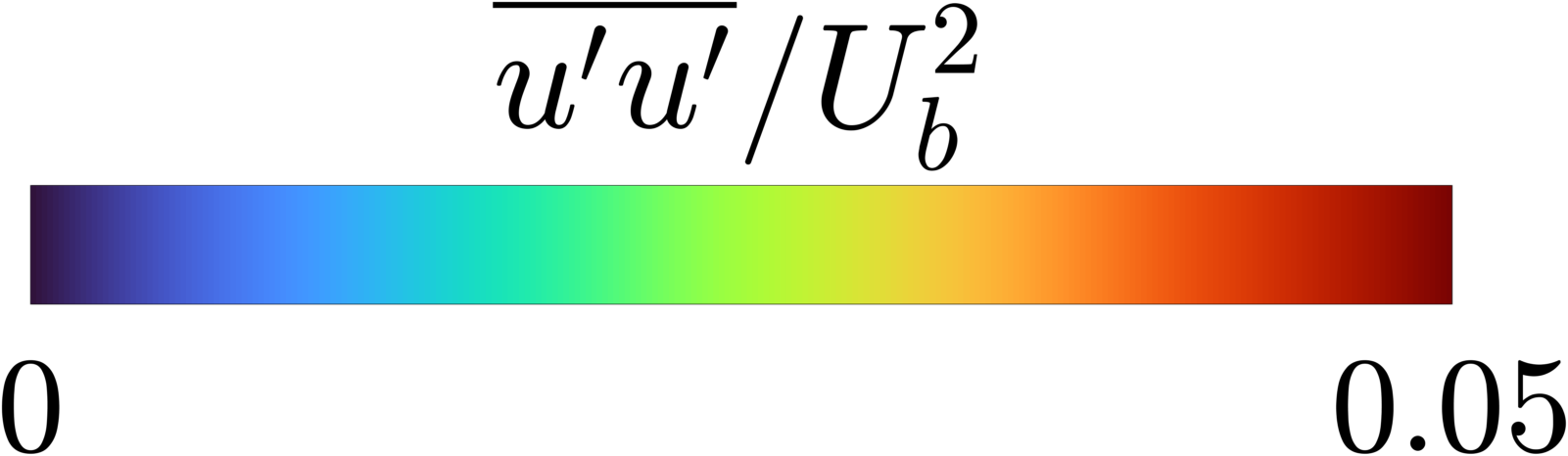}
      \caption{Contours of $\overline{u'u'}$.  \solid Separating streamline; \dotted $S=-1.0$; \dashdot $S=-0.5$. $S = 0$ contour line at the peak streamwise velocity is not plotted for clarity.}
    \label{fig:uu}
\end{figure}

\begin{figure}
    \centering
    \includegraphics[width=0.67\textwidth]{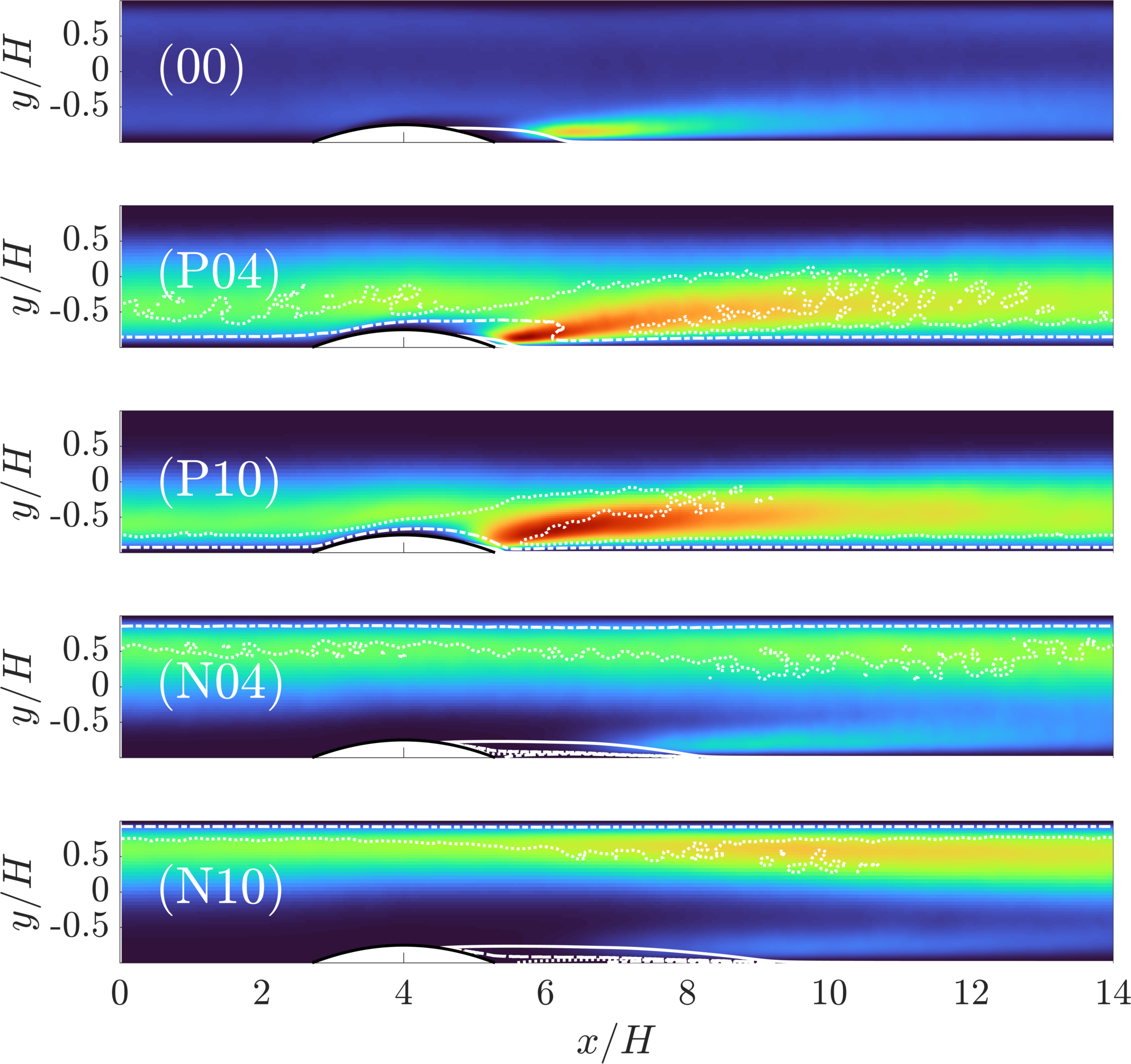}
    \\
    \centering
    \hspace{0.44cm}
     \includegraphics[width=0.2\textwidth]{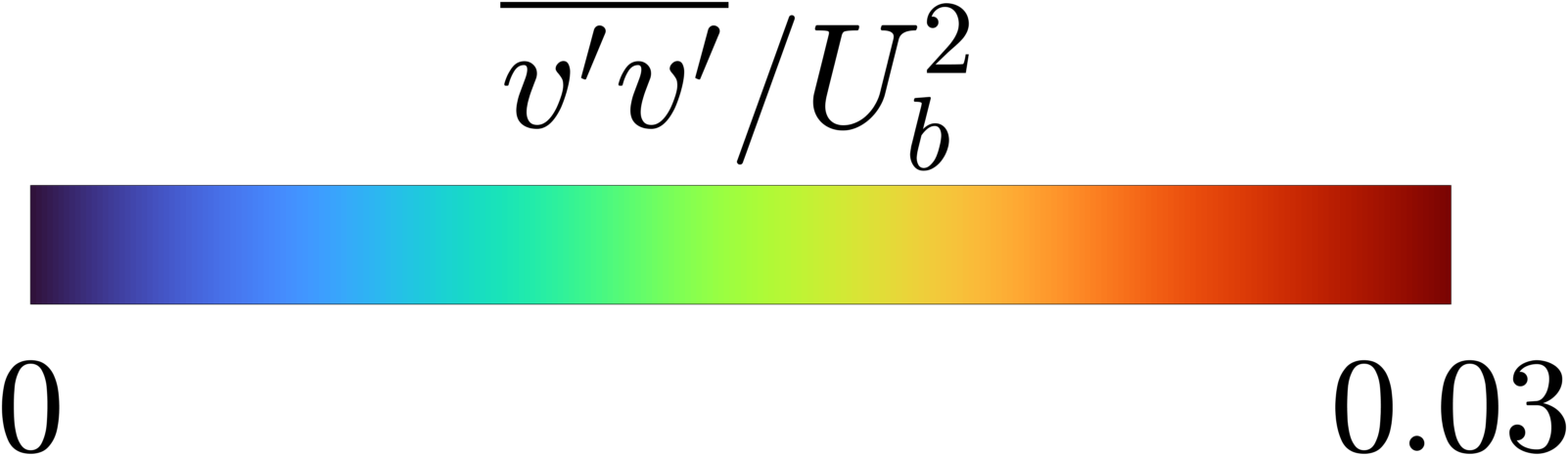}
      \caption{Contours of $\overline{v'v'}$.  For linestyle legend, see caption of figure \ref{fig:uu}.}
    \label{fig:vv}
\end{figure}

In the wake of the bump, the stabilization regimes become two-dimensional and are observed to be correlated with the distribution of the Reynolds stresses (figure \ref{fig:uu}-\ref{fig:uv}). 
Due to the APG, the flow over the lee side of the bump is decelerated. A reduced-velocity layer forms near the bump regardless of whether the flow separates or not. In case P10, the separation is nearly completely suppressed yet the momentum deficit remains significant. The slow-moving layer decreases the distance over which the velocity increases to its peak, therefore increasing the velocity gradient. Because $S$ is inversely proportional to the velocity gradient in magnitude, it becomes less negative along the lee-side of the anti-cyclonic bump.
Figure \ref{fig:SLines} shows that the region approximately up to the height of the bump has $-1<S<0$ for fully-recovered flow. Due to the mentioned decelerated layer, in the wake of the bump this region becomes $S>-0.5$.

The $S=-0.5$ threshold receives significantly less attention than the $S=-1$ neutral condition in rotating separating flows. As discussed in \S \ref{sec:Intro}, several linear stability analyses have shown that maximum destabilization of coherent structures and shear flows 
occurs when $S=-0.5$ \citep{Yanase93, Cambon94, Metias95, Salhi97, Brethouwer05}. Since our results show a strong correlation between the contour of $S=-0.5$ and the Reynolds stress variations, we look further into its physical implications through the production of Reynolds stresses. The full budgets will be further discussed in later sections.
When combining the shear production and rotational production for $\overline{u'u'}$ (refer to the Appendix), the total production of \uu reads: 
\begin{equation}
P_{uu, \mathrm{tot}} \approx 4\Omega \overline{u'v'} - 2\overline{u'v'}\frac{\partial U}{\partial y} = 2\overline{u'v'}\left( 2\Omega - \frac{\partial U}{\partial y} \right) = -2\overline{u'v'}\frac{\partial U}{\partial y}\left( S+1\right).
\label{eqn:Puutot}
\end{equation}
Here, $\partial U/\partial x$ is considered negligible compared with the wall-normal shear. Typically, $\overline{u'v'}$ has the opposite sign of ${\partial U/\partial y}$ through $P_{uv}$. Thus the sign of $P_{uu, \mathrm{tot}}$ depends on $(S+1)$.
Considering the bottom wall of the channel under positive rotation (anti-cyclonic), $P_{11,\mathrm{tot}}$ is a source of $\overline{u'u'}$ when $-1<S<0$ (\textit{i.e.}, $\partial U/\partial y > 2\Omega$) and a sink when $S<-1$ ($\partial U/\partial y < 2\Omega$). Therefore, $S=-1$ has been identified in the literature as the threshold of the net change of $\overline{u'u'}$, delimiting between gain from the mean flow and loss to $\overline{v'v'}$. Because the generation of wall turbulence originates from $\overline{u'u'}$, the sign of this net change determines whether rotation creates (\textit{i.e}., destabilizes, $-1<S<0$) or prevents (\textit{i.e.}, stabilizes, $S<-1$) turbulence when $\Omega$ and $\Omega_s$ are anti-cyclonic. 
When $S>0$ (cyclonic), on the other hand, $P_{uu,\mathrm{tot}}>0$, meaning there is net production of $\overline{u'u'}$. However, $G_{vv} = -2\overline{u'v'}{\partial U}/{\partial y}\left( -S\right)$ will be negative such that the TKE redistributed from \uu will be transferred back. This suppresses the turbulence generation cycle and relaminarizes the flow. These mechanisms have been well characterized in the literature.

Here we justify the $S=-0.5$ threshold by considering the concurrent role of $G_{uv} := -2\Omega(\overline{u'u'}-\overline{v'v'})$.
Under positive rotation near the bottom wall, $G_{uv}$ makes \uv more (less) negative when \uu is greater (smaller) than $\overline{v'v'}$. A change of $\overline{u'v'}$, then in return, will affect $P_{uu, \mathrm{tot}}$ and $G_{vv}$. Consider a $\delta\overline{u'v'}<0$ in the anti-cyclonic region,
$\delta P_{uu,\mathrm{tot}}> \delta G_{vv}>0$ when $(S+1)>-S>0$, that is $-0.5<S<0$. The resultant more significant production of \uu than \vv in such scenarios may augment the (\uu-$\overline{v'v'}$) term in $G_{uv}$ and thus further enhance the negative $\delta \overline{u'v'}$. This feedback is self-exciting. On the contrary, $\delta G_{vv} > \delta P_{uu,\mathrm{tot}} > 0$ when $-1<S<-0.5$, which decreases the difference between \uu and $\overline{v'v'}$, {changing the impact of $G_{uv}$ on $G_{uu}$ from beneficial to detrimental
(\textit{i.e.}, self-restraining). } 
Therefore, we propose that $S=-0.5$ is a threshold that differentiates these two states within the $-1<S<0$ destabilization regime. 
Note that, $B:=S(S+1)$ has been used extensively in the domain of hydrodynamic stability, called the Bradshaw–Richardson number~\citep{Bradshaw69,Pedley69}. Our interpretation indicates the physical self-exciting (self-restraining) regime corresponds to $0>B>-0.25$ ($B<-0.25$) within the $B<0$ destabilized regime.

While $S = -0.5 \; (B=-0.25)$ has been found to lead to the maximum growth rate of disturbances for various types of one-dimensional mean flows \citep{Yanase93, Cambon94, Metias95, Brethouwer05}, the flow in the current study has a spatial variation of the background mean shear. We, therefore, consider 
\begin{equation}
\dpr{B}{\Omega_s} = -\frac{2S^2}{\Omega}(S+0.5) \;\; \mathrm{and}  \;\; 
\dpr{B}{\Omega} = \frac{2B}{\Omega}.
\label{eq:dBdom}
\end{equation}
That is, the change of the stabilization regime with respect to the variation of the mean shear or rotation rate. $S=-0.5$ differentiates the change of stabilization/destabilization as the rotation/mean shear vorticity varies. For example, for the anti-cyclonic side of the channel, the region with $S<-0.5$ ($S>-0.5$) will become more (less) unstable if there is an increase in $\partial U/\partial y$. Therefore, the change from $S<-1$ towards $S=-0.5$ is sustainable,  while further increasing $S$ beyond -0.5 is subject to inherent constraints.

\section{Turbulence statistics}\label{sec:RS}
\subsection{Reynolds stress distribution}\label{sec:Restr}
\begin{figure}
    \centering
    \includegraphics[width=0.67\textwidth]{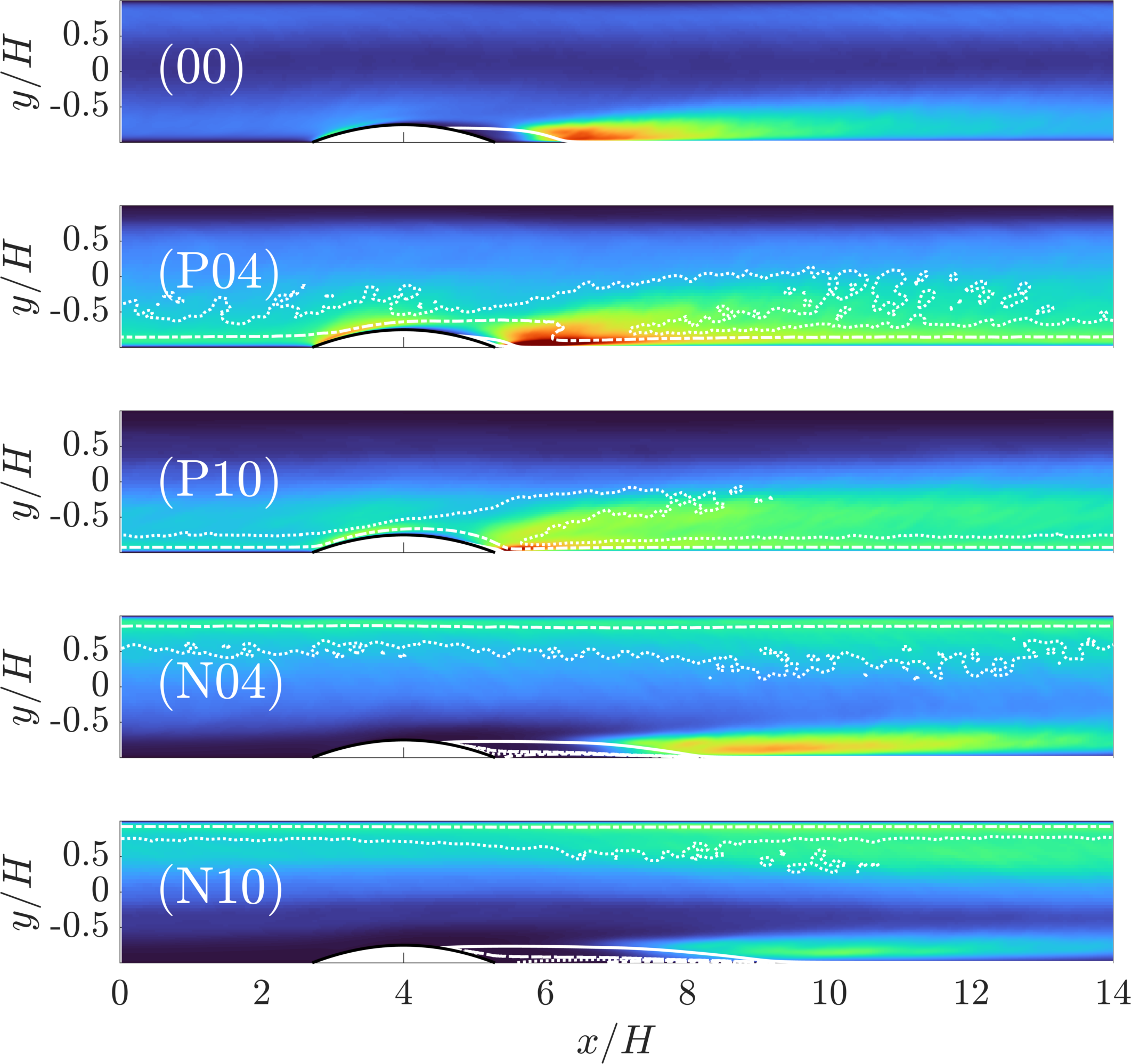}
    \\
    \centering
    \hspace{0.44cm}
      \includegraphics[width=0.2\textwidth]{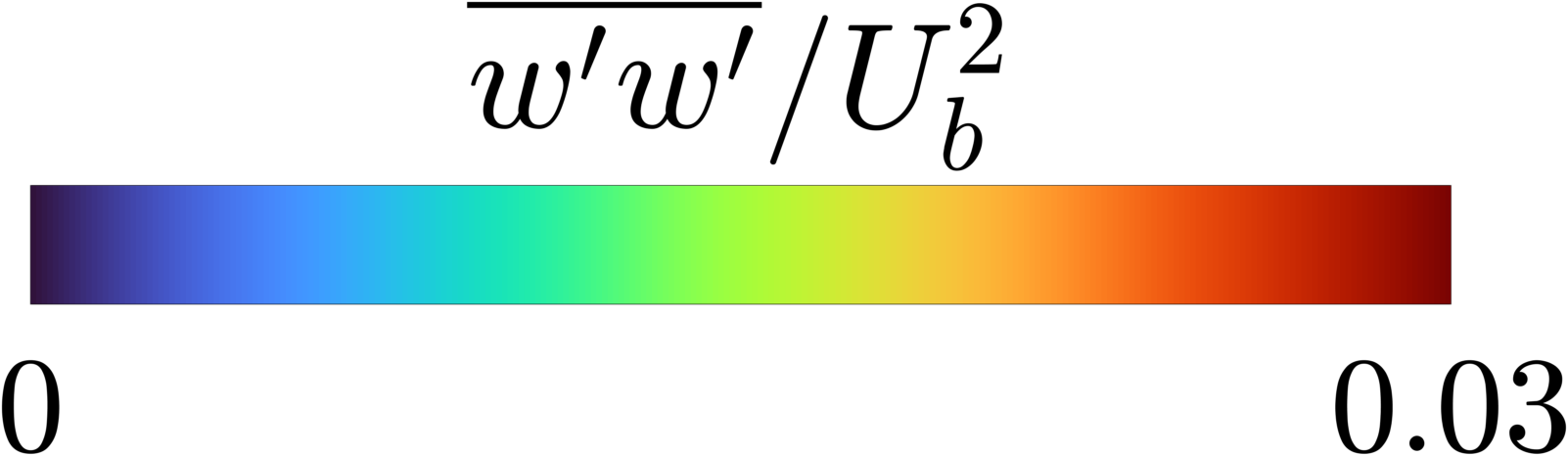}
      \caption{Contours of $\overline{w'w'}$.  For linestyle legend, see caption of figure \ref{fig:uu}.}
    \label{fig:ww}
\end{figure}

\begin{figure}
    \centering
    \includegraphics[width=0.67\textwidth]{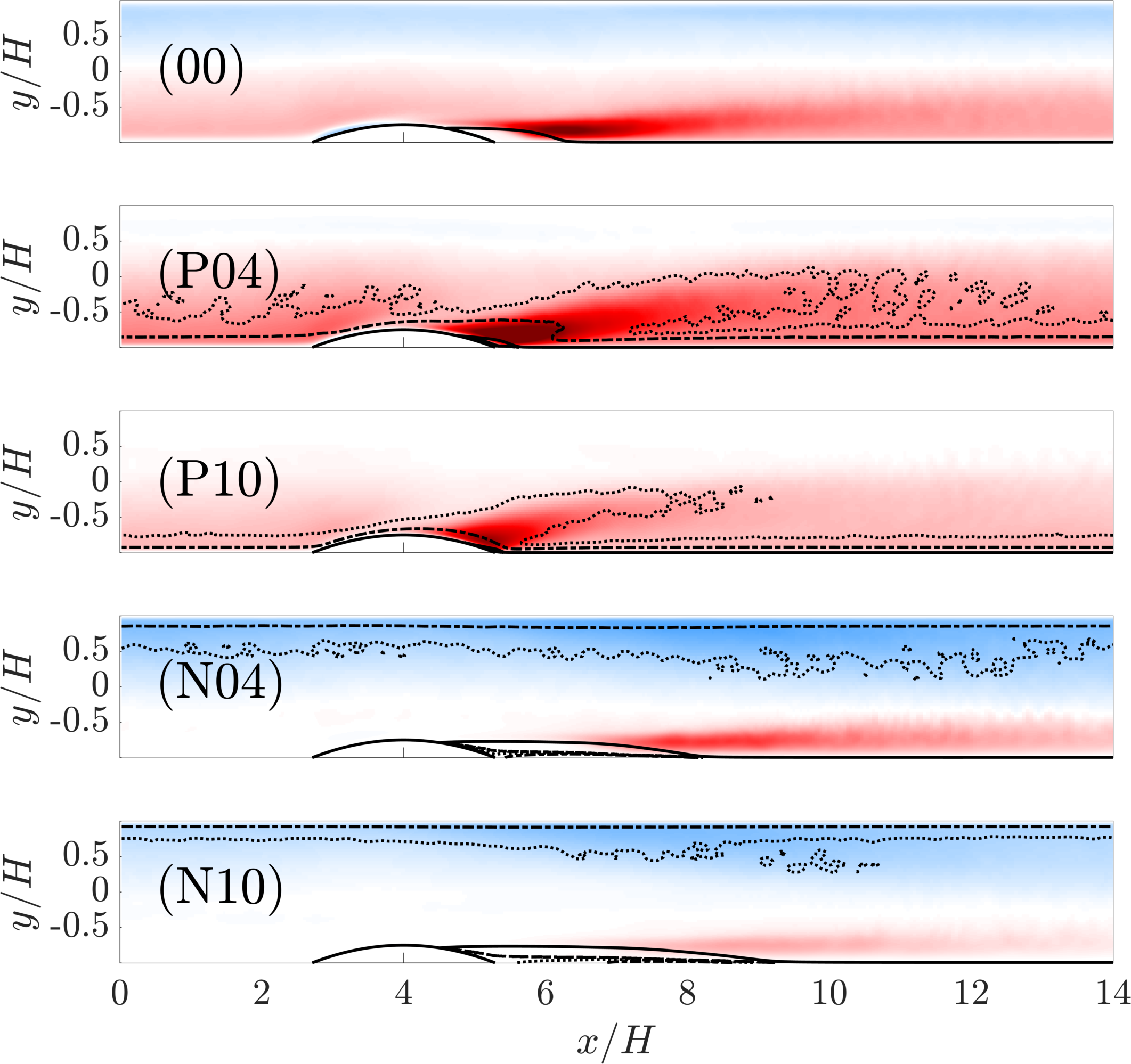}
    \\
    \hspace{0.44cm}
   \includegraphics[width=0.2\textwidth]{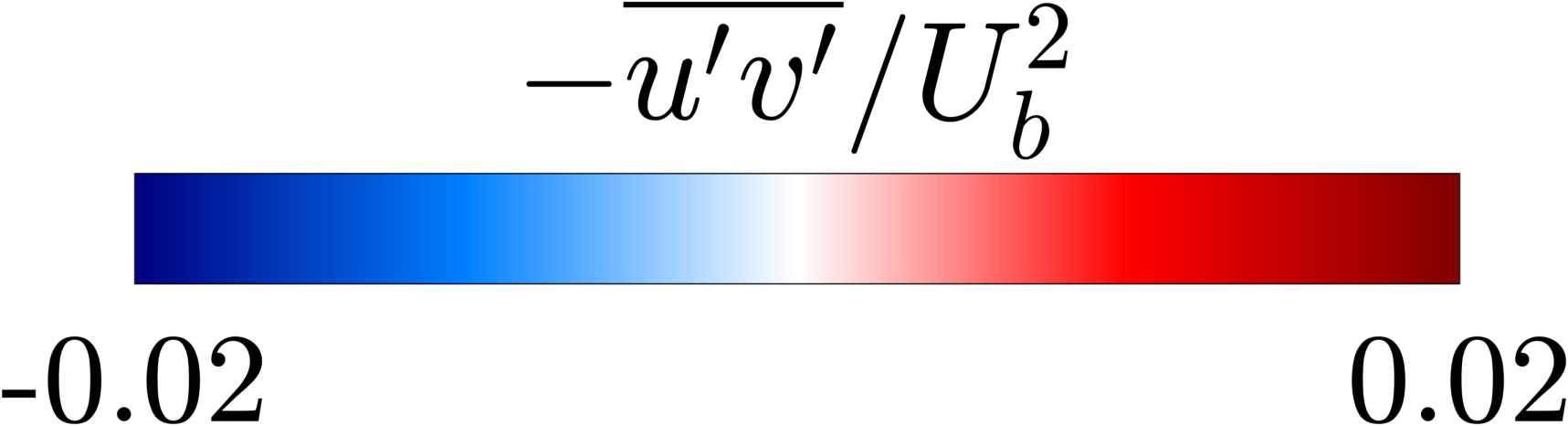}
      \caption{Contours of $-\overline{u'v'}$. For linestyle legend, see caption of figure \ref{fig:uu}.}
    \label{fig:uv}
\end{figure}

The spatial distribution of Reynolds stresses associated with the reattachment and wake recovery is discussed in this section. The variation in stability regimes discussed above will be incorporated to understand the mechanisms responsible for the observed phenomena.
Contours of the Reynolds normal stresses ($\overline{u'u'}, \overline{v'v'}$, and $\overline{w'w'}$) along with the Reynolds shear stress ($-\overline{u'v'}$) are given in figures \ref{fig:uu}-\ref{fig:uv}. The solid line represents the separating streamline. The dotted and dash-dot contour lines signify where $S=-1.0$ and $S=-0.5$, respectively. 

Using case 00 as a benchmark, the onset of streamwise fluctuations ($\overline{u'u'}$) occurs first near the separation point, while significant wall-normal ($\overline{v'v'}$) and spanwise ($\overline{w'w'}$) fluctuations develop downstream near the reattachment point. $-\overline{u'v'}$ develops an appreciable magnitude at the streamwise center of the separation bubble. These high-stress regions all occur in the forward-flow side of the separating shear layer. The augmented stresses begin to lose intensity within a few $H$ of reattachment, displaying a nearly full recovery (\textit{i.e.}, little streamwise variation) by $x/H=12$. Throughout the shear layer and wake recovery, the wall-normal position of peak stresses changes little, forming a wake that extends parallel to the walls. This spatial distribution aligns with the established understanding of Reynolds stress distribution in APG-induced separation, representing the development of the turbulent separating shear layer and its recovery after reattachment, as discussed in \cite{Na98} (among others). On the top wall, the only notable change due to the bump is a decrease in $\overline{u'u'}$ where the flow area is reduced by the bump, an expected suppression of turbulence due to the FPG. Yet, this effect is insignificant as the bump is designed to provide little blockage and generate mild pressure gradients.

When subject to positive rotation (cases P04 and P10), the spatial distribution of Reynolds stresses exhibits significant variations compared with case 00. Relative to the separation point, $\overline{u'u'}$ develops appreciable values earlier than in case 00. Additionally, its region of elevated magnitude is remarkably reduced in size and confined nearer the bump surface. Similar spatial variations apply to $-\overline{u'v'}$. While the onset of $\overline{v'v'}$ also occurs earlier relative to reattachment, there is a stark increase in the size of its enhanced region when subject to positive rotation, despite the considerably smaller mean separation region in these two cases. The increase of $\overline{w'w'}$, which is conventionally used to represent the three-dimensionalization of the separated shear layer, remains highly correlated with the reattachment. However, the peak $\overline{w'w'}$ occurs near the wall rather than in the shear layer. 

Another qualitative difference with the intense-stress regions from the non-rotating case is that those in the positive rotating cases show an upward shift of their peak towards the opposite wall along the wake of the bump.
This peak line does not occur along any mean streamline. This behavior is most prominent for $\overline{v'v'}$ in which it extends over $3H$ downstream of the bump. For $\overline{u'u'}$ and $-\overline{u'v'}$ this region rapidly diminishes approximately $0.5-1.5H$ downstream of the bump. For $\overline{w'w'}$, the inclined augmented region is present, however not comparable in magnitude to its near-wall peak. This phenomenon has been observed in rotating separation literature. Specifically, in the rotating sudden expansion study of \cite{Lamballais14}, a region of augmented $\overline{v'v'}$ originating on the lee side of the anti-cyclonic separation bubble extends toward the opposite cyclonic wall, while $\overline{u'u'}$ is significantly attenuated and extends at the same angle (see \cite{Lamballais14} figure  13\textit{d, f}). An explanation for this spurious spatial distribution was not provided, however.

We found that the discussed high Reynolds stress regions in the positive rotating cases exhibit a correlation with the thresholds of $S$ that differentiate the stability regimes. In the fully recovered channel flow of each case (refer to figure \ref{fig:SLines}), the destabilized, neutral, and stabilized regions are well-defined as wall-parallel layers. On the lee side of the bump, the mean velocity in the wake of the bump and its recovery leads to a two-dimensional variation of the stability regions. Specifically, the diminishing boundaries of the high \uu and \uv regions align surprisingly well with $S= -0.5$ contour line, while for \vv and \ww it shows general agreement with $S=-1.0$. Because $S$ is a representation of the interplay between the mean shear vorticity and the system rotation, the observed correlation between the Reynolds stresses and specific $S$ values suggests a strong influence of the Coriolis force on the Reynolds stresses. Additionally, the spatial sequence, with elevated \vv and \ww regions downstream of the peak \uu and $-\overline{u'v'}$, implies a causal relationship between these variations.

As discussed in the previous section, the change of the region behind the bump from $S\sim-1$ (neutral) to $S>-1$ (destablized) indicates the mean shear vorticity becomes more negative than -2$\Omega$. Upon examination of $\partial V/\partial x$ and $\partial U/\partial y$, it was found that the latter has a similar spatial variation as the inclined $S=-1$ contour (not shown). Therefore, an increasing $\partial U/\partial y$, caused by the deceleration of the flow that peaks near the wall, is the primary reason for the increasing $S$.
It alters the flow near the bump from near-neutral/self-restraining to more destabilized, leading to a remarkable modulation of Reynolds stresses. A highly augmented velocity gradient may cause $S$ to be greater than -0.5, leading to the self-exciting regime. However, further increasing $S$ towards zero is restrained.
Note that this stability regime change is highly sensitive to the mean velocity profile: referring to the mean velocity profiles at $x/H=$5.5, 6 and 8 in figure \ref{fig:Ucont}, the misalignment between the reference lines of $2\Omega$ slope and the mean velocity is only notable to approximately the bump height, yet this slight variation into the unstable regime leads to a prominent change of the Reynolds stresses up to the centerline of the channel. 

Comparing the two positive rotation rates, case P10 has significantly lower Reynolds stresses than case P04 upstream of the bump. This represents the suppressed turbulence at the high rotation rate. 
Similarly, in the wake of the bump, $\overline{u'u'}$ and $-\overline{u'v'}$ are smaller in magnitude in case P10, even when compared to case 00. However, the augmentation of \vv in the inclined region along the $-1<S<0$ region remains remarkable and more pronounced than in case P04. $\overline{w'w'}$, while qualitatively similar to case P04, exhibits a smaller magnitude in case P10. 

The Reynolds stress distribution in the negative rotating cases is much as expected. The separating shear layer exhibits the typical features of a laminar separating shear layer. Intense Reynolds stresses only appear considerably downstream near the reattachment point, corresponding to the formation and decay of two-dimensional roller vortices. The stabilization ($S>0$) effect of the Coriolis force is evident: the stresses are much lower in case N10 than the ones in case N04. 
Note that, the increased $\partial U/\partial y$ behind the bump does not lead to a change in the stability regime in the negative rotation cases as it only makes $S$ less positive. The prolonged separating shear layers cause an extended blockage of the channel. This leads to slight increases of \uu along the top wall followed by augmented $\overline{v'v'}$ and $\overline{w'w'}$. The APG induced by the expansion as the flow reattaches leads to an increased $\partial U/\partial y$ along the top (anti-cyclonic) wall. Thus, this region becomes more unstable, leading to the observed stress augmentation. 
\subsection{Reynolds stresses budget: separating shear layer}
\label{sec:Restrbud}
\begin{table}
  \begin{center}
    \def~{\hphantom{0}}
    \begin{tabular}{lcccc}
      Term      & $i=1, j=1$ & $i=2, j=2$ & $i=3, j=3$ & $i=1, j=2$ \\ \\
      $G_{ij}$  & $4\Omega \overline{u'v'}$ & $-4\Omega \overline{u'v'}$    & 0   & $-2\Omega (\overline{u'u'}-\overline{v'v'})$ \\\\   
      $P_{ij}$ & $-2 \left( \overline{u'u'}\frac{\partial U}{\partial x} + \overline{u'v'}\frac{\partial U}{\partial y} \right)$ & $-2\left( \overline{u'v'}\frac{\partial V}{\partial x} + \overline{v'v'}\frac{\partial V}{\partial y}  \right)$ & 0 & $-\left(  \overline{u'u'}\frac{\partial V}{\partial x} + \overline{v'v'}\frac{\partial U}{\partial y}\right)$
      \\
      \end{tabular}
    \caption{Production and Coriolis terms in the Reynolds stress budgets.}
    \label{tab:Prod}
  \end{center}
\end{table}

 \begin{figure}
    \centering
    \includegraphics[height=0.76\textwidth, trim={0cm 0 0 0}, clip]{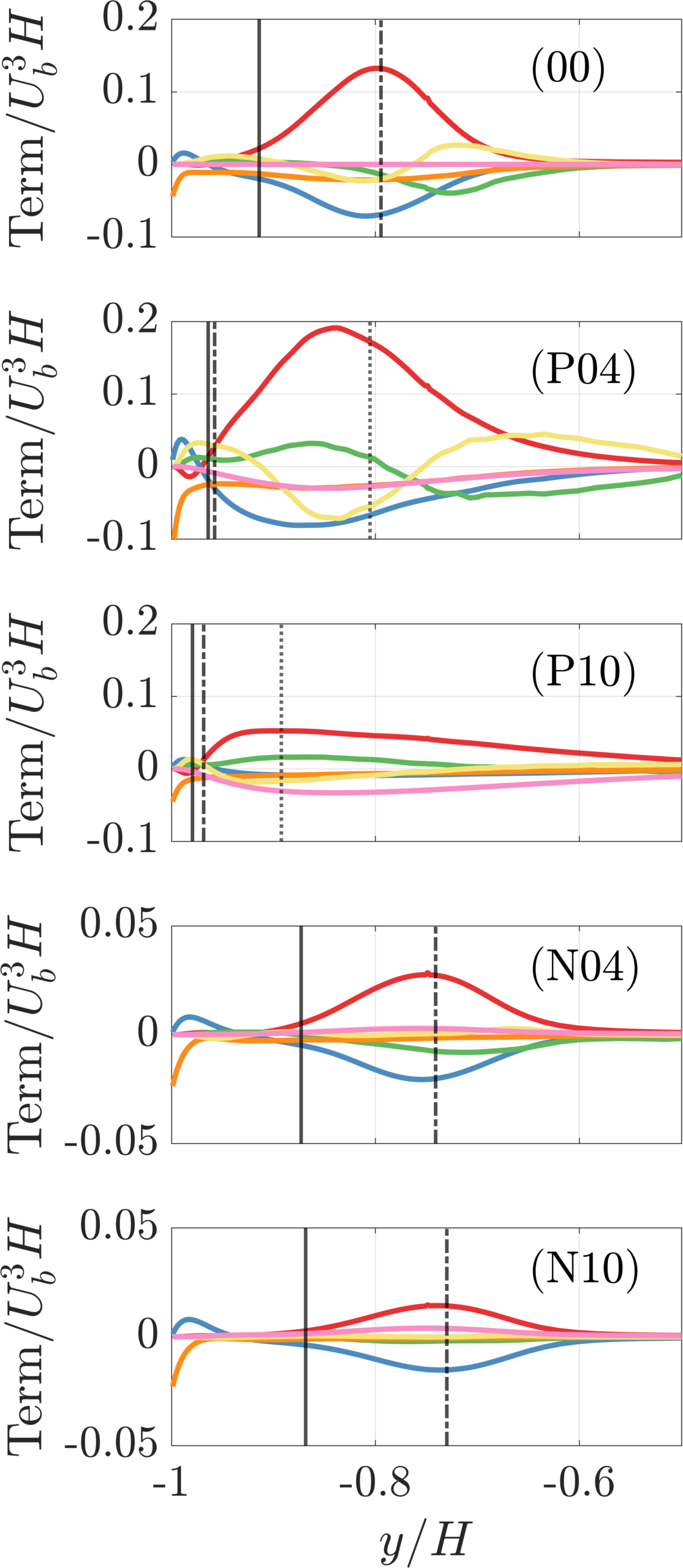}$\;\;\;$
    \includegraphics[height=0.76\textwidth, trim={1.12cm 0 0 0}, clip]{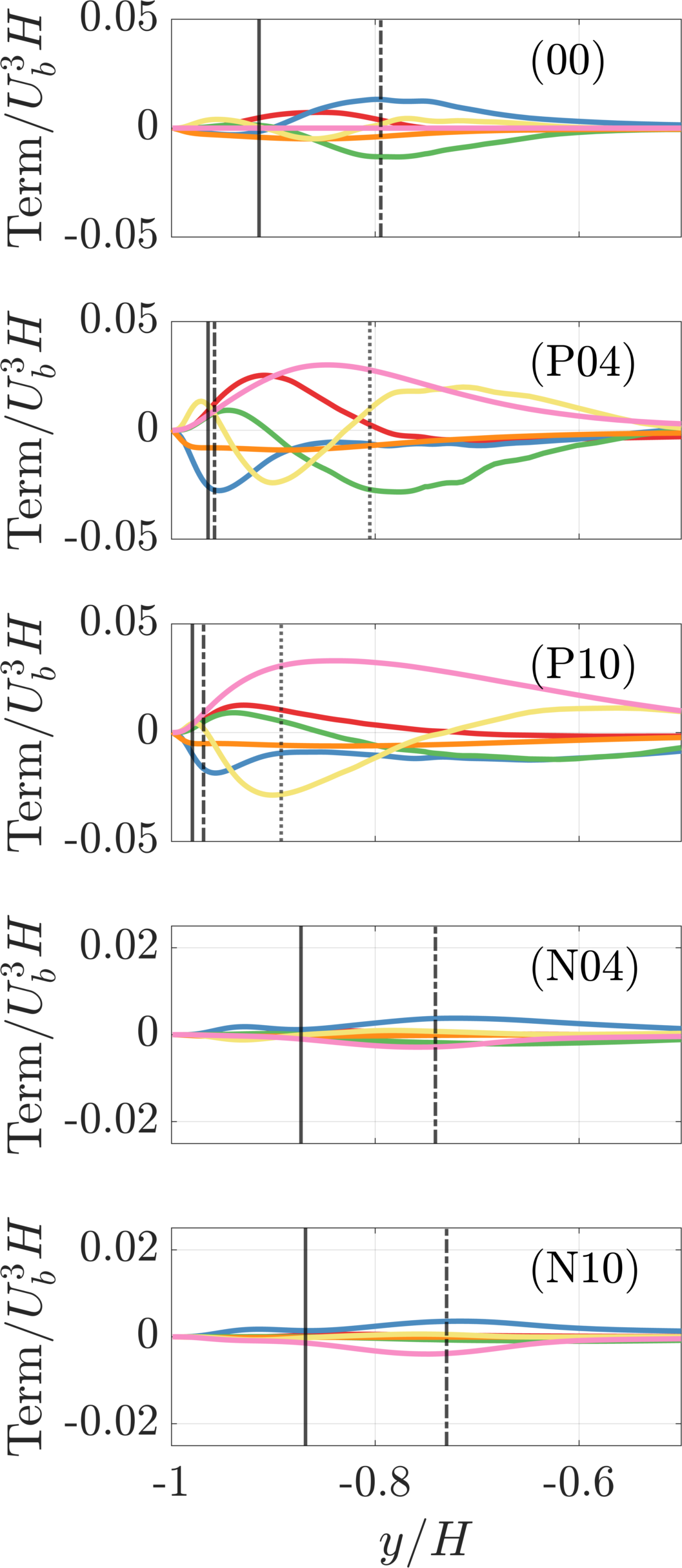}$\;\;\;$
     \includegraphics[height=0.76\textwidth, trim={1.12cm 0 0 0}, clip]{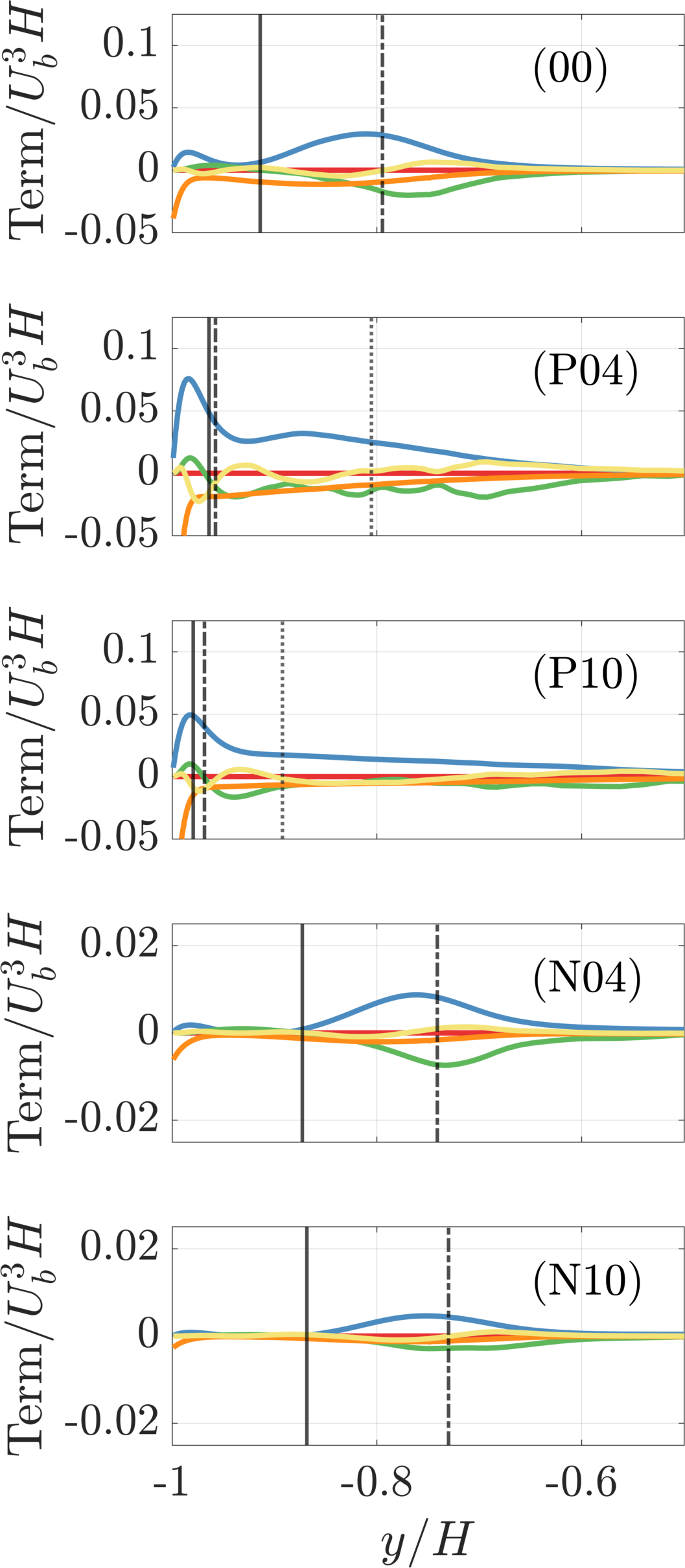}\\
         \includegraphics[width=0.4\textwidth]{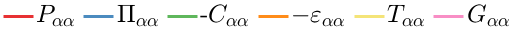}
    
      \caption{Budgets of $\overline{u'u'}$ (left), $\overline{v'v'}$ (middle), and $\overline{w'w'}$ (right) at the respective streamwise center of the separation bubble for each case. Note the difference in axis scaling between terms and cases. \solid $y\vert_{U=0}$; \dashdot $y\vert_{\mathrm{max}(\partial U/\partial y)}$; \dotted $y\vert_{S=-0.5}$.}
    \label{fig:budget_bubble}
\end{figure}

The budgets of $\overline{u'u'}$, $\overline{v'v'}$, and $\overline{w'w'}$ at the center of the separation region of each case are shown in figure \ref{fig:budget_bubble}. Production ($P_{ij}$) and Coriolis ($G_{ij}$) terms are given in table \ref{tab:Prod} and the conventional terms in the Appendix. Without rotation, the $\overline{u'u'}$ budget is dominated by the gain from shear production and loss through the fluctuating velocity-pressure-gradient correlation, both of which reach their peak magnitudes at the inflection point in the separating shear layer. The latter is the primary source of $\overline{v'v'}$ and $\overline{w'w'}$. This agrees with previous studies of plane mixing layers and separating shear layers \citep{Rogers94, Na98}. When rotation is cyclonic for the flow near the bottom wall (cases N04 and N10), the budgets of the Reynolds stresses are qualitatively similar to shear-layer dominant dynamics in case 00. Note that the terms corresponding to these two cases are depicted with reduced $y$-axis ranges in figure \ref{fig:budget_bubble} as the flow is quasi-laminarized. 
$G_{uu}$ is extracting energy from \vv and preventing the formation of the roller vortices in the shear layer.

When rotation is anti-cyclonic, the Coriolis term acts as a sink for $\overline{u'u'}$ and a source for $\overline{v'v'}$, as expected. One notable change to the $\overline{u'u'}$ budget in case P04 is that the production does not occur at the inflection point in the shear layer. Rather, it appears where $G_{uu}$ reaches its peak magnitude. Examining the sub-terms of $P_{uu} = -\overline{u'u'}\partial U/\partial x -\overline{u'v'}\partial U/\partial y$, we found that both $\overline{u'u'}$ and $\overline{u'v'}$ obtain their peak magnitudes at this location, yet the mean shear peaks at the inflection point of the velocity profile nearer to the wall. This indicates that the Coriolis effect replaces the canonical shear layer mechanisms of the $\overline{u'u'}$ dynamics in this case. Referring to \S\ref{sec:stabreg}, we notice that the peak production of \uu is in the self-exciting stability regime where $-0.5<S<0$: more energy is extracted from the mean flow to \uu than redistributed to $\overline{v'v'}$. The increased difference between \uu and \vv enhances $\overline{u'v'}$, enabling the self-exciting cycle. This is further proven by our data as the magnitude of $\overline{u'v'}$ is doubled at the peak, and $\overline{u'u'}$ remains comparable to case 00 despite that \vv is extracting energy from it.
For $\overline{v'v'}$, the $G_{vv}$ term becomes a leading source. Similar to $\overline{u'u'}$, $P_{vv}$ is misaligned with the inflection point where the maximum shear occurs. However, it is shifted less towards the peak of the $G_{vv}$ because the leading shear ($\partial V/\partial y$) is more skewed towards the separating shear layer than $\partial U/\partial y$ for \uu (not shown). Another remarkable qualitative change in the \vv budget is that the velocity-pressure correlation term becomes a sink that peaks in the vicinity of the wall. Therefore, the TKE redistribution from \uu to \vv and \ww in the non-rotating case (and canonical TBLs) is altered such that the fluctuating pressure extracts energy from both \uu and \vv and redistributes it to $\overline{w'w'}$. 
This modified redistribution is the most significant near the inflection point of the shear layer, where the transfer is primarily from \vv to \ww. It leads to an increased gain in the \ww budget compared to case 00. Summarizing the observation for all three Reynolds normal stresses, it is evident that the total gain is increased for all components compared to the non-rotating case. This is the reason responsible for the enhanced Reynolds stresses (see figure \ref{fig:uu}-\ref{fig:uv}), as well as the early reattachment observed in case P04.

At the higher positive rotation rate, most of the terms in the \uu budgets reduce in magnitude. It is found that the mean shear is still substantial in the shear layer, yet the Reynolds stresses are reduced. The consequence of the decreased fluctuations is also exhibited in the $G$ terms which are nearly equivalent to those in case P04 despite that the rotation rate is more than doubled. 
We again refer to the stability regime to justify this change: the deceleration changed the neutral ($S\sim -1$)region to self-exciting ($-0.5<S<0$) in case P04, yet only to self-restraining ($-1<S<-0.5$) in P10. This is because $\partial B/\partial \Omega$ decreases with an increase of $\Omega$, as shown in Eq. (\ref{eq:dBdom}).
In such a self-restraining regime, the net $P_{uu,\mathrm{tot}}$ can still extract energy from the mean flow. However, this extraction is counteracted by the Coriolis influence through $G_{uv}$. As \vv extracts a significant amount of energy from \uu and surpasses the latter, the $G_{uv}$ term becomes positive, mitigating the negative \uv and resulting in a reduction of  $P_{uu,\mathrm{tot}}$. For $\overline{v'v'}$, the Coriolis term is now the dominant gain.  Therefore, the enhanced \vv region shown in figure \ref{fig:vv} for case P10 is mainly through rotation.  The redistribution by fluctuating pressure is now only between \vv and \ww, likely a cause of the weakened \uu due to extraction by $G_{uu}$.

\begin{figure}
\centering
\includegraphics[width=0.49\textwidth]{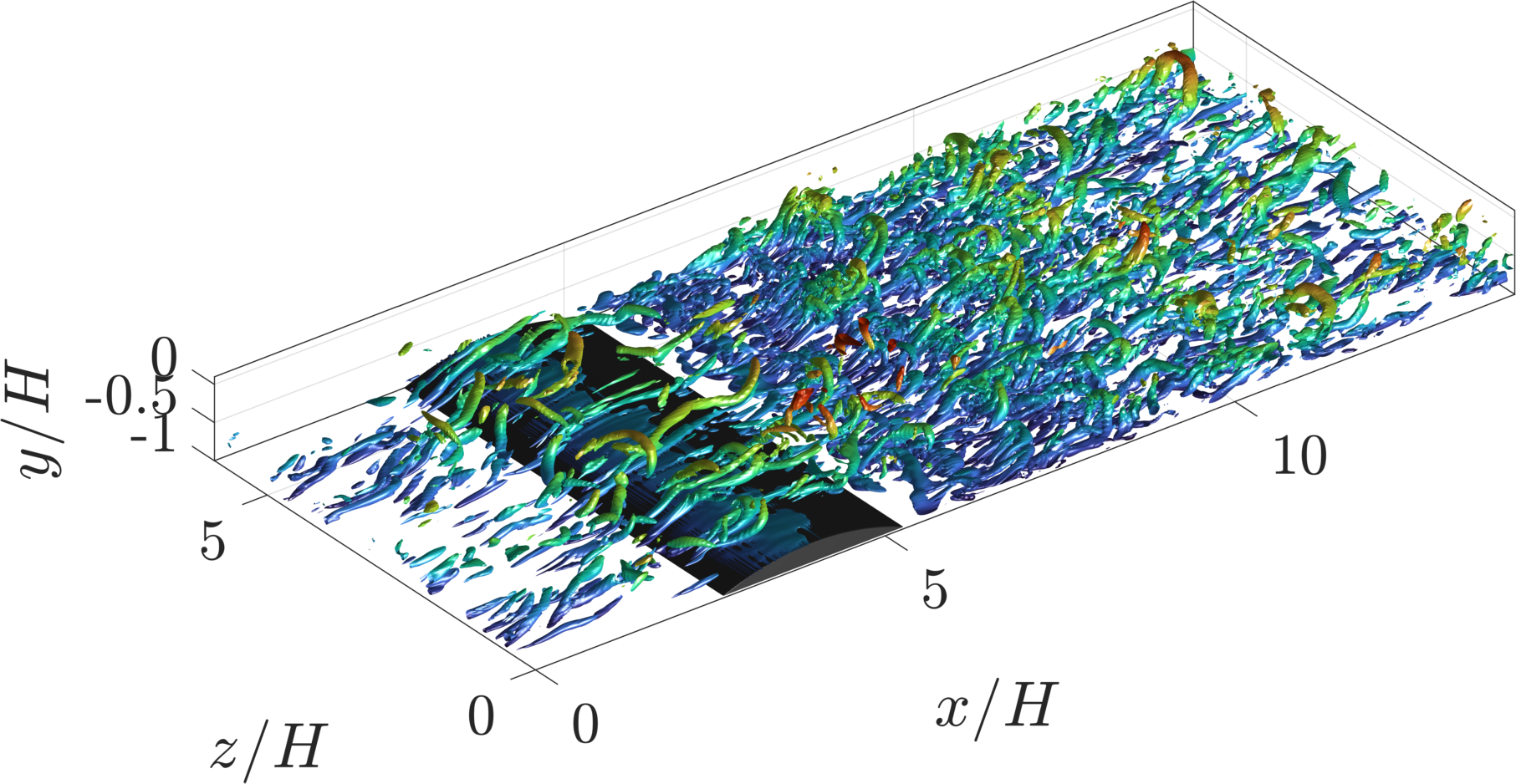}
\includegraphics[width=0.49\textwidth]{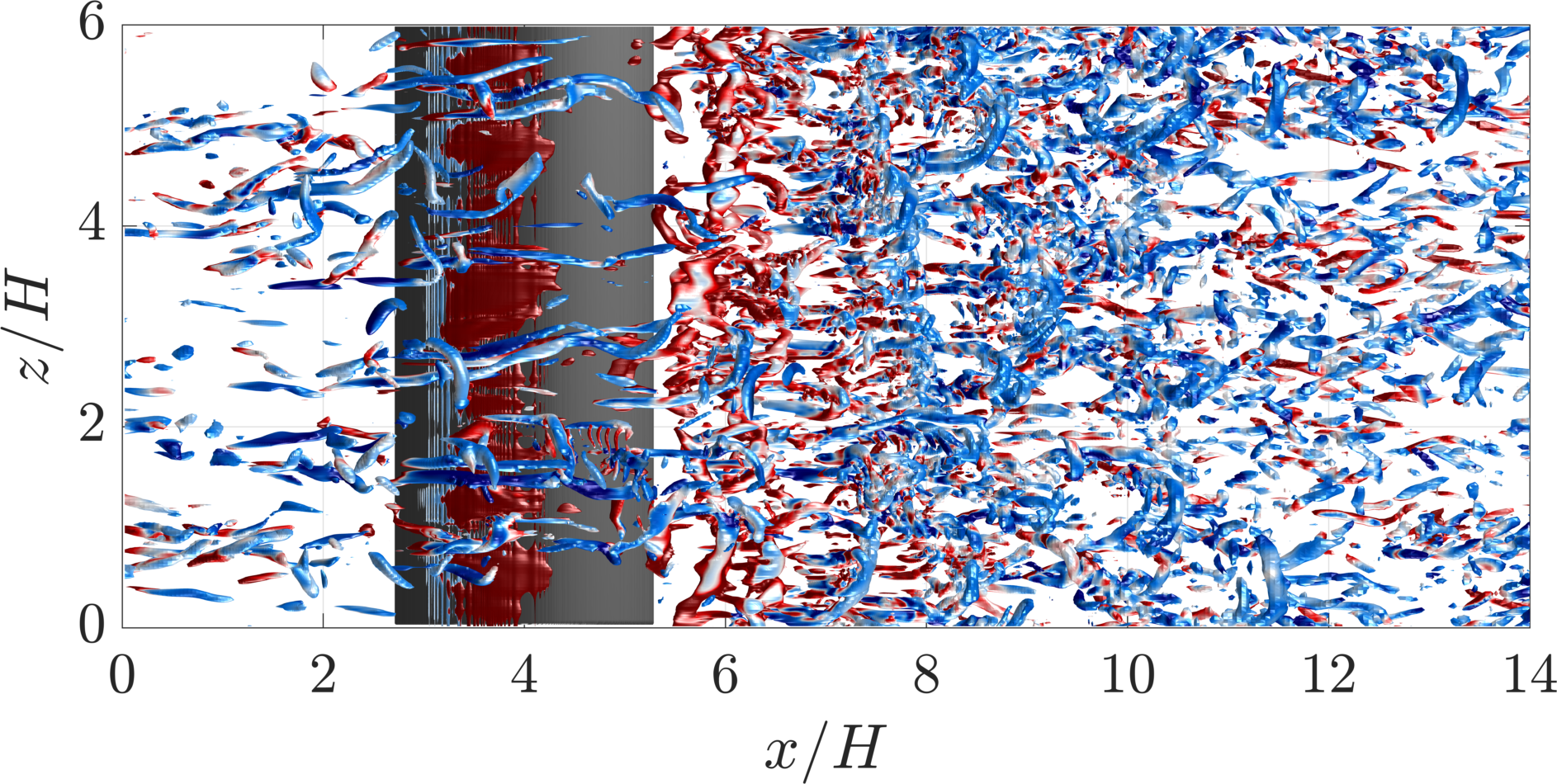}
\includegraphics[width=0.49\textwidth]{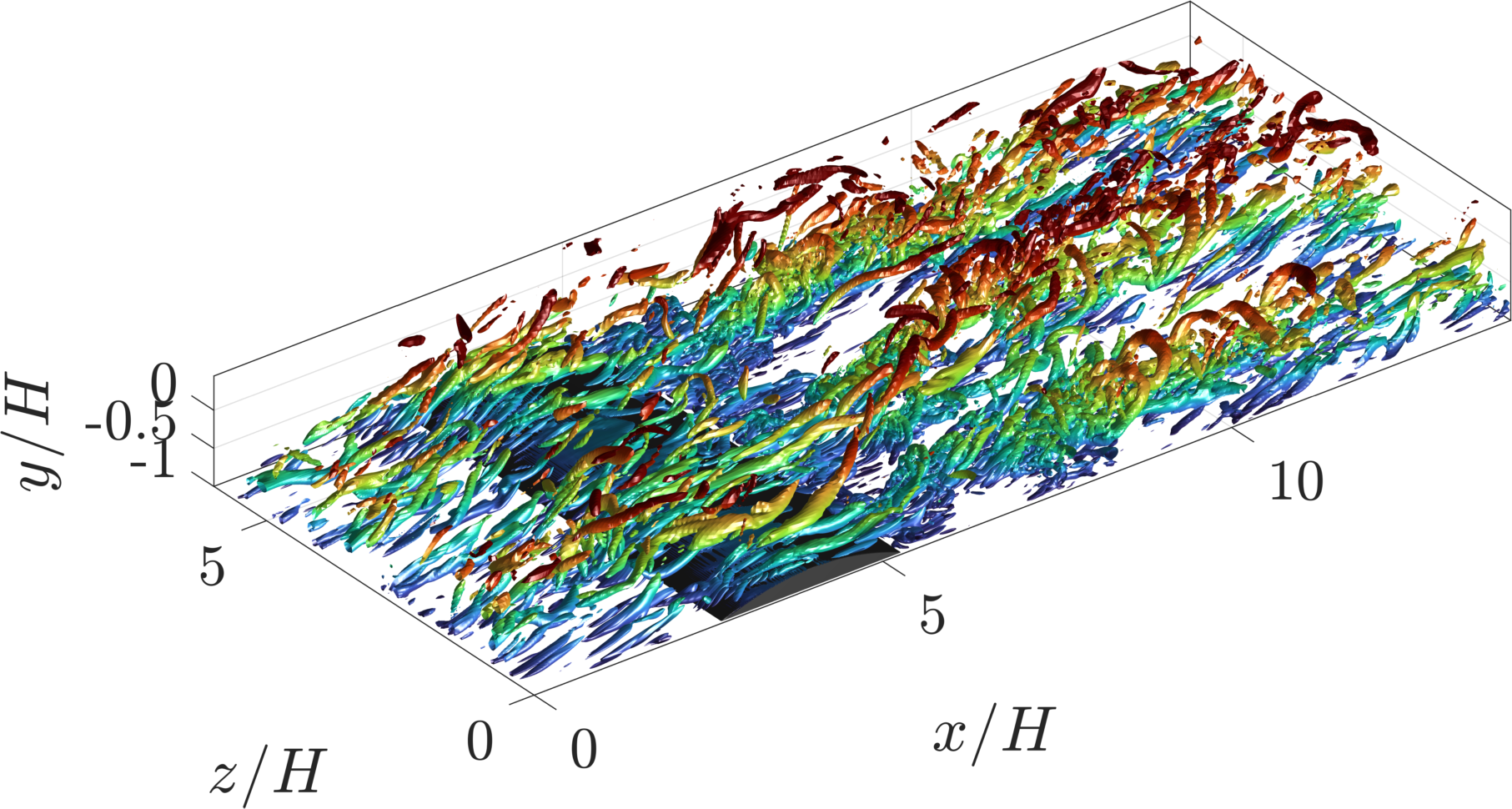}
\includegraphics[width=0.49\textwidth]{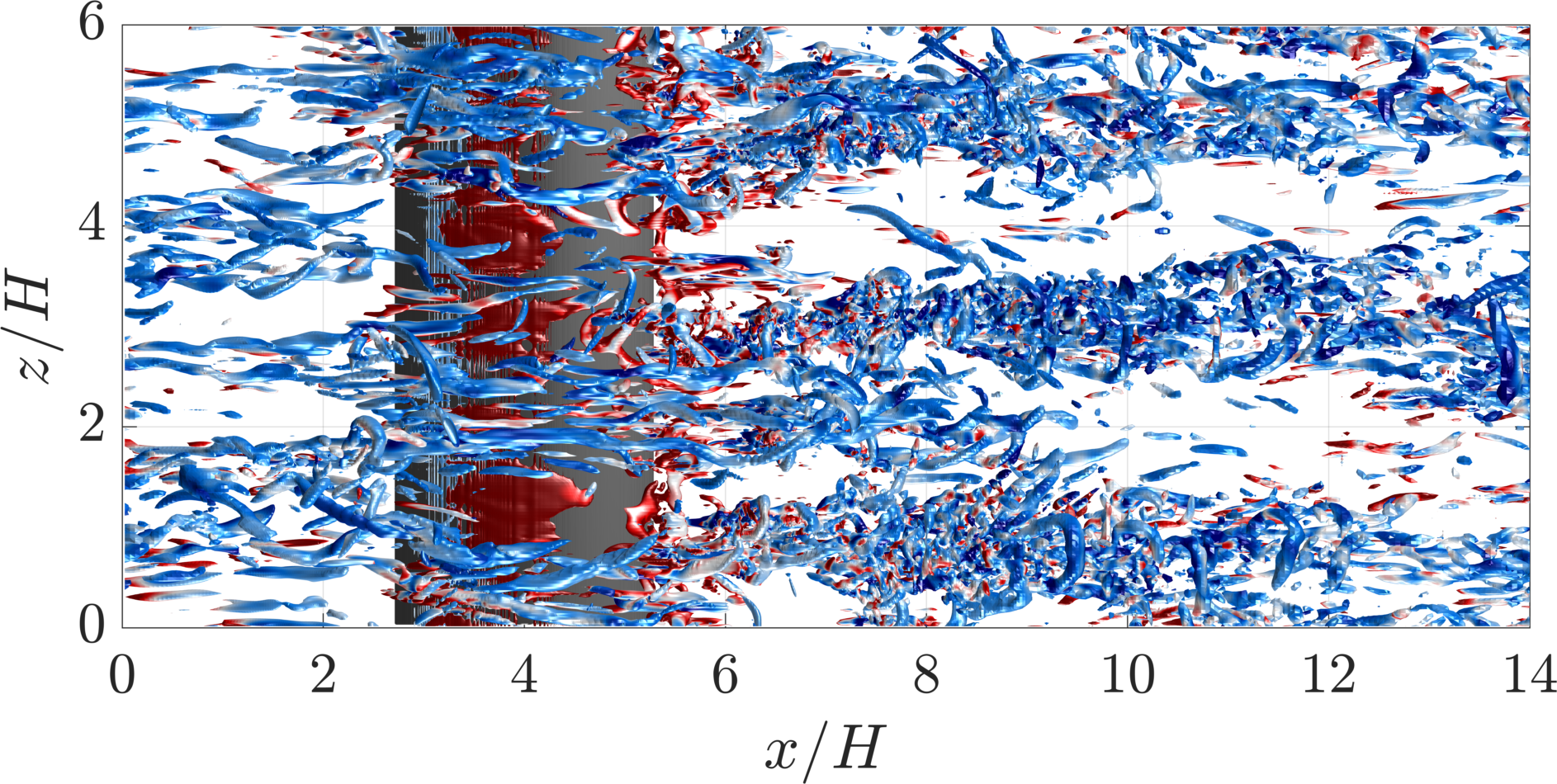}
\includegraphics[width=0.49\textwidth]{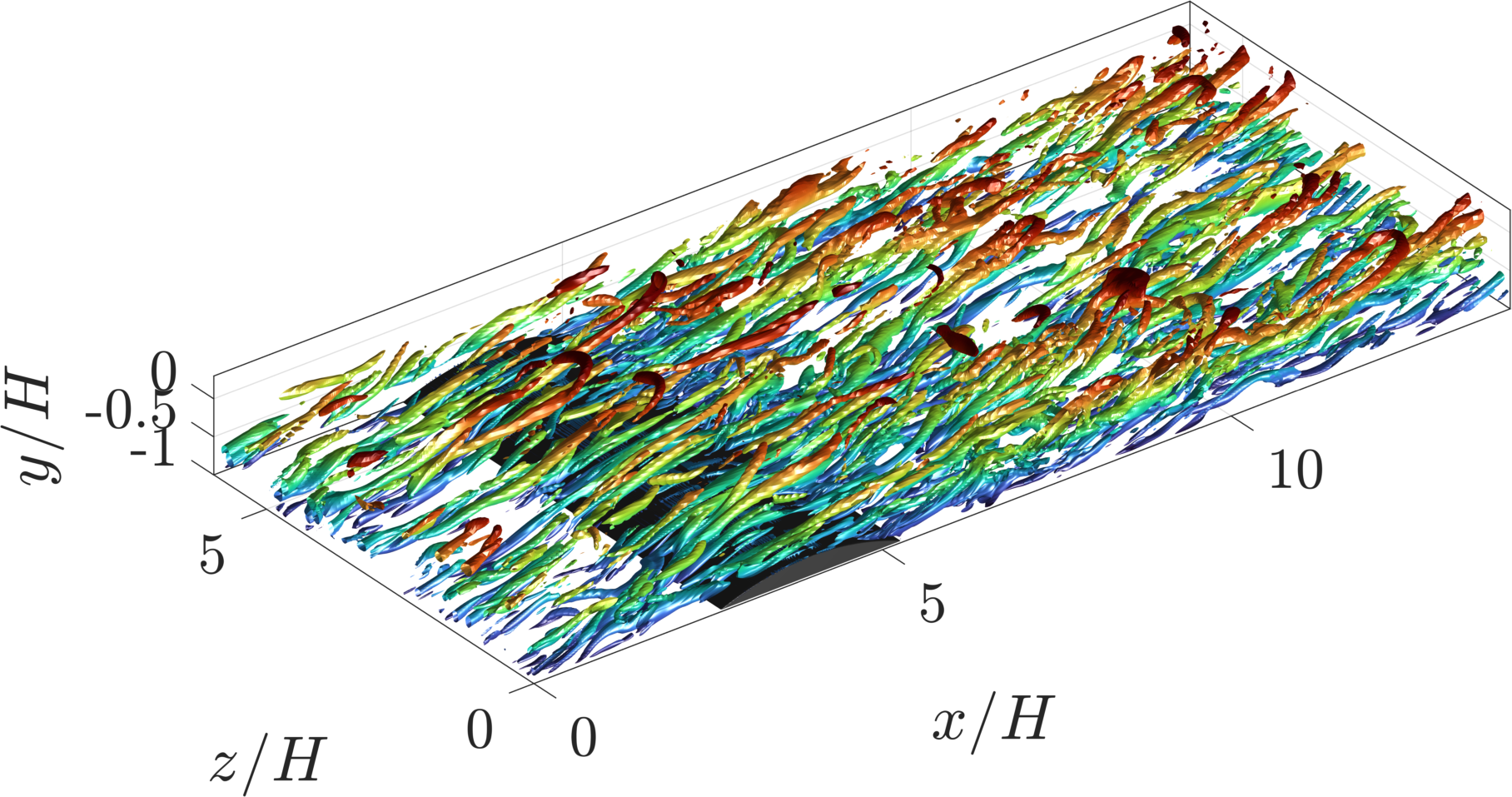}
\includegraphics[width=0.49\textwidth]{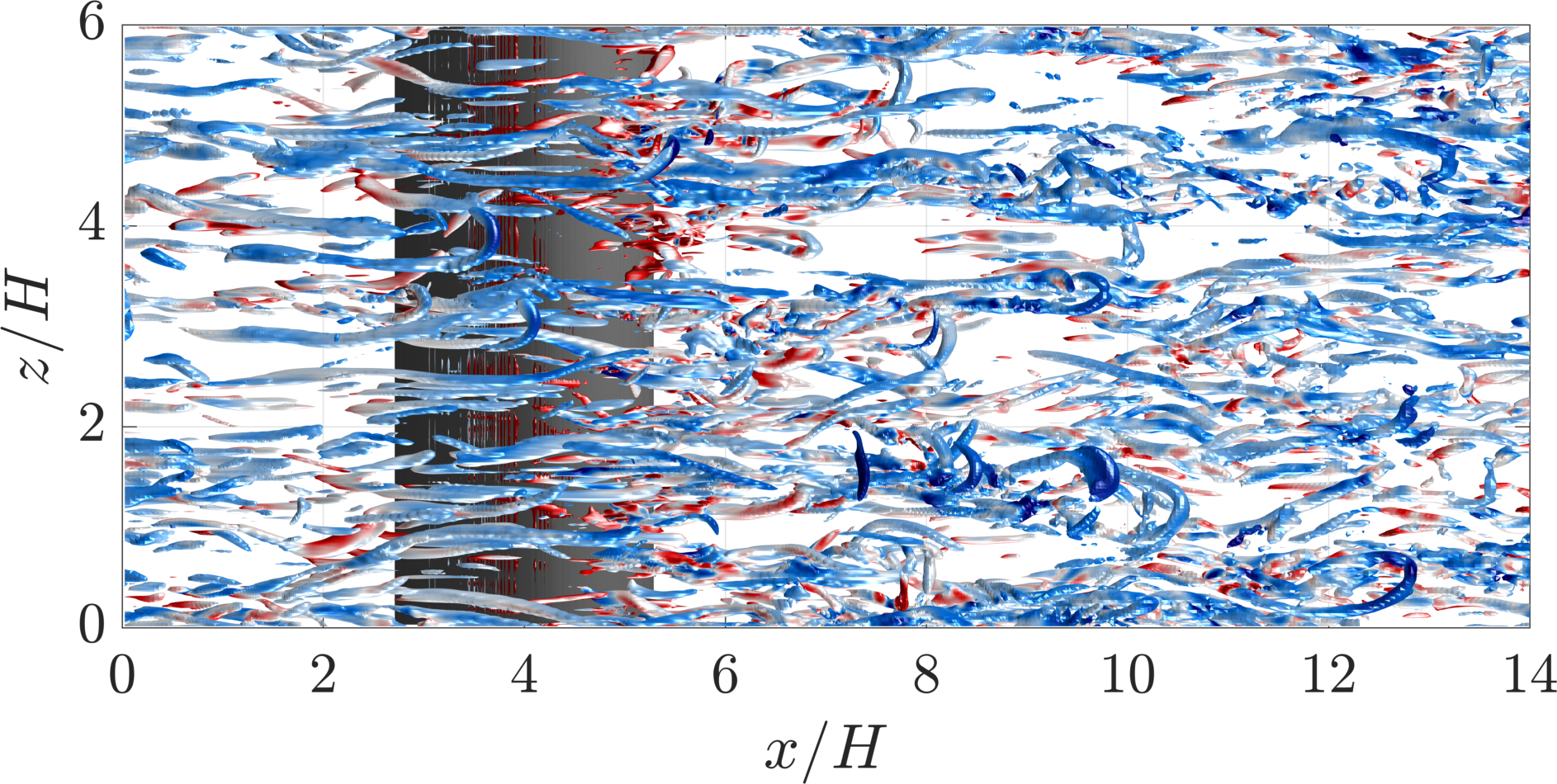}
\includegraphics[width=0.49\textwidth]{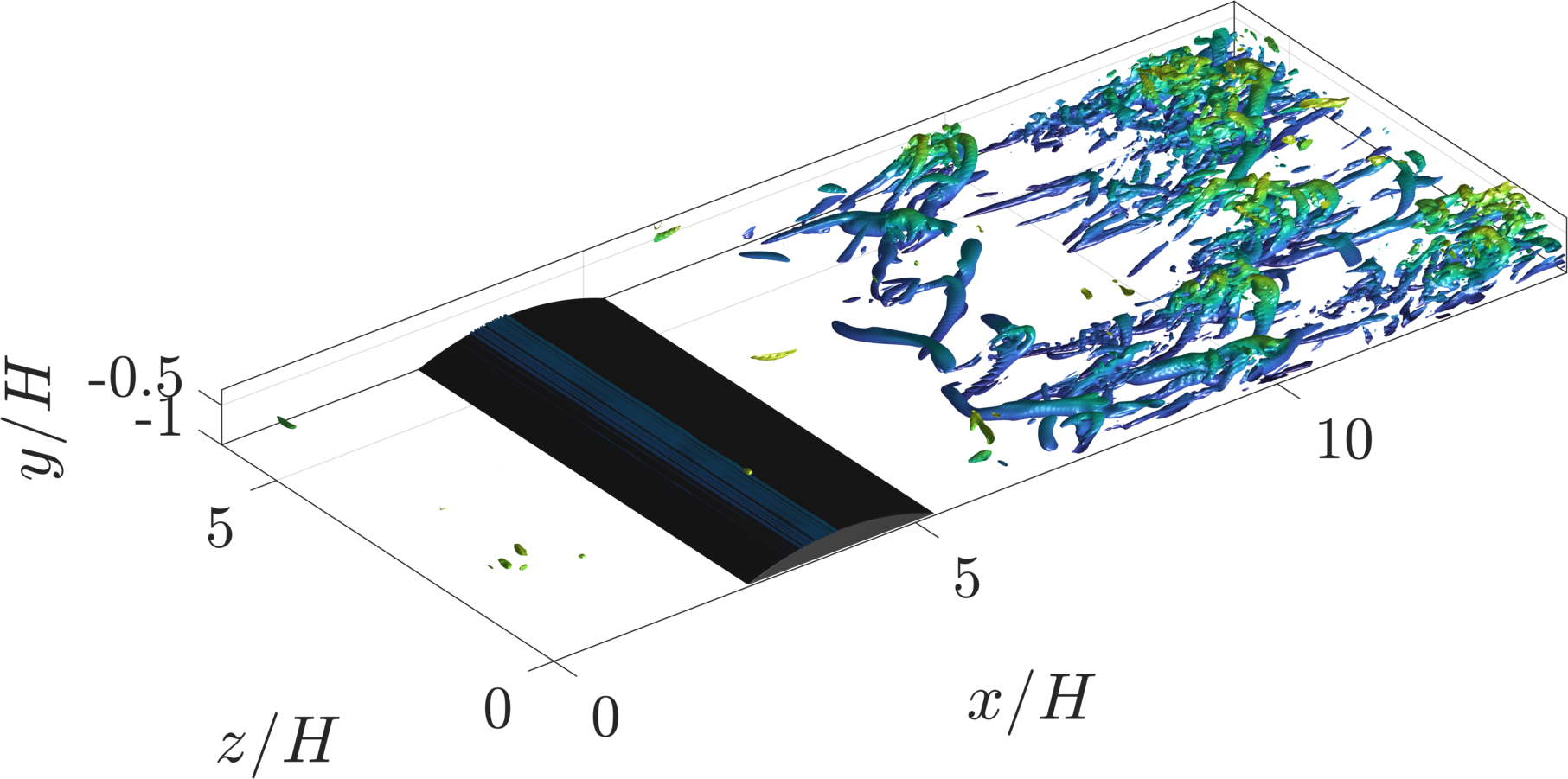}
\includegraphics[width=0.49\textwidth]{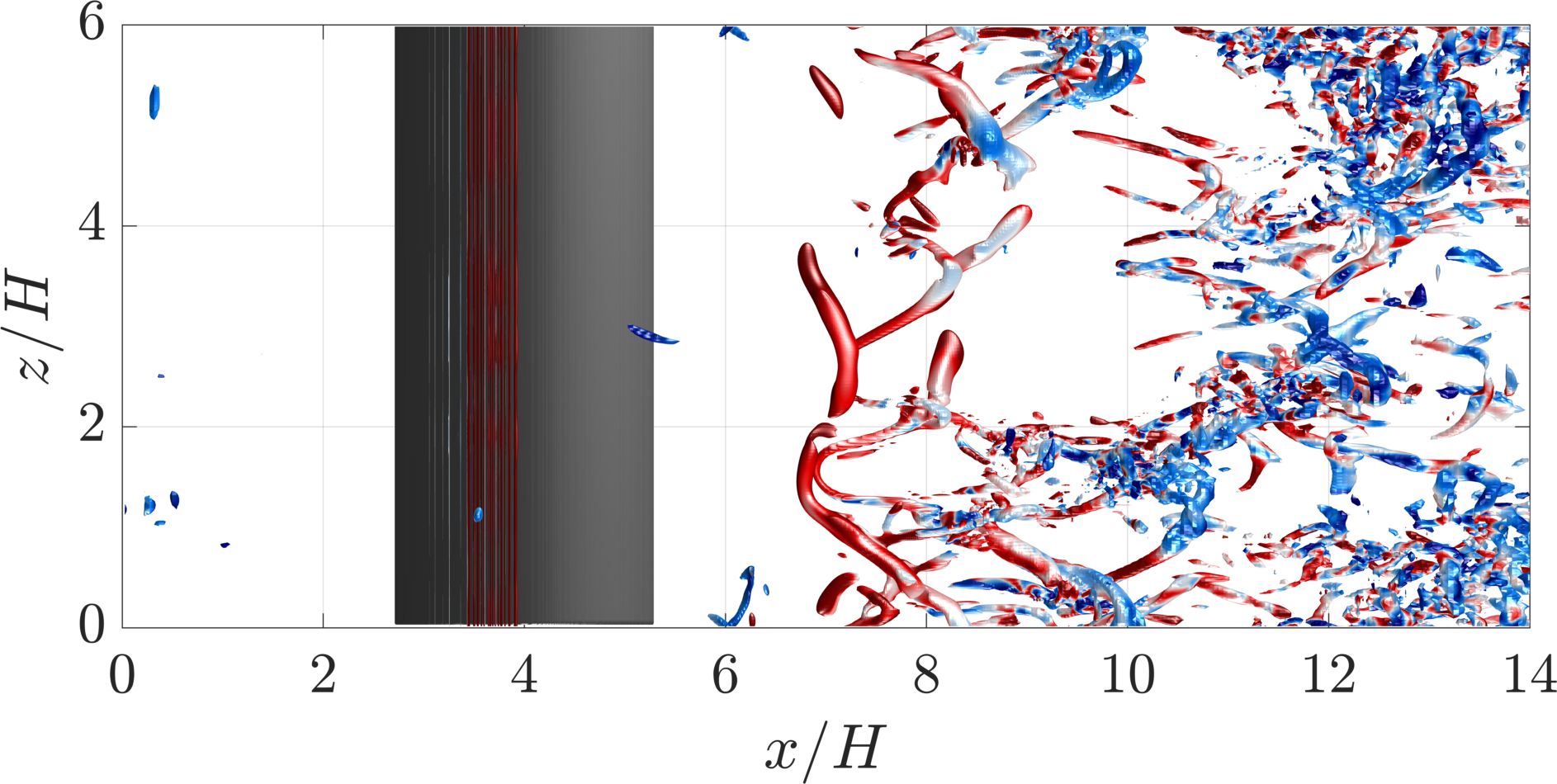}
\includegraphics[width=0.49\textwidth]{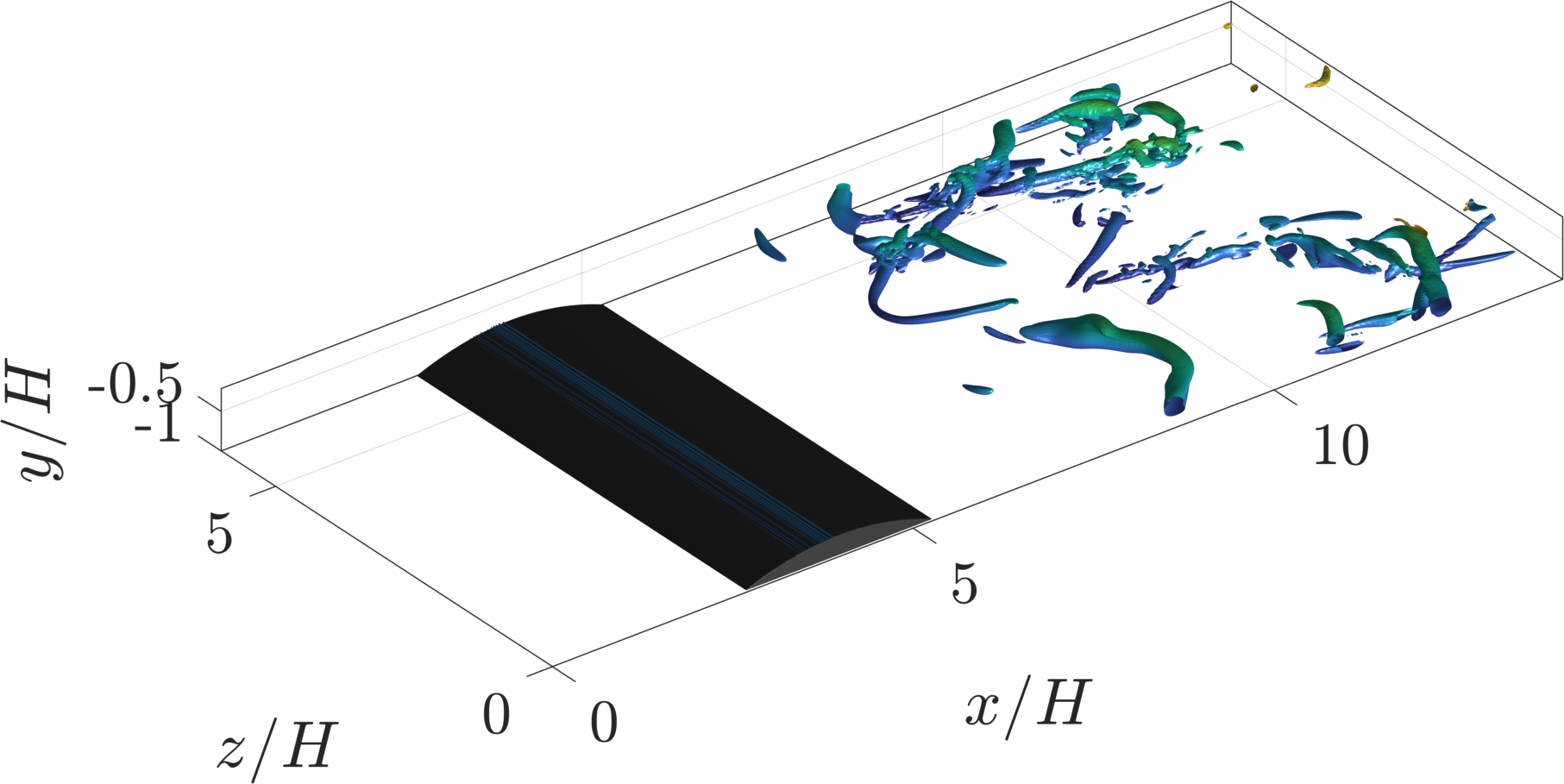}
\includegraphics[width=0.49\textwidth]{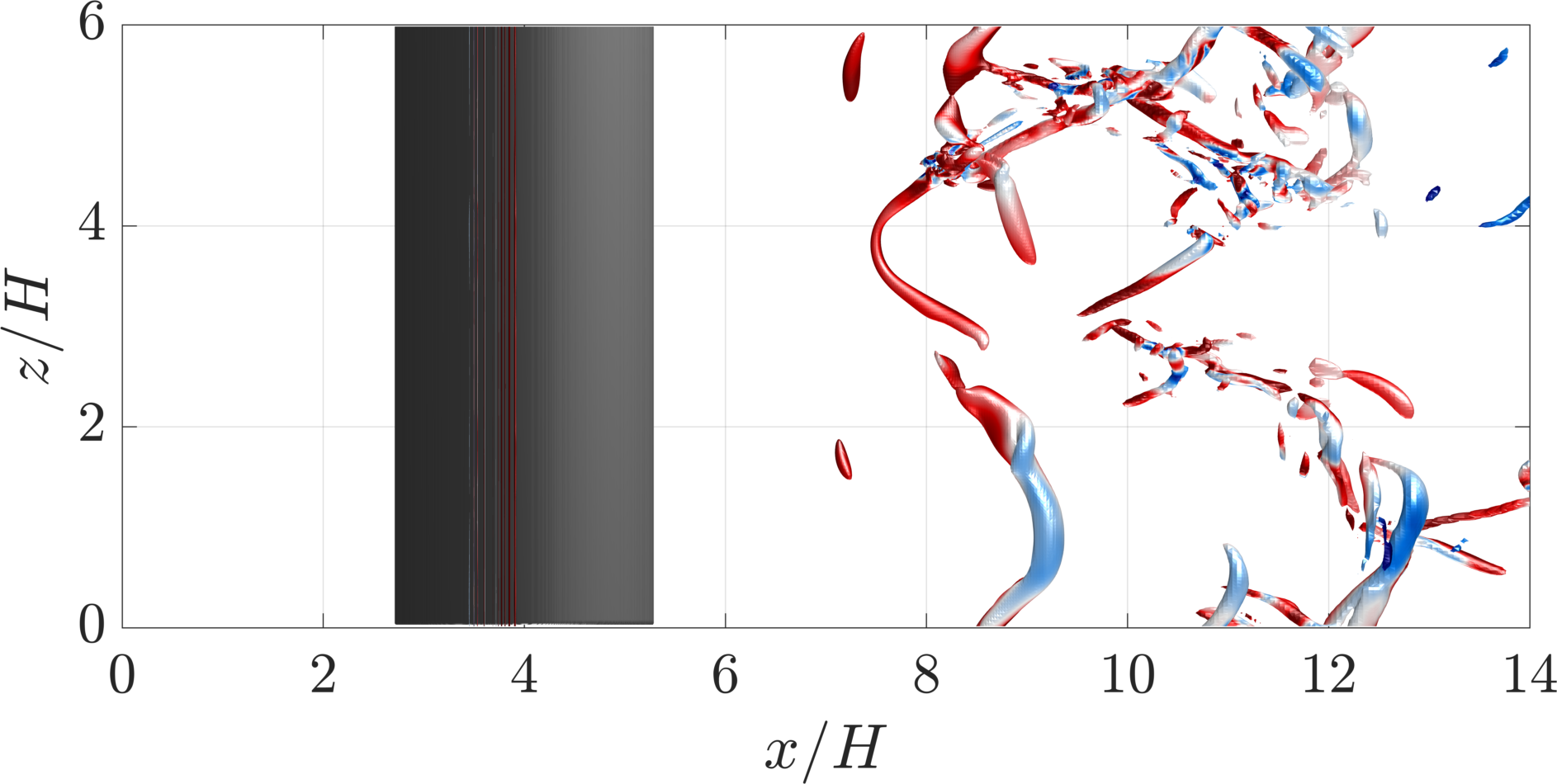}
\includegraphics[width=0.2\textwidth]{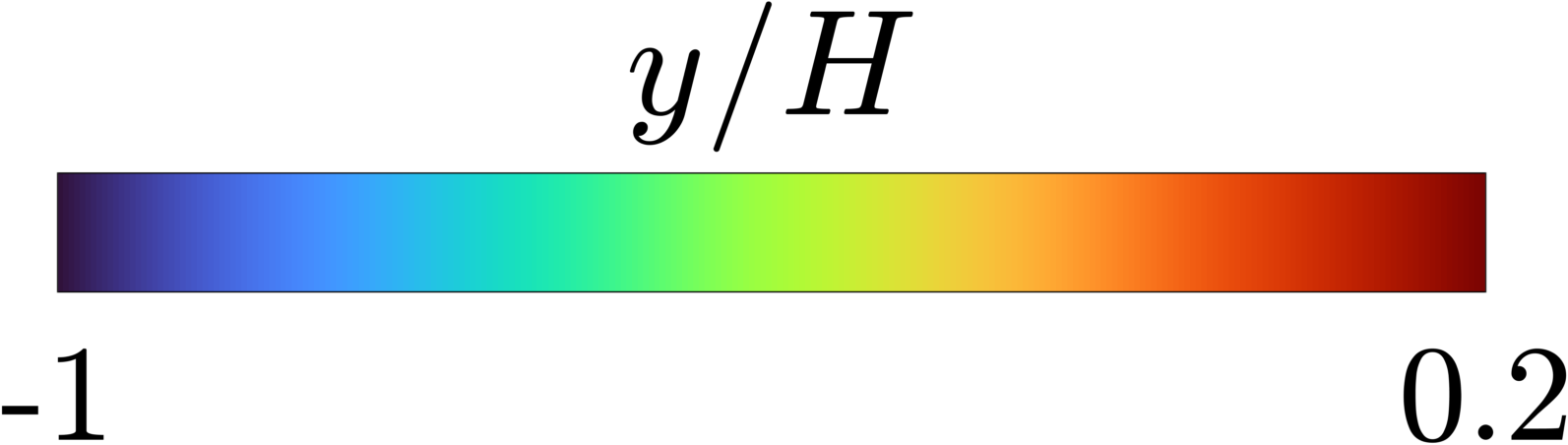}
\includegraphics[width=0.2\textwidth]{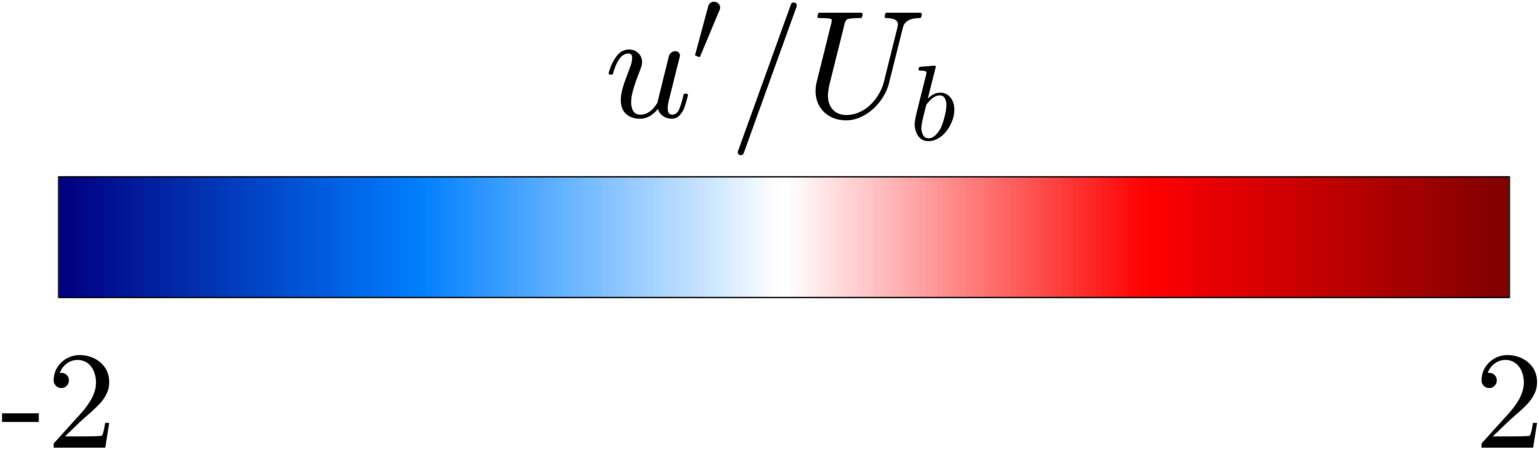}
\caption{Isosurfaces of the second invariant of the velocity gradient tensor visualized at a level of $Q=2.0 U_b^2/H^2$. Left column: isometric view colored by wall-normal distance to the bottom wall ($y/H$). Right column: top view colored by instantaneous streamwise velocity fluctuation ($u'/U_b$). From top to bottom, rows correspond to cases 00, P04, P10, N04, N10, as in preceding figures.}
\label{fig:q_iso_3d}
\end{figure} 

\section{Characterization of embedded structures}\label{sec:struct}
In this section, we aim to relate the rotation effects on Reynolds stresses to the dynamic structures in the flow, \textit{i.e.}, the hairpin vortices of turbulence, Taylor-G\"ortler (TG) vortices, roller vortices in the separating shear layer, and the oblique waves on the stable side.

\subsection{Instantaneous flow field}
\begin{figure}
\centering
\includegraphics[width=0.49\textwidth]{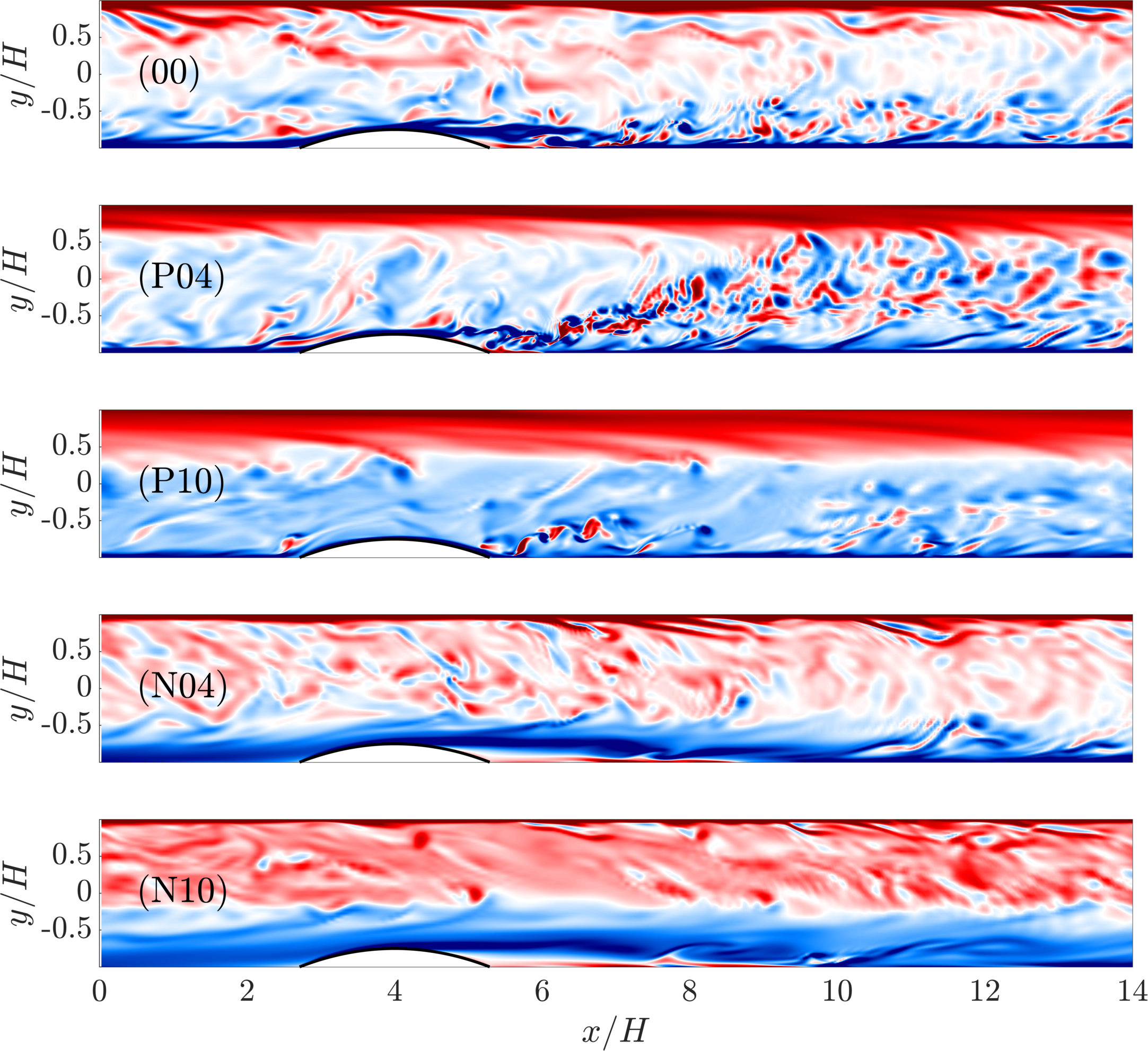}
\includegraphics[width=0.49\textwidth]{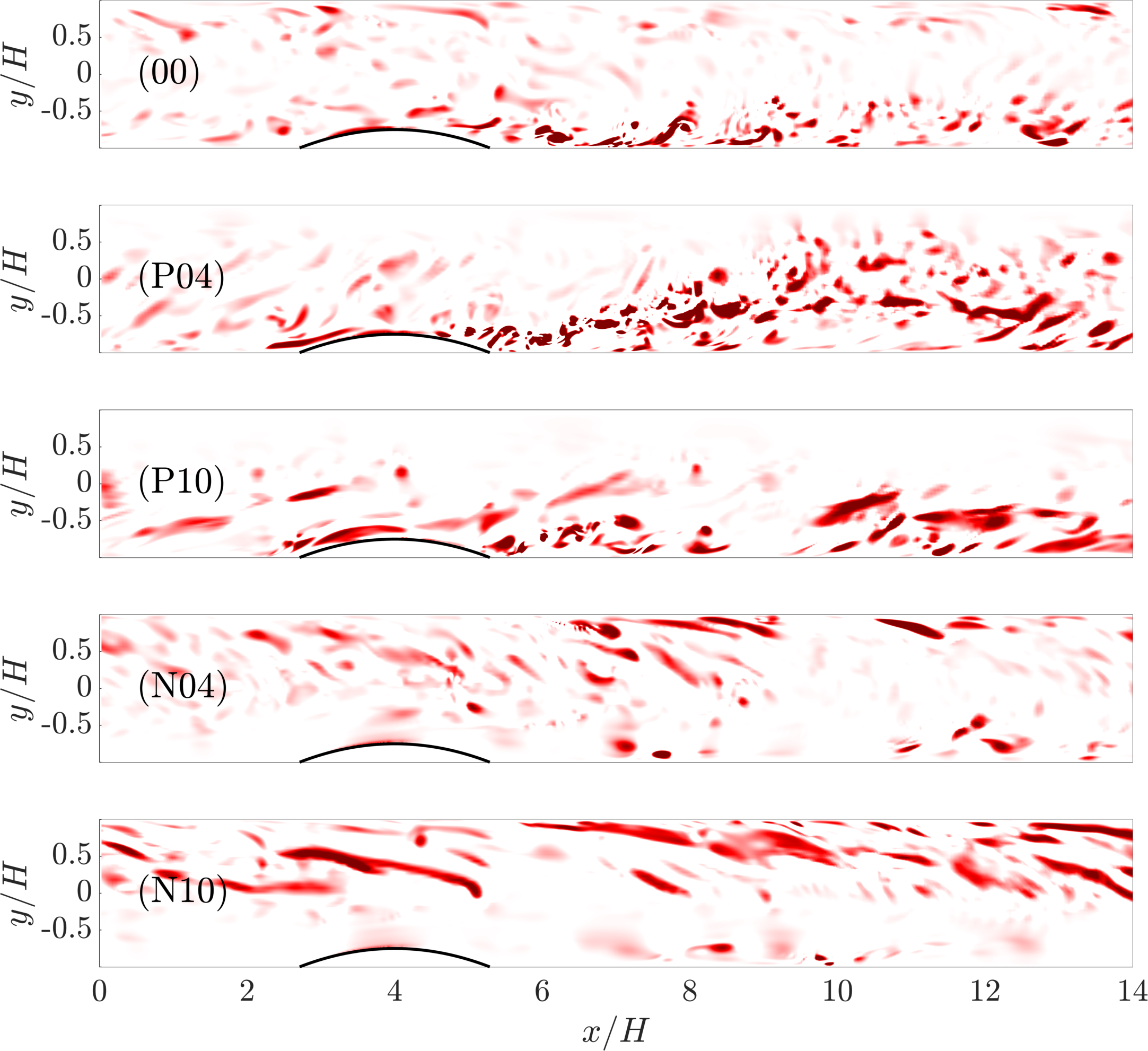}
\hspace * {0.4cm}  \includegraphics[width=0.15\textwidth]{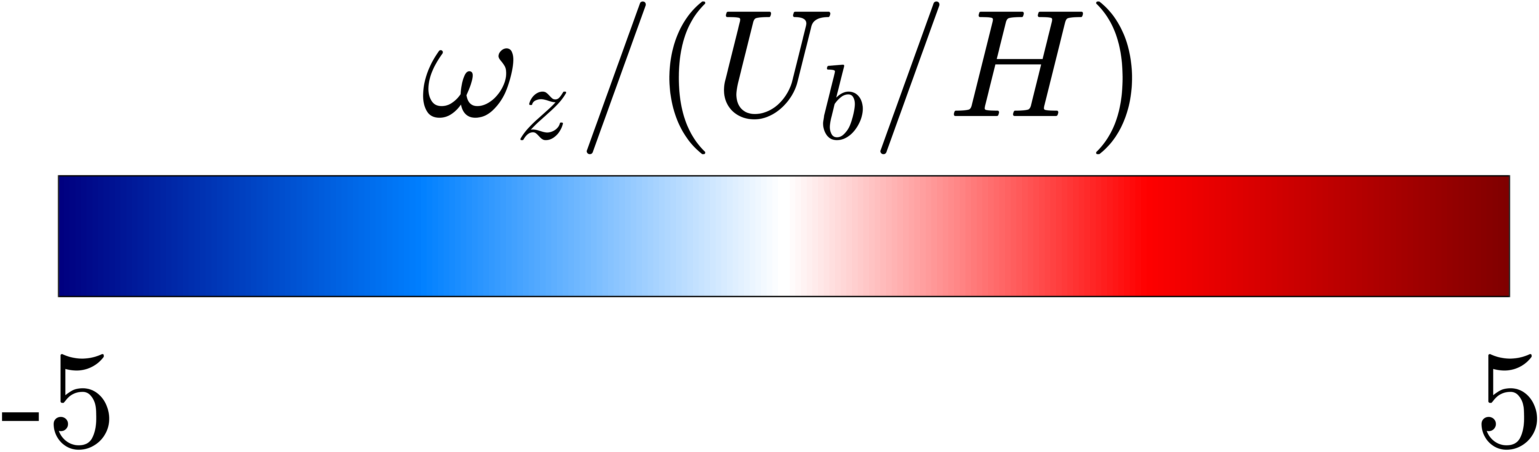}
\hspace * {3.5cm}\includegraphics[width=0.15\textwidth]{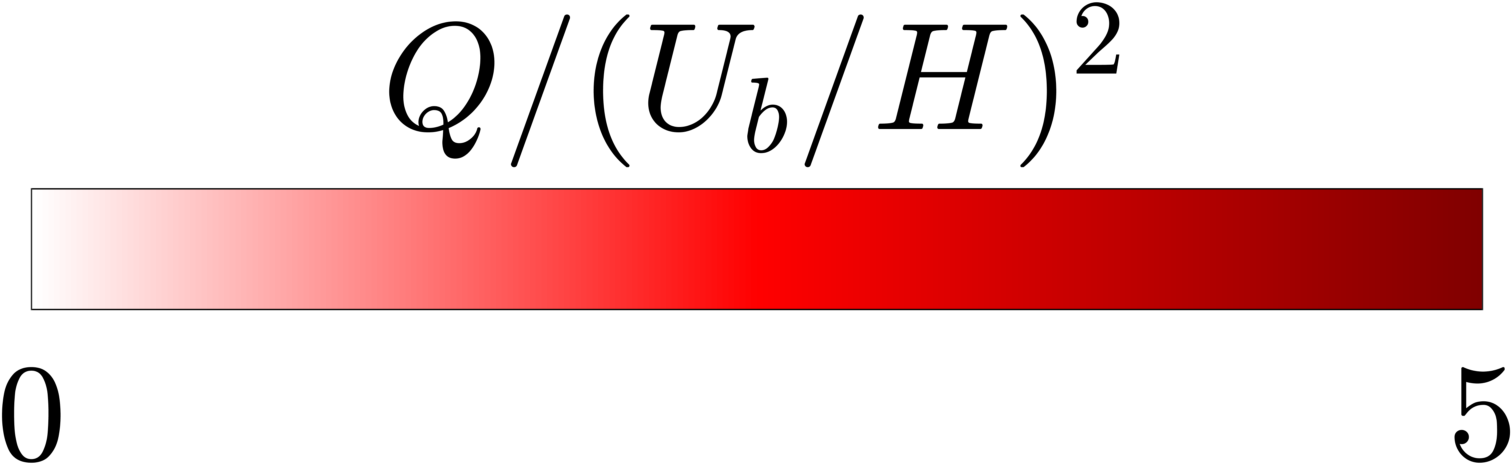}
\caption{Contours of instantaneous flow in the channel midspan. Left: spanwise vorticity ($\omega_z$). Right: second invariant of the velocity gradient tensor ($Q$, see text).}
\label{fig:inst_xyplanes}
\end{figure}

\begin{figure}
\centering
\includegraphics[width=0.4\textwidth]{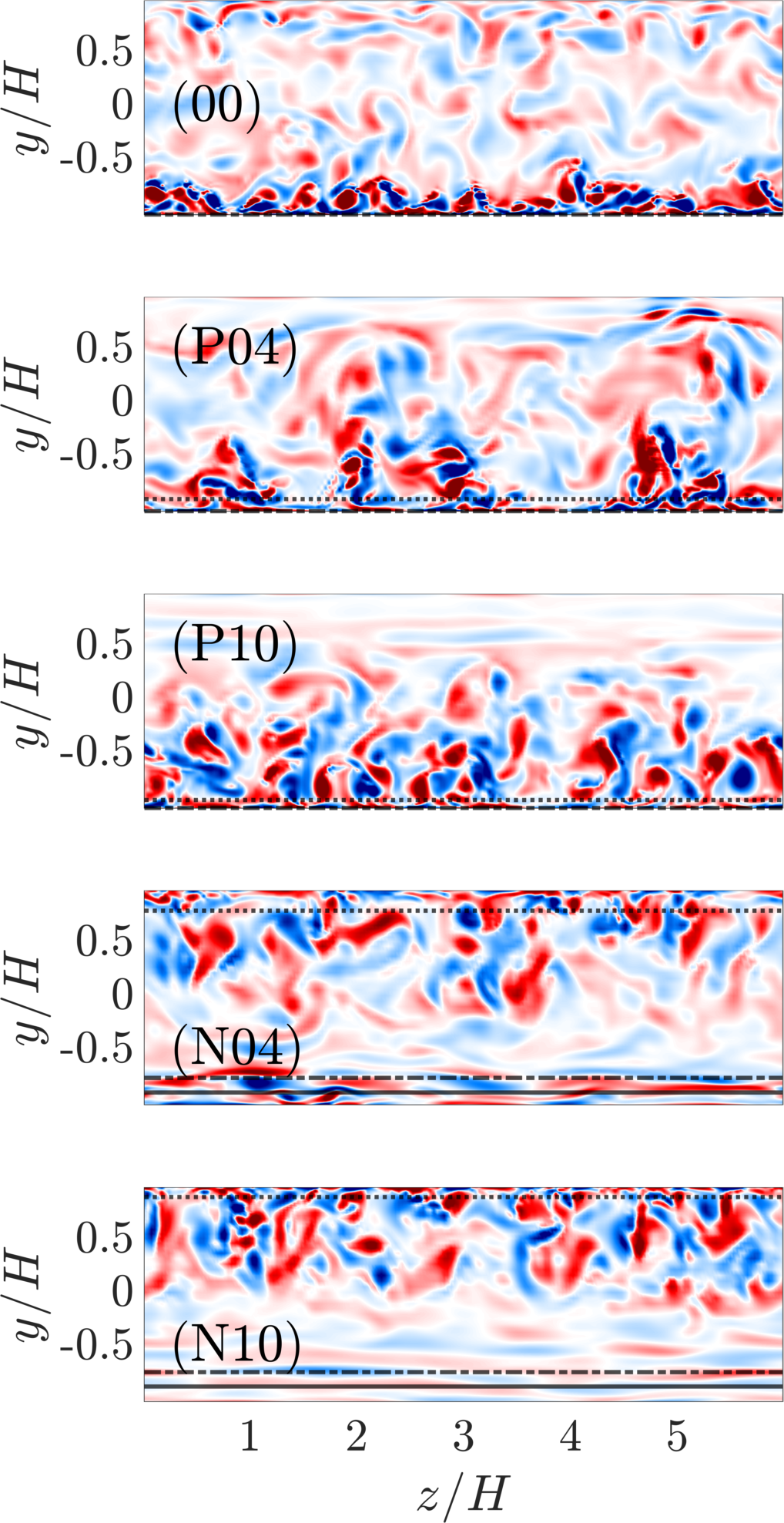}
\includegraphics[width=0.4\textwidth]{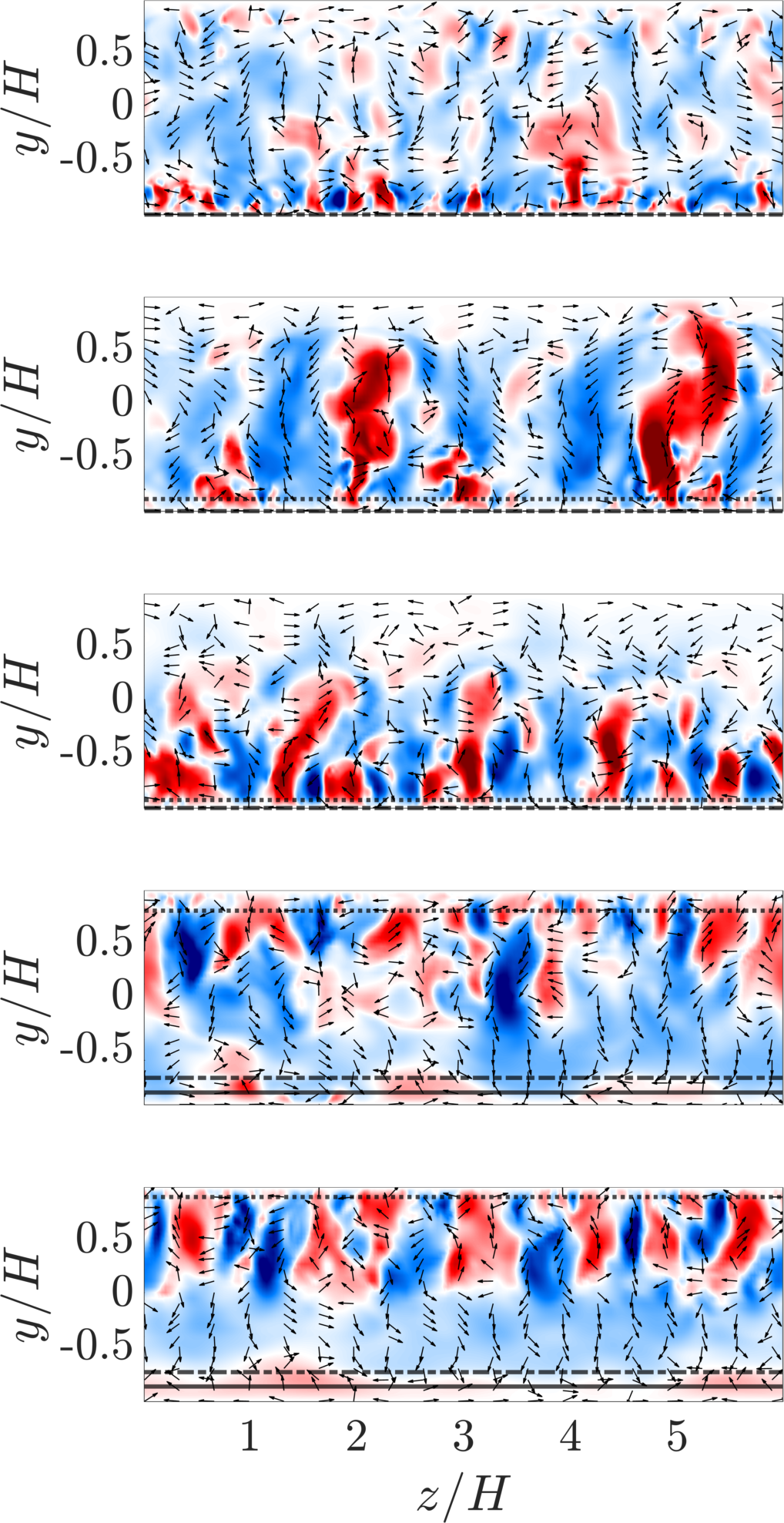}\\
\hspace * {0.4cm} \includegraphics[width=0.15\textwidth]{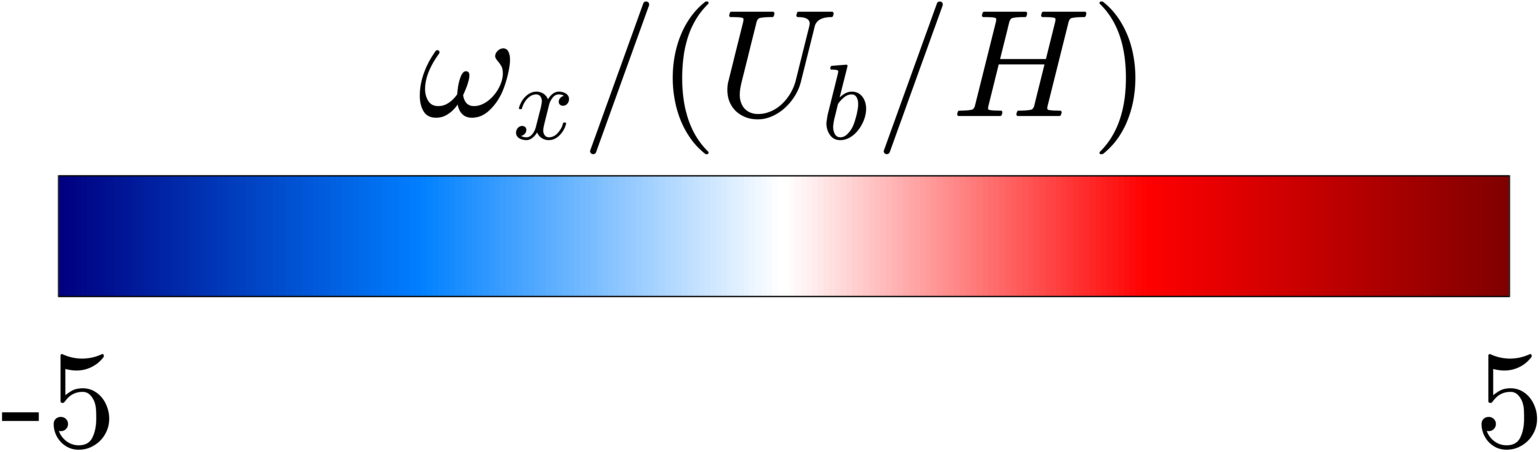}
\hspace * {3.5cm} \includegraphics[width=0.15\textwidth]{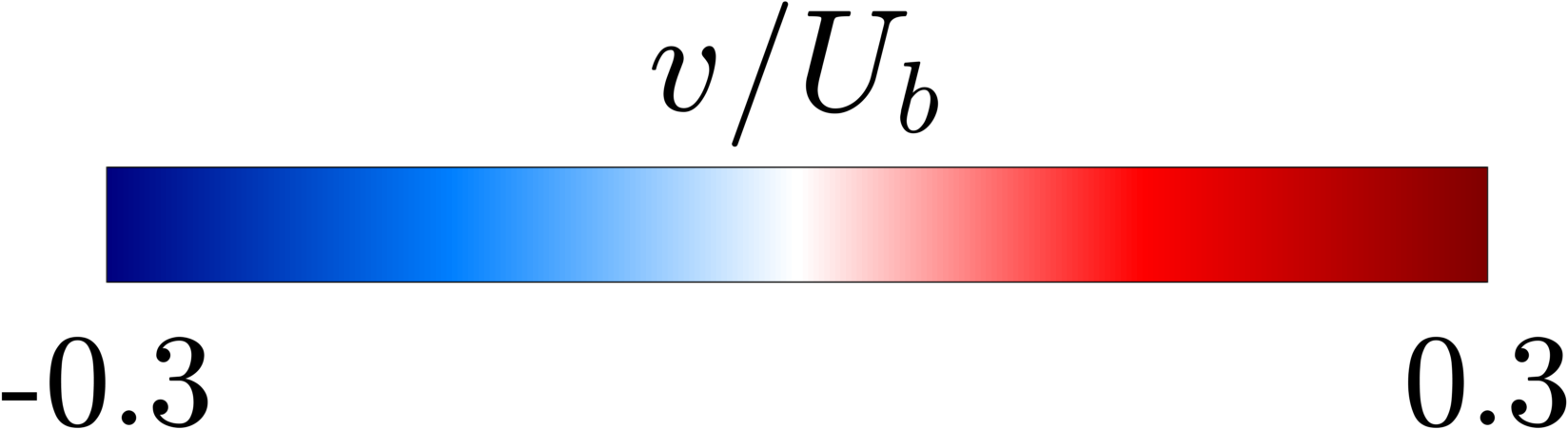}
\caption{Contours of instantaneous flow in the cross-flow plane at $x/H=7.0$. Left, streamwise vorticity $\omega_x$; Right: wall-normal velocity and velocity vectors. The vectors are shown every $0.3H$ in $z$ and $0.1h$ in $y$, normalized by their magnitude for clarity. Horizontal lines represent (if applicable): \dotted  $y\vert_{S=-0.5}$; \dashdot $y\vert_{\mathrm{max}(\partial U/\partial y)}$; \solid $y\vert_{U=0}$. }
\label{fig:inst_zyplanes}
\end{figure}

The small-scale structures are shown by the isosurfaces of the second invariant of the velocity gradient tensor ($Q:= 
- 1/2 \partial u_j/\partial x_i \partial u_i/\partial x_j$), 
in figure \ref{fig:q_iso_3d}. In figure \ref{fig:inst_xyplanes}, spanwise vorticity ($\omega_z$) and $Q$ are extracted in the streamwise-wall normal ($x-y$) plane at the channel mid-span to visualize the shear layer and the wall-normal distribution of the vortices. Notably, TG vortices are not able to be visualized by these quantities due to their large-scale coherence. Thus, instantaneous flow fields of streamwise vorticity ($\omega_x$) and wall-normal velocity in the cross-flow ($z-y$) plane in the wake of the bump are provided in figure \ref{fig:inst_zyplanes} to visualize them.  They are observed by the large-scale counter-rotating swirls, indicating alternating regions of upwash and downwash. The absence of vortices in the streamwise-elongated regions in the top view of the $Q$ isosurfaces (figure \ref{fig:q_iso_3d}) serves as an additional depiction of TG vortices.

When the channel is not rotating, a roller vortex can be seen underneath a nest of hairpin vortices in the separating shear layer. The latter dominates the downstream region with several well-defined arched heads appearing randomly across the domain.
In the positive rotating cases, the separating shear layer is hardly noticeable in the $Q$ isosurfaces, although the spanwise vorticity contour in figure \ref{fig:inst_xyplanes} captures the formation of one roller vortex in case P04. The hairpin vortices in the incoming flow are elongated in the streamwise direction as shown by the $Q$ isosurfaces. They are longer and more parallel to the wall in case P10 than in P04.
This agrees with the findings in the literature that rotational instability favors the growth of streamwise vortices \citep{Lesieur91, Lamballais96, Yanase93, Metias95} and that hairpin legs are enhanced in rotating flows \citep{Iida08, Yang12, Brethouwer17}.

The wake of the bump exhibits rich dynamics represented by the evolution of the hairpin vortices. In the spanwise vorticity and $Q$ contours at the channel midspan (figure \ref{fig:inst_xyplanes}), it can be seen that the series of hairpin vortices are distributed along a path that extends from the separating shear layer towards the channel centerline. This trajectory aligns with the extended $-1<S<0$ region and where \vv and \ww exhibit enhancement. This trend can also be observed in the $Q$ isosurfaces (figure \ref{fig:q_iso_3d}) and the cross-flow contours (figure \ref{fig:inst_zyplanes}), where a clear correlation of the region between TG vortex pairs (\textit{i.e.}, extreme upwash) and the ejection of hairpin vortices can be identified.
At the time instants shown in these figures, four ejection gaps can be seen in case P04 with more visible in case P10. This agrees with the literature that TG vortices are smaller and closer to the anti-cyclonic wall at high rotation rates.
Correlating the cross-flow contours, the $Q$ isosurfaces, and the Reynolds stresses contours (figures \ref{fig:uu} - \ref{fig:uv}), the eddies being ejected have large streamwise vorticity near the wall, thus corresponding to the legs of the hairpin vortices. The bottom of these legs creates the high \ww in the vicinity of the wall, while their rotation appears as the strong \vv and $\overline{u'u'}$. Beyond $y/H\sim$ -0.6 to -0.7, the streamwise vorticity decreases, forming arched structures in the cross-flow contours. This is where the head of the hairpin is formed.  The vortex filaments generated by the breakdown of the hairpin heads can be ejected up to the cyclonic side of the channel around $y/H\sim 0.5$ (refer to 2D $Q$ contour). Due to the smaller TG vortices at the higher rotation rate, the width and ejection height of the hairpin vortices in the wake of the bump are smaller in case P10 than in P04.

We interpret the correlation between the $-1<S<0$ region and the trajectory of the hairpin vortices as follows. 
Recall that the lee-side of the bump is in the destablized regime, as discussed in \S\ref{sec:stabreg}. Therefore, hairpins here are ejected into a region where they are augmented and able to extract more significant energy from the mean flow (versus being ejected into a neutrally-stable regime, as occurs in a planar rotating channel). Furthermore, 
the ejection regions between TG vortices are found to have negative $u^\prime$, as clearly visible in the top view of the $Q$ isosurfaces.
Thus, the ejection mechanism is similar to the ejection events between low-speed streaks in turbulent boundary layers: a local low streamwise velocity will increase the wall-normal velocity by mass conservation. However, the causation is the opposite here since the upwash generated by the TG vortex pair initiates the local deceleration.
The reduced streamwise velocity increases the $\partial u/\partial y$ in the wake of bump (\textit{i.e.}, away from the wall). As discussed in the stability regime section (\S \ref{sec:stabreg}), an increased velocity gradient leads to a less negative $S$ that makes such regions unstable. The extraction of kinetic energy from the mean flow to \uu and redistributed to \vv via the Coriolis terms is the statistical description of the augmentation of the hairpin vortices mentioned above. Moreover, we attribute this process to the inclined extension of the destabilization region that reaches up to the channel center line. On one hand, it can be viewed as a faster recovery of the near-wall region to the neutral $S=-1$ regime, compared to the channel centerline. On the other hand, it also represents the consequence of the continuous mean momentum extraction as the enhanced hairpin vortices are ejected. Both of these two processes are characterized by the Reynolds stress divergence in the mean momentum budget and the production and turbulent diffusion in Reynolds stress budgets. 
The physical picture corresponding to this process is that as the hairpin vortices are lifted in the ejection regions, their (enhanced) legs bring high momentum to the region underneath by their rotation to facilitate the recovery of the near-wall flow. Yet, they keep draining the mean momentum at their height through the Coriolis terms. Thus, the region away from the wall attains a prolonged modulation. This process continues until the TG vortices cannot further sustain the ejection of the enhanced hairpin vortices, and/or the hairpin vortices break down such that they no longer extract significant energy from the mean flow.

In the negative rotating cases, the hairpin and TG vortices are on the opposite side of the bump and do not directly impact the separating shear layer. The $Q$ isosurfaces show the footprints of the oblique waves on the roller vortices formed far downstream of the bump. A spanwise wavelength about half of the spanwise domain size can be seen. Case N04 exhibits significantly more distorted vortices than N10. This is consistent with the literature that turbulent spots on the stable side diminish as the rotation rate increases \citep{Grudenstam08, Brethouwer17}. Note that, there are diverse opinions about the origin of these turbulent spots. Local cyclic turbulent bursts by the growth and breakdown of Tollmien-Schlichting (TS) waves and turbulent fluctuations penetrating into the stable region are two main mechanisms proposed \citep{Kristoffersen93, Brethouwer14, Dai16, Brethouwer16}. For the flow in the current study, we are inclined to the latter because: 1) there is a prominent ejection of the hairpin vortices excited by the extended unstable region in the wake of the bump; and 2) the relatively low $Re_\tau$ and $Ro_\tau$ prevents the TS waves from reaching very large amplitude \citep{Brethouwer14}.
The cross-flow contours of these two cases exhibit that the head of the hairpin vortices ejected from the opposite side indeed reaches the separating shear layer in case N04, while it does not seem so in case N10. This is again because the TG vortices are smaller and nearer the anti-cyclonic wall as the rotation rate increases.

The dynamics of the structures elaborated above indicate that, as long as there is an APG in the rotating system (or, a mainstream acceleration/surge), the near-wall region will have an increased $\partial U/\partial y$ that changes the neutrally stable region in the bulk anti-cyclonic flow to hydrostatically unstable. In such a condition, the hairpin vortices in the boundary layer will be augmented and sustain longer as they are ejected by the TG vortices, reaching further into the core of the flow (than in the same rotating system under a zero pressure gradient, in which the hairpin also get ejected by the TG vortices). 
Therefore, this process could also occur in attached flows, including on the anti-cyclonic side of our negative rotating cases. Indeed, we observed an increased $S$ near the top anti-cyclonic wall and enhanced \vv and \ww in cases N04 and N10 (figures \ref{fig:vv}, \ref{fig:ww}). Therefore, the turbulent patches superposed to the oblique wave and disturbing the laminar separating shear layer in case N04 are likely energized filaments of the ejected hairpins from the opposite (attached) side. 
Particularly, the increased \vv and \ww near the anti-cyclonic wall in cases N04 and N10 are most prominent not immediately downstream of the bump, but where the wall-parallel laminar shear layer starts to curve towards the wall, creating an expansion of the nominal cross-sectional area and thus an APG. This supports our justification that deceleration is what initiates a series of dynamic changes. 

\begin{figure}
\centering
\includegraphics[width=0.99\textwidth,trim={0 1.2cm 0 0},clip]{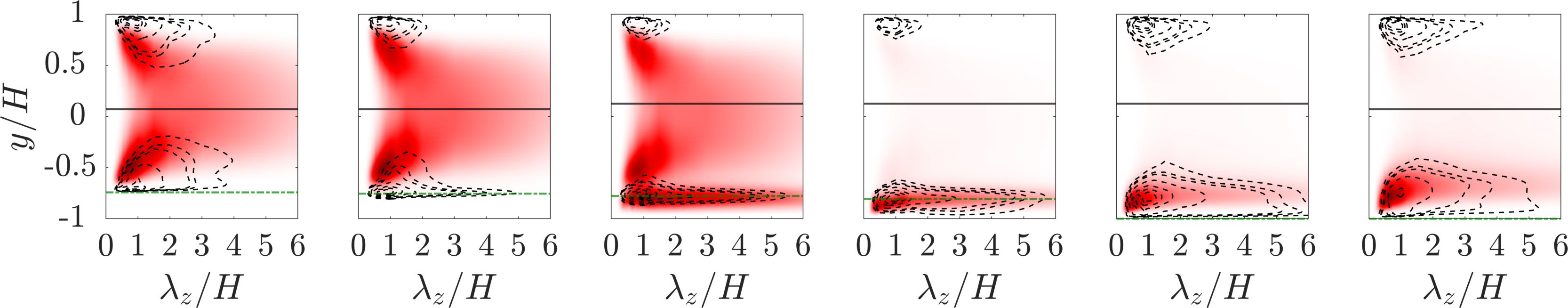}
\includegraphics[width=0.99\textwidth,trim={0 1.2cm 0 0},clip]{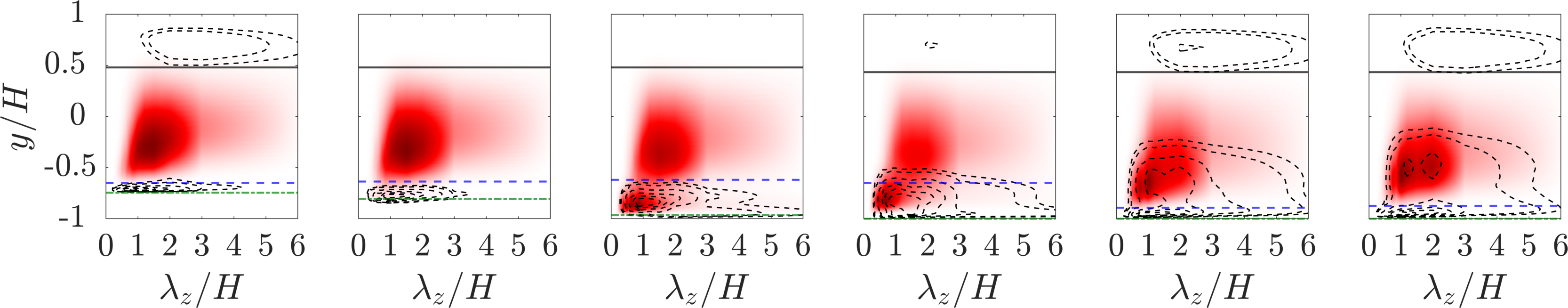}
\includegraphics[width=0.99\textwidth,trim={0 1.2cm 0 0},clip]{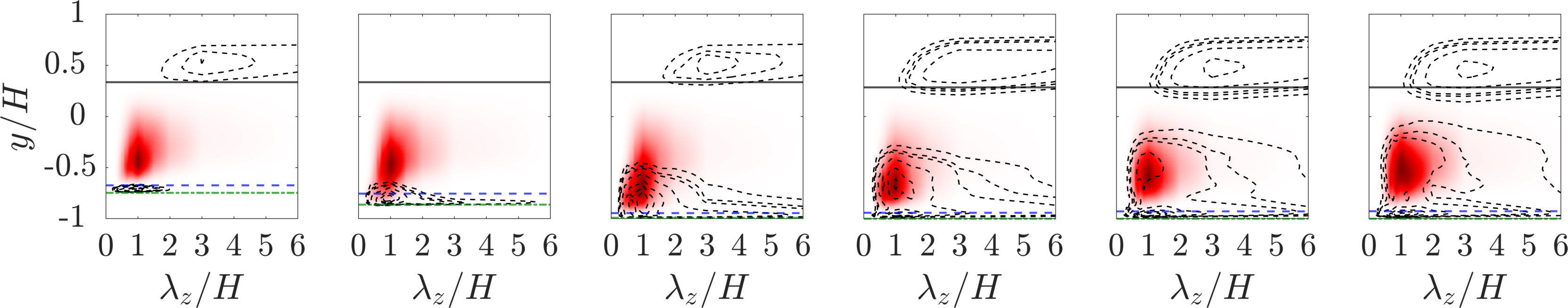}
\includegraphics[width=0.99\textwidth,trim={0 1.2cm 0 0},clip]{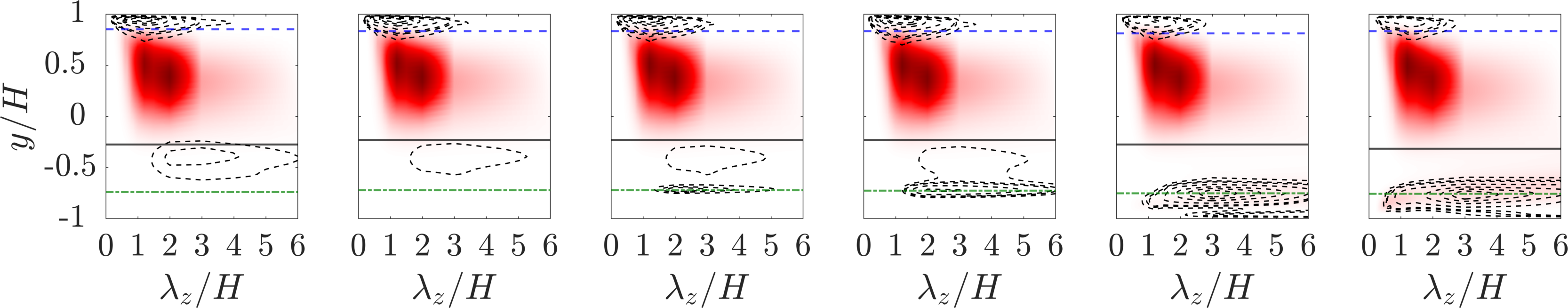}
\includegraphics[width=0.99\textwidth]{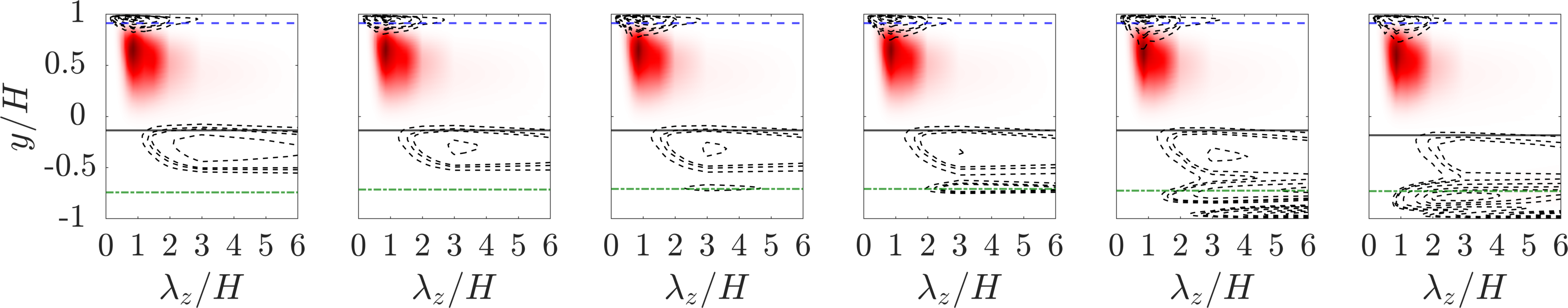}
\caption{Spanwise premultiplied energy spectra of the wall-normal ($k_z\Phi_{vv}$ and streamwise ($k_z\Phi_{uu}$) fluctuating velocity at (from left to right) $x/H = $ 4, 5, 5.5, 6, 7, and 8. Rows from top to bottom correspond to cases 00, P04, P10, N04, and N10, as in preceding figures. Contour represents $k_z\Phi_{vv}$, while the dashed contour lines represent $k_z\Phi_{uu}$. Spectra are normalized by their maximum values. Contour levels of 0.05, 0.08, 0.1, 0.2, 0.4, 0.6, and 0.8 are displayed for $k_z\Phi_{uu}$. The horizontal solid (\solid) black line represents the wall-normal location of maximum velocity, the dashed (\dashed) blue line represents the wall-normal location of $S=-0.5$, and the dash-dot (\dashdot) green line represents the wall-normal location of the maximum velocity gradient.}
\label{fig:premulti}
\end{figure}

\subsection{Premultiplied spanwise spectrum}
The spanwise wavelength and wall-normal location of the embedded structures are quantified by the premultiplied energy spectrum of velocity fluctuations in figure \ref{fig:premulti}. Note that each spectrum is normalized by its peak magnitude. Without rotation, the peaks of the \uu and \vv spectrum collapse at the location of the inflection point, representing the generation of roller vortices. The primary wavelength $\sim0.7H$. In wall units, it corresponds to $\lambda_z^+\sim 110$, the spanwise-spacing of near-wall streaks in canonical channel flows. This indicates the interaction between low-speed streaks and the separating shear layer.  After the flow reattaches, the spanwise wavelength is reduced as three-dimensionality develops.

For the four rotating cases, the \vv spectrum shows a peak away from the anti-cyclonic wall, which represents the TG vortices. The top and bottom limits of the TG vortices can be determined by the wall-normal locations where the outer peak of the \vv spectrum diminishes. Our results show that the former is represented by the location of the peak streamwise velocity, and the latter is correlated to where $S=-0.5$. The spectrum indicates the TG vortices persist across the entire channel including over the bump where they get displaced. They have a characteristic spanwise wavelength of $1.5H$ at the lower rotation rate (\textit{i.e.}, four pairs over the spanwise extent of the domain) and $1.0H$ at the higher rotation rate (\textit{i.e}., six pairs). 
The core of the TG vortices, denoted by the location of the peak of the spectrum, moves closer to the anticyclonic wall and the wall-normal extent reduces as the rotation number increases. These are consistent with previous studies \citep{Kristoffersen93, Dai16, Brethouwer17}. 

The augmentation and streamwise evolution of the hairpin vortices associated with the ejection mechanism described above is quantified by streamwise development of the spectra.
Focusing initially on case P04, near-wall peaks of \uu and \vv spectra develop at $x/H=5$ and $x/H = 5.5$, respectively. These peaks lie between the inflection point of the shear layer and beneath the TG vortices, indicating that they represent the hairpin vortices and not rollers associated with the separation. The peaks occur in the self-exciting regime ($S>-0.5$), indicating that \uu and \vv will be enhanced. At this streamwise location, the characteristic wavelength of the inner peaks is lower than that of the TG vortices. Downstream of reattachment ($x/H= 6$), the inner peaks of \uu and \vv shift away from the wall. Further downstream, the \vv spectrum merges into a single peak, the \uu spectra overlies that of $\overline{v'v'}$, indicating a strong correlation between the two.  The characteristic wavelengths of the \uu and \vv spectra are now corresponding to that of the TG vortices. This continuous trajectory of the \uu and \vv peaks away from the wall moving downstream from the bump corresponds to the ejection of the hairpin filaments by the TG vortices. 
For case P10, the dynamics are similar, yet the wall-normal extent of the ejection process is less than P04 (as expected due to the smaller, more confined TG vortices at the high rotation rate). Due to the lower characteristic wavelength of the TG vortices in case P10, there is no scale separation between them and the hairpin vortices. However, the ejection process described by spreading and upward trajectory of the \uu remains qualitatively consistent with case P04, indicating that the same ejection process is occurring.

Also noteworthy is that on the cyclonic side, the \uu spectrum shows a peak at $\lambda_z\sim 3H$, representing the oblique wave. The center of this wave is near the peak streamwise velocity around $y/H=0.5$.  The characteristic wavelength of the oblique wave is consistent at the high rotation rate.

When subject to negative rotation, the TG vortices appear on the top (anti-cyclonic) side of the channel, and therefore do not interact with the separating shear layer. The separating shear layer is also not at the same wall distance as the oblique waves. In case N04, the two are slightly nearer since peak velocity occurs closer to the cyclonic wall. Yet, the spanwise wavelength of the roller vortices in the laminar separating shear layer matches that of the oblique waves. It suggests that the latter is the primary source of disturbance for the three-dimensionality of the shear layer. 

\section{Discussion and concluding remarks}
The present flow configuration is designed to permit the investigation of rotation effects on the onset of pressure-induced flow separation and the full recovery process, complementing existing literature on fixed-point geometry-induced separation. 
Changes in the mean separation region, drag, and recovery of the bump wake with the rotation direction and rate are reported.  
Special emphasis is placed on justifying the factors that cause these changes. Additionally, the structural and dynamical mechanisms underlying the variations in turbulent statistics during the separation and recovery of the wake of the bump are discussed. 

Several changes of the mean separation region observed in this study align with existing studies of rotating geometry-induced separating flows. Specifically, when separation occurs on the anti-cyclonic side the flow reattaches earlier than the non-rotating case. The opposite happens when the bump is on the cyclonic side. The mild curvature of the bump in this study allows the onset of separation to vary with the rotation rate. We observed that the mean separation occurs earlier in cyclonic rotation and is delayed under anti-cyclonic rotation. Under negative rotation, the separating flow features a laminar separating shear layer more than twice longer than the one in the non-rotating case.  At the highest positive rotation rate, the separation region is nearly completely diminished. This is different from the geometry-induced separation reported in the literature which always separates at a fixed point and cannot be monotonically suppressed through rotation. 

Quantifying the initiation of separation through the mean momentum budget reveals that the skewed velocity profile, characterized as a mean momentum deficit (MMD) in comparison to the non-rotating case, is the key factor influencing the separation for the current flow.
In all rotation rates explored in this study, there is a reduction in the mean streamwise velocity on both sides of the channel when compared to the non-rotating case. The constant bulk velocity is maintained through the increased peak velocity near channel centerline.
The augmented MMD indicates that the near-wall flow is more prone to deceleration and detachment. However, the MMD reduces the adverse pressure gradient (APG), the primary decelerating force. The interplay between these two counteracting effects is a critical determinant in the delay or promotion of the onset of flow separation. In terms of turbulent mixing, the relaminarization by the cyclonic rotation disables the fluctuating momentum transfer, thereby promoting flow separation.  
The modulation of the turbulence by anti-cyclonic rotation, on the contrary, delays separation by bringing high momentum from the outer flow to the near-wall region. Yet, this role is only significant at high rotation rate when the APG is remarkably reduced by the extreme MMD to the comparable low magnitude.  

The MMD also impacts the drag generated by the bump. Conventional notion is that a smaller separation region implies improved pressure recovery and, consequently, reduced drag. Our results challenge this assumption. In our cases, a lower drag is observed for the long cyclonic separation regions compared to the small anti-cyclonic ones.
Analysis of the force balance highlights the importance of considering the drag produced by a protrusion as a net outcome between the drag on the wind side and the thrust on the lee side--both of which are subject to variation with the flow conditions. In the present study, the MMD leads to a reduction in the drag on the wind side as well. This reduction emerges as the leading factor resulting in a reduction in the total drag produced by the bump. 
Conversely, the variation in thrust on the lee side, associated with the changes in the separation region size, contributes to the total drag to a lesser extent.

The evolution of the separated shear layer and the reattached flow are significantly influenced by the stability regime of the flow, which shows a 2D spatial variation due to the deceleration and recovery. 
Cyclonic rotation stabilizes the flow and results in a quasi-laminar separating shear layer. Because the Coriolis force counteracts the redistribution of turbulent kinetic energy (TKE) from \uu to $\overline{v'v'}$, the rolling up of the separating shear layer by the Kelvin–Helmholtz (KH) instability is suppressed, resulting in delayed reattachment. 
When the rotation is anti-cyclonic, the deceleration by APG increases $S$ behind the bump, leading to the establishment of a self-exciting destabilization region characterized by $-0.5<S<0$: the Coriolis force effectively extracts energy from the mean flow and only a portion of this extra gain of \uu is redistributed to $\overline{v'v'}$. The enhanced turbulence promotes the diffusion of momentum and thus reattachment of the flow. The key structural change in this region is the augmentation of the legs of the hairpin vortices who lay between the Taylor-G\"ortler (TG) vortices and the separated shear layer. They enhance the diffusion of momentum via rapid distortion of the separating shear layer and thus promote the reattachment of the flow. These hairpin vortices experience a prolonged enhancement downstream of the bump as the mean flow gradually recovers (and $S$ decreases such that the detablization transits to a self-restraining regime). 
As a result, they are ejected further into the outer flow by the upwash between TG vortex pairs. These processes are exhibited as elevated \vv and \ww regions expanding from the separating shear layer near the bottom wall to the channel centerline more than 15 bump heights downstream. The ejected hairpin vortices are more efficient in mixing the momentum than the conventional near-wall ones, thus make the wall skin friction recover within a shorter distance.

In this process, the TG vortices appears to be passive structures that are not affected by the hairpin vortices and the separating shear layer which lie underneath. At the higher rotation rate, TG vortices reside closer to the wall and becomes weaker. However, in the highest rotation rate examined in this study, they still enable a substantial ejection of the hairpin vortices and the associated augmentation of \vv and $\overline{w'w'}$.   

While these mechanisms are initiated by the deceleration and flow separation, flow reversal does not appears as a necessary condition. Regardless of whether separation occurs, any change of the mean shear in a rotating flow will result in flow stability modulation. The neutrally stable region in the anti-cyclonic side will become hydrodynamically unstable when the velocity gradient is increased via an APG or a freestream gust. Then, the ejection of hairpin vortices by the TG vortices and their prolonged augmentation will follow, leading to enhanced turbulence.
This has the potential to influence skin-friction, heat transfer, and scalar transport in a variety of engineering settings, including centrifugal pumps and hydroturbines. Moreover, as rotating mean shear flows are analogous to thermal convection and boundary layer over concave walls, these mechanism may also augment the thermal plumes and hairpin vortices in such applications.

The present study is limited to a single geometrical configuration and relatively low Reynolds number. The quantitative behavior of flow separation behind a protrusion in rotating systems is likely to depend on the geometry of the protrusion. Despite that the role of mean momentum deficit in modulating the pressure gradient and flow stability is expected to be valid in general, the quantitative effects may vary in different flow configurations and conditions. 
Further work could focus on these areas.

\section*{Acknowledgement}
The authors acknowledge the support from NSF EPSCoR Track-4 Research Fellowship Grant OIA-2131942, monitored by Dr H. Luo. B.S. appreciates the support of the NSF GRFP Award 2235036. The simulations were performed at the Texas Advanced Computing Center (TACC) Stampede-2 cluster and the San Diego Supercomputer Center (SDSC) Expanse cluster.

\section*{Declaration of Interests}
The authors report no conflict of interest.

\section*{Appendix: Mean momentum and Reynolds stress budgets} 
\label{sec:app_RS}
The mean momentum equation considering homogeneous flow in the spanwise direction, as given in \S\ref{sec:mombud}, reads:
\begin{equation}
    0 = -\frac{\partial P}{\partial x_i} + A_i + D_i + R_i + G_i + F_i.
    \label{eqn:meanMomApp}
\end{equation}
The terms from left to right are the mean pressure gradient ($-\partial P/\partial x_i$), mean convection ($A_i$), mean viscous diffusion ($D_i$), mean Reynolds stress divergence ($R_i$), mean Coriolis force ($G_i$), and mean immersed boundary method force ($F_i$). The terms (excluding pressure gradient and immersed boundary force) read as follows (non-dimensionalized by $U_b^2/H$):
\begin{equation}
    A_i = -U_k \frac{\partial U_i}{\partial x_k}
    \label{eqn:Amean}
\end{equation}
\begin{equation}
    D_i = \frac{1}{Re_b} \dpr{}{x_k} \frac{\partial  U_i}{\partial x_k} 
    \label{eqn:Dmean}
\end{equation}
\begin{equation}
    R_i = -\frac{\partial \overline{u'_iu'_k}}{\partial x_k}
    \label{eqn:Rmean}
\end{equation}
\begin{equation}
    G_i = -Ro_b \epsilon_{i3k} U_k.
    \label{eqn:Gmean}
\end{equation}
Here, $\epsilon_{i3k}$ is the Levi-Civita symbol.

The budget of time-averaged Reynolds stresses in a rotating flow reads:
\begin{equation}
0 = -C_{ij} + P_{ij} + \Pi_{ij} + T_{ij} + D_{ij} - \varepsilon_{ij} + G_{ij}.
\label{eqn:Bud}
\end{equation}
The terms from left to right represent convection (note the negative as this term arises from the material derivative on the left-hand-side (LHS) of the transport equation for time-averaged Reynolds stresses), production, velocity-pressure-gradient correlation, turbulent transport, viscous diffusion, dissipation, and the rotational term which results as a product of the Coriolis term in (\ref{eq:ns2}). The terms read as follows (non-dimensionalized by $U_b^3/H$):

\begin{equation}
-C_{ij} = -U_k\frac{\overline{\partial u'_i u'_j}}{\partial x_k}
\label{eqn:C}
\end{equation}
\begin{equation}
P_{ij} = -\left( \overline{u'_i u'_k}\frac{\partial U_j}{\partial x_k} + \overline{u'_j u'_k}\frac{\partial U_i}{\partial x_k} \right)
\label{eqn:P}
\end{equation}
\begin{equation}
\Pi_{ij} = -\left( \overline{u'_i\frac{\partial p'}{\partial x_j} + u'_j \frac{\partial p'}{\partial x_i} }\right)
\label{eqn:Pi}
\end{equation}
\begin{equation}
T_{ij} = -\frac{\partial \left( \overline{u'_i u'_j u'_k}\right)}{\partial x_k}
\label{eqn:T}
\end{equation}
\begin{equation}
D_{ij} = \frac{1}{Re_b} \dpr{}{x_k} \frac{\partial  \left( \overline{u'_i u'_j} \right)}{\partial x_k}
\label{eqn:D}
\end{equation}
\begin{equation}
\varepsilon_{ij} = \frac{2}{Re_b}\overline{\frac{\partial u'_i}{\partial x_k} \frac{\partial u'_j}{\partial x_k}  }
\label{eqn:eps}
\end{equation}
\begin{equation}
G_{ij} = - Ro_b \left( \epsilon_{i3k} \overline{u'_j u'_k} + \epsilon_{j3k} \overline{u'_i u'_k} \right).
\label{eqn:G}
\end{equation}
A widely reported feature of the rotational effects on the Reynolds stresses budgets is that the rotational production term ($G_{ij}$) appears in the $\overline{u'u'}$, $\overline{v'v'}$, and $\overline{u'v'}$ budgets (see table \ref{tab:Prod}). In particular, on the anti-cyclonic side of the channel, $G_{uu} = -G_{vv} < 0$, redistributing energy from $\overline{u'u'}$ to $\overline{v'v'}$ whereas on the cyclonic side the opposite occurs. This redistribution has been used to characterize the stabilizing and destabilizing effects of rotation on turbulent channel flows \citep{Tafti91, Andersson95, Barri10, Brethouwer17, Wu19}.

\bibliographystyle{jfm}
\bibliography{ref_ben}

\end{document}